\documentclass{emulateapj}
\usepackage{graphicx}
\usepackage[flushleft]{threeparttable}

\shorttitle{ALMA view of UV-selected $z\geq2$ galaxies}
\shortauthors{Bouwens et al.}

\def\lsim{\mathrel{\rlap{\lower 3pt \hbox{$\sim$}} \raise 2.0pt \hbox{$<$}}}
\def\gsim{\mathrel{\rlap{\lower 3pt \hbox{$\sim$}} \raise 2.0pt \hbox{$>$}}}

\begin{document}

\title{ALMA Spectroscopic Survey in the Hubble Ultra Deep Field: The
  Infrared Excess of $UV$-selected $z=2$-10 galaxies as a function of
  $UV$-continuum Slope and Stellar Mass}

\author{
Rychard Bouwens\altaffilmark{1}, Manuel Aravena\altaffilmark{2}, Roberto Decarli\altaffilmark{3}, Fabian Walter\altaffilmark{3,4,5}, Elisabete da Cunha\altaffilmark{6,7}, Ivo Labb{\'e}\altaffilmark{1}, Franz Bauer\altaffilmark{8,9,10}, Frank Bertoldi\altaffilmark{11}, Chris Carilli\altaffilmark{5,12}, Scott Chapman\altaffilmark{13}, Emanuele Daddi\altaffilmark{14}, Jacqueline Hodge\altaffilmark{1}, Rob Ivison\altaffilmark{15,16}, Alex Karim\altaffilmark{11}, Olivier Le Fevre\altaffilmark{17}, Benjamin Magnelli\altaffilmark{11}, Kazuaki Ota\altaffilmark{12}, Dominik Riechers\altaffilmark{18}, Ian Smail\altaffilmark{19}, Paul van der Werf\altaffilmark{1}, Axel Weiss\altaffilmark{11}, Pierre Cox\altaffilmark{20}, David Elbaz\altaffilmark{14}, Jorge Gonzalez-Lopez\altaffilmark{8}, Leopoldo Infante\altaffilmark{8}, Pascal Oesch\altaffilmark{21}, Jeff Wagg\altaffilmark{22}, Steve Wilkins\altaffilmark{23}}
\altaffiltext{1}{Leiden Observatory, Leiden University, NL-2300 RA Leiden, Netherlands}
\altaffiltext{2}{N\'{u}cleo de Astronom\'{\i}a, Facultad de Ingenier\'{\i}a, Universidad Diego Portales, Av. Ej\'{e}rcito 441, Santiago, Chile}
\altaffiltext{3}{Max-Planck Institut f\"{u}r Astronomie, K\"{o}nigstuhl 17, D-69117, Heidelberg, Germany.}
\altaffiltext{4}{Astronomy Department, California Institute of Technology, MC105-24, Pasadena, California 91125, USA}
\altaffiltext{5}{NRAO, Pete V.\,Domenici Array Science Center, P.O.\, Box O, Socorro, NM, 87801, USA}
\altaffiltext{6}{Centre for Astrophysics and Supercomputing, Swinburne University of Technology, Hawthorn, Victoria 3122, Australia}
\altaffiltext{7}{Research School of Astronomy and Astrophysics, Australian National University, Canberra, ACT 2611, Australia}
\altaffiltext{8}{Instituto de Astrof\'{\i}sica, Facultad de F\'{i}sica, Pontificia Universidad Cat\'{o}lica de Chile, Casilla 306, Santiago 22, Chile} 
\altaffiltext{9}{Millennium Institute of Astrophysics (MAS), Nuncio Monse\~{n}or S\'{o}tero Sanz 100, Providencia, Santiago, Chile} 
\altaffiltext{10}{Space Science Institute, 4750 Walnut Street, Suite 205, Boulder, Colorado 80301} 
\altaffiltext{11}{Argelander Institute for Astronomy, University of Bonn, Auf dem H\"{u}gel 71, 53121 Bonn, Germany}
\altaffiltext{12}{Cavendish Laboratory, University of Cambridge, 19 J J Thomson Avenue, Cambridge CB3 0HE, UK}
\altaffiltext{13}{Dalhousie University, Halifax, Nova Scotia, Canada}
\altaffiltext{14}{Laboratoire AIM, CEA/DSM-CNRS-Universite Paris Diderot, Irfu/Service d'Astrophysique, CEA Saclay, Orme des Merisiers, 91191 Gif-sur-Yvette cedex, France}
\altaffiltext{15}{European Southern Observatory, Alonso de Cordova 3107, Casilla 19001, Vitacura Santiago, Chile}
\altaffiltext{16}{Institute for Astronomy, University of Edinburgh, Royal Observatory, Blackford Hill, Edinburgh EH9 3HJ, UK}
\altaffiltext{17}{IRAM, 300 rue de la piscine, F-38406 Saint-Martin d'H\`eres, France}
\altaffiltext{18}{Cornell University, 220 Space Sciences Building, Ithaca, NY 14853, USA}
\altaffiltext{19}{Centre for Extragalactic Astronomy, Department of Physics, Durham University, South Road, Durham, DH1 3LE, UK}
\altaffiltext{20}{Joint ALMA Observatory - ESO, Av. Alonso de Cordova,
  3104, Santiago, Chile}
\altaffiltext{21}{Astronomy Department, Yale University, New Haven, CT 06511, U
SA}
\altaffiltext{22}{SKA Organisation, Lower Withington, UK}
\altaffiltext{23}{Astronomy Centre, Department of Physics and Astronomy, University of Sussex, Brighton, BN1 9QH, UK}

\begin{abstract}
We make use of deep $1.2\,$mm-continuum observations (12.7$\mu$Jy/beam
RMS) of a 1 arcmin$^2$ region in the Hubble Ultra Deep Field (HUDF) to
probe dust-enshrouded star formation from 330 Lyman-break galaxies
spanning the redshift range $z=2$-10 (to $\sim$2-3 M$_{\odot}$/yr at
$1\sigma$ over the entire range).  Given the depth and area of ASPECS,
we would expect to tentatively detect 35 galaxies extrapolating the
Meurer $z\sim0$ IRX-$\beta$ relation to $z\geq2$ (assuming $T_d\sim35$
K).  However, only 6 tentative detections are found at $z\gtrsim2$ in
ASPECS, with just three at $>$3$\sigma$.  Subdividing $z=2$-10
galaxies according to stellar mass, $UV$ luminosity, and
$UV$-continuum slope and stacking the results, we only find a
significant detection in the most massive ($>$$10^{9.75}$ $M_{\odot}$)
subsample, with an infrared excess (IRX=$L_{IR}/L_{UV}$) consistent
with previous $z\sim2$ results.  However, the infrared excess we
measure from our large selection of sub-$L^*$ ($<$10$^{9.75}$
$M_{\odot}$) galaxies is 0.11$_{-0.42}^{+0.32}$$\pm$0.34 (bootstrap
and formal uncertainties) and 0.14$_{-0.14}^{+0.15}$$\pm$0.18 at
$z=2$-3 and $z=4$-10, respectively, lying below even an SMC
IRX-$\beta$ relation (95\% confidence).  These results demonstrate the
relevance of stellar mass for predicting the IR luminosity of
$z\gtrsim2$ galaxies.  We find that the evolution of the IRX-stellar
mass relationship depends on the evolution of the dust temperature.
If the dust temperature increases monotonically with redshift
($\propto (1+z)^{0.32}$) such that $T_d \sim 44$-50 K at $z\geq4$,
current results are suggestive of little evolution in this
relationship to $z\sim6$.  We use these results to revisit recent
estimates of the $z\geq3$ star-formation rate density.


\end{abstract}

\section{Introduction}

One particularly interesting and long-standing focus of galaxy studies
has been quantifying the total energy output and stellar birth rate in
galaxies across cosmic time.  The first significant investigations
became possible following systematic searches for galaxies to
intermediate and high redshifts $z\sim3$-4 (Lilly et al.\ 1996; Madau
et al.\ 1996).  The immediate picture was that the star formation rate
density likely peaked around $z\sim2$-4 (Madau et al.\ 1996; Steidel
et al.\ 1999).  Using photometric searches for Lyman-break galaxies at
$z>4$, it later became clear that there was a significant drop in the
star formation density to $z>4$ (Dickinson 2000; see review by Madau
\& Dickinson 2014).  Rough constraints now exist on the star formation
rate density to $z\sim10$ (e.g., Bouwens et al.\ 2011; Ellis et
al.\ 2013; Oesch et al.\ 2014, 2015; Bouwens et al.\ 2015; McLeod et
al.\ 2015; Laporte et al.\ 2016).

\begin{figure*}
\epsscale{1.15}
\plottwo{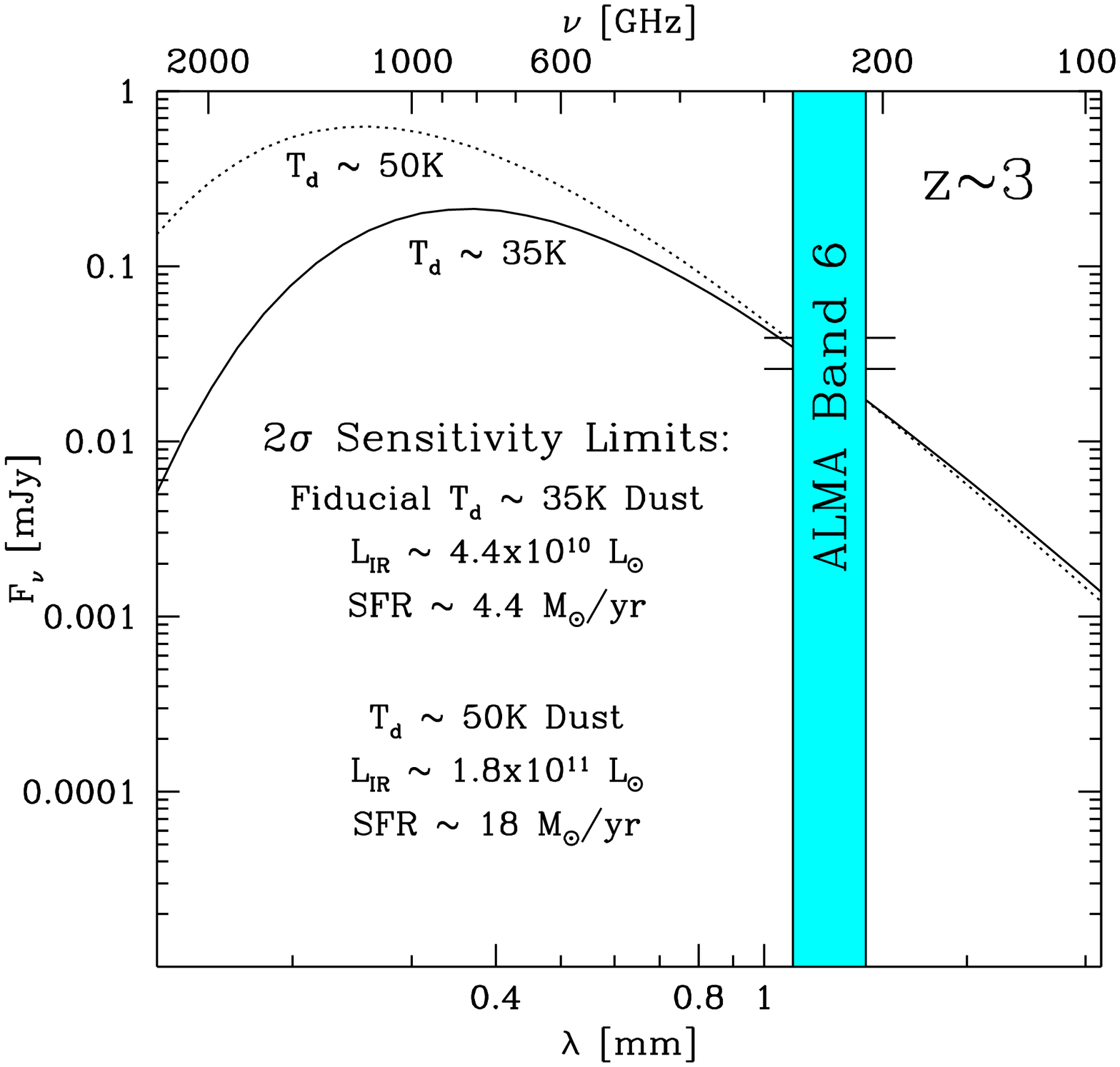}{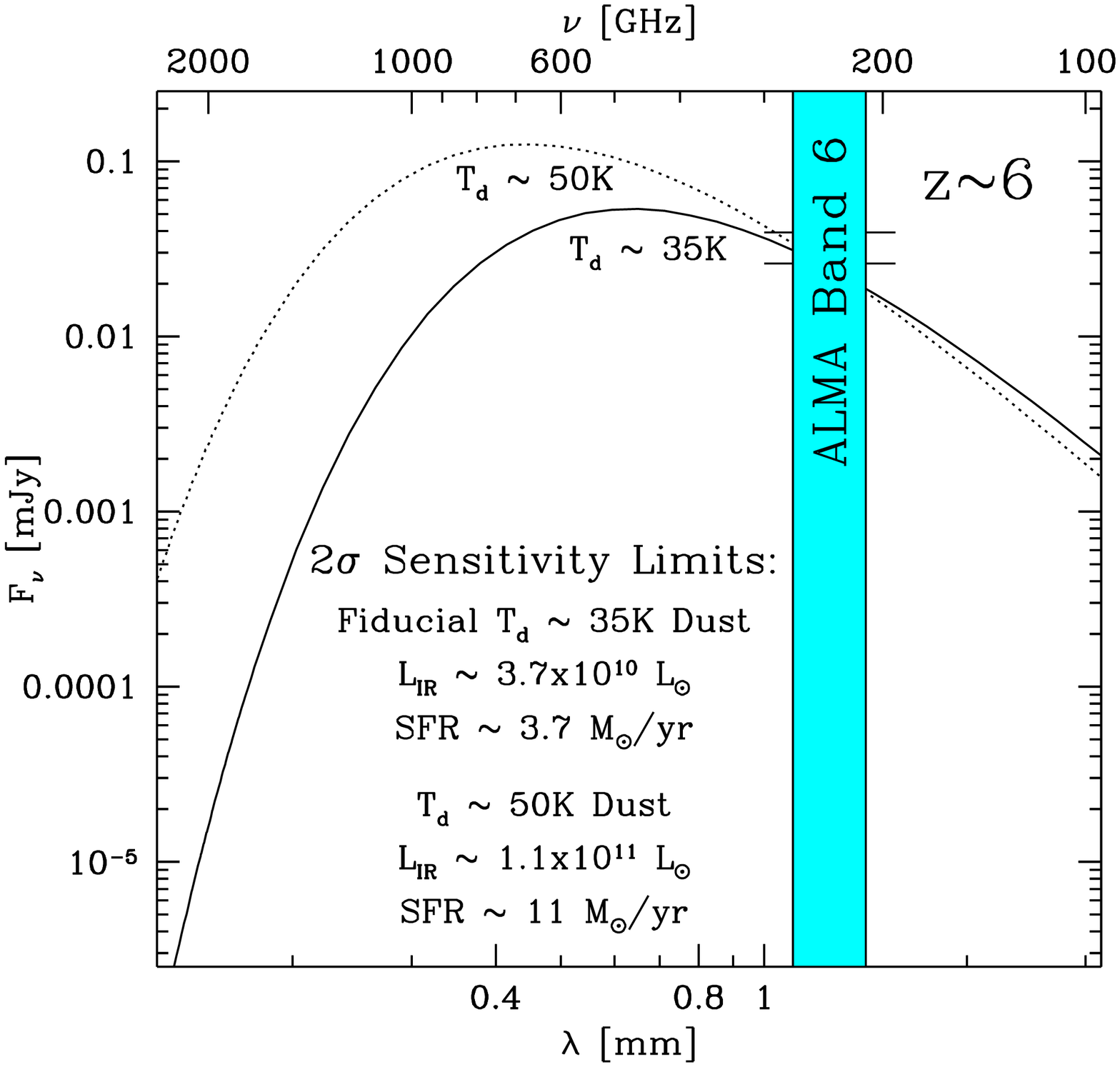}
\caption{Model SEDs in the far IR for the faintest $z\sim3$ and
  $z\sim6$ galaxies we would be able to obtain tentative individual
  detections ($>2\sigma$) in our deep ALMA band-6 observations.  The
  plotted SED are modified blackbody SEDs with a dust temperature of
  35 K and a dust emissivity power-law spectral index of $\beta_d =
  1.6$ (Eales et al.\ 1989; Klaas et al.\ 1997).  The two short
  horizontal lines at 1.2mm show our 2$\sigma$ and $3\sigma$
  sensitivity limits.  At both $z\sim3$ and $z\sim6$, we would expect
  to tentatively detect individual galaxies to approximately the same
  IR luminosity $\sim4$$\times10^{10}$ $L_{\odot}$, which is
  equivalent to an obscured star-formation rate (SFR) of $\sim$4
  M$_{\odot}$/yr.  For galaxies with dust temperatures of 50 K, we
  would expect to tentatively galaxies to a IR luminosity
  $\sim1.8\times10^{11}$ $L_{\odot}$ and $\sim1.1\times10^{11}$
  $L_{\odot}$ at $z\sim3$ and $z\sim6$, equivalent to obscured SFRs of
  $\sim18$ and $\sim11$ $M_{\odot}$/yr, respectively.  These
  sensitivities account for the impact of the CMB (da Cunha et
  al.\ 2013).
\label{fig:showsed36}}
\end{figure*}

\begin{deluxetable*}{cccccccccc}
\tablewidth{0cm}
\tablecolumns{10}
\tabletypesize{\footnotesize}
\tablecaption{$2\sigma$ Sensitivity Limits for our Probe of Obscured Star Formation from Individual $z\gtrsim2$ Galaxies and the Dependence on SED\label{tab:tempdep}}
\tablehead{\colhead{Far-Infrared} & \multicolumn{9}{c}{2$\sigma$ Sensitivity Limits ($10^{10}$ $L_{\odot}$)}\\
\colhead{SED Model} & \colhead{$z$$\sim$2} & \colhead{$z$$\sim$3} & \colhead{$z$$\sim$4} & \colhead{$z$$\sim$5} & \colhead{$z$$\sim$6} & \colhead{$z$$\sim$7} & \colhead{$z$$\sim$8} & \colhead{$z$$\sim$9} & \colhead{$z$$\sim$10}}
\startdata
35K modified blackbody\tablenotemark{a} (\textit{fiducial}) & 5.0 & 4.4 & 4.0 & 3.7 & 3.7 & 3.8 & 4.2 & 4.8 & 6.0 \\
Modified blackbody with evolving $T_d$\tablenotemark{b} & 6.3 & 7.8 & 8.8 & 9.5 & 10.1 & 10.2 & 9.5 & 9.2 & 9.1 \\
25K modified blackbody\tablenotemark{a} & 1.4 & 1.4 & 1.4 & 1.6 & 1.9 & 2.7 & 4.1 & 6.9 & 12.3 \\
30K modified blackbody\tablenotemark{a} & 2.7 & 2.5 & 2.4 & 2.4 & 2.5 & 2.9 & 3.6 & 4.9 & 7.0 \\
40K modified blackbody\tablenotemark{a} & 8.5 & 7.3 & 6.3 & 5.7 & 5.3 & 5.2 & 5.3 & 5.7 & 6.4 \\
45K modified blackbody\tablenotemark{a} & 13.9 & 11.6 & 9.8 & 8.5 & 7.7 & 7.3 & 7.1 & 7.2 & 7.5 \\
50K modified blackbody\tablenotemark{a} & 21.6 & 17.6 & 14.6 & 12.4 & 11.0 & 10.1 & 9.5 & 9.3 & 9.3 \\
NGC6946\tablenotemark{c} & 1.3 & 1.3 & 1.4 & 1.6 & 1.9 & 2.5 & 3.5 & 5.4 & 8.7 \\
M51\tablenotemark{c} & 1.4 & 1.4 & 1.5 & 1.6 & 1.9 & 2.5 & 3.4 & 5.2 & 8.2 \\
Arp220\tablenotemark{c} & 7.4 & 6.6 & 6.1 & 5.6 & 5.4 & 5.5 & 5.7 & 5.9 & 6.6 \\
M82\tablenotemark{c} & 11.8 & 10.4 & 9.4 & 8.7 & 8.4 & 8.4 & 8.6 & 9.1 & 9.9 \\\\
  & \multicolumn{9}{c}{2$\sigma$ Limit for Probes of the Obscured SFR (M$_{\odot}$/yr)\tablenotemark{d}}\\
 SED Model & $z$$\sim$2 & $z$$\sim$3 & $z$$\sim$4 & $z$$\sim$5 & $z$$\sim$6 & $z$$\sim$7 & $z$$\sim$8 & $z$$\sim$9 & $z$$\sim$10 \\
\tableline
35K modified blackbody\tablenotemark{a} (\textit{fiducial}) & 5.0 & 4.4 & 4.0 & 3.7 & 3.7 & 3.8 & 4.2 & 4.8 & 6.0 \\
Modified blackbody with evolving $T_d$\tablenotemark{b} & 6.3 & 7.8 & 8.8 & 9.5 & 10.1 & 10.0 & 9.4 & 9.0 & 8.9 \\
25K modified blackbody\tablenotemark{a} & 1.4 & 1.4 & 1.4 & 1.6 & 1.9 & 2.7 & 4.1 & 6.9 & 12.3 \\
30K modified blackbody\tablenotemark{a} & 2.7 & 2.5 & 2.4 & 2.4 & 2.5 & 2.9 & 3.6 & 4.9 & 7.0 \\
40K modified blackbody\tablenotemark{a} & 8.5 & 7.3 & 6.3 & 5.7 & 5.3 & 5.2 & 5.3 & 5.7 & 6.4 \\
45K modified blackbody\tablenotemark{a} & 13.9 & 11.6 & 9.8 & 8.5 & 7.7 & 7.3 & 7.1 & 7.2 & 7.5 \\
50K modified blackbody\tablenotemark{a} & 21.6 & 17.6 & 14.6 & 12.4 & 11.0 & 10.1 & 9.5 & 9.3 & 9.3 \\
NGC6946\tablenotemark{c} & 1.3 & 1.3 & 1.4 & 1.6 & 1.9 & 2.5 & 3.5 & 5.4 & 8.7 \\
M51\tablenotemark{c} & 1.4 & 1.4 & 1.5 & 1.6 & 1.9 & 2.5 & 3.4 & 5.2 & 8.2 \\
Arp220\tablenotemark{c} & 7.4 & 6.6 & 6.1 & 5.6 & 5.4 & 5.5 & 5.7 & 5.9 & 6.6 \\
M82\tablenotemark{c} & 11.8 & 10.4 & 9.4 & 8.7 & 8.4 & 8.4 & 8.6 & 9.1 & 9.9 
\enddata
\tablenotetext{a}{Standard modified blackbody form (e.g., Casey 2012)
  with a dust emissivity power-law spectral index of $\beta_d = 1.6$
  (Eales et al.\ 1989; Klaas et al.\ 1997).}
\tablenotetext{b}{Assuming dust temperature $T_d$ evolves as
  (35 K)$((1+z)/2.5)^{0.32}$ (Bethermin et al.\ 2015) such that $T_d
  \sim 44$-50 K at $z\sim4$-6.  See \S3.1.3.}
\tablenotetext{c}{Empirical SED template fits to specific galaxies
  in the nearby universe (Silva et al.\ 1998).}
\tablenotetext{d}{Using the conversion $\textrm{SFR} = L_{IR} /
  (10^{10} L_{\odot})$ appropriate for a Chabrier IMF (Kennicutt 1998;
  Carilli \& Walter 2013).}
\end{deluxetable*}

In spite of huge progress in mapping the SFR density from $z\sim0$ to
$z\sim11$ from surveys in the rest-frame $UV$, our understanding of
the energy output from $UV$-selected galaxies at far-infrared
wavelengths is most developed over the redshift range $z\sim0$ to
$z\sim3$, with increasing maturity (Reddy et al.\ 2008; Daddi et
al.\ 2009; Magnelli et al.\ 2009, 2011, 2013; Karim et al.\ 2011;
Cucciati et al.\ 2012; {\'A}lvarez-M{\'a}rquez et al.\ 2016).  This
general picture of the energy output from galaxies at rest-frame $UV$
and far-IR wavelengths has been confirmed by many independent probes
at X-ray and radio wavelengths and using the H$\alpha$ emission line,
with consistent results (e.g., Erb et al.\ 2006b; Reddy et al.\ 2004,
2006, 2010; Daddi et al.\ 2007).

The observational situation becomes much more uncertain when one
considers dust-enshrouded star formation at $z>3$, as use of standard
techniques or facilities becomes more difficult (owing to the PAH
features shifting out of the {\it Spitzer}/MIPS 24$\mu$m band or
increasing source confusion in {\it Spitzer} or {\it Herschel}
70-500$\mu$m observations), effectively limiting studies to the
brightest, most intensely star-forming sources at $z>3$ (e.g., HFLS3:
Riechers et al.\ 2013).  As a result of these challenges, various
researchers (e.g., Meurer et al.\ 1999 [M99]; Daddi et al.\ 2007;
Reddy et al.\ 2006; Bouwens et al.\ 2007, 2009, 2012) have made use of
well-known $z\sim0$ correlations, i.e., the infrared excess (IRX =
$L_{IR}/L_{UV}$) $UV$-continuum slope ($\beta)$ relationship, to
estimate dust extinction in more typical $z\geq 3$ galaxies based on
the $UV$ slopes measured from the observed $UV$ data.  Such
correlations have been confirmed to apply out to $z\sim2$ using a
variety of multi-wavelength data (Reddy et al.\ 2006, 2008; Daddi et
al.\ 2007; Daddi et al.\ 2009; Pannella et al.\ 2009), but it is
unclear if they apply at $z\geq 3$.

Despite the power of stacking, e.g., with the SCUBA2 or {\it Herschel}
data, more direct measurements of the dust-enshrouded star formation
have required the advent of some new or upgraded facilities (e.g.,
Atacama Large Millimeter Array [ALMA]; NOrthern Extended Millimeter
Array [NOEMA]).  The first results from these facilities indicated
that there was no detectable dust-enshrouded star formation in normal
or even extremely bright $UV$-selected $z>5$ galaxies (Walter et
al.\ 2012; Ouchi et al.\ 2013; Ota et al.\ 2014;
Gonz{\'a}lez-L{\'o}pez et al.\ 2014; Riechers et al.\ 2014; Maiolino
et al.\ 2015).  Later, some bright $z\sim5$-6 galaxies were weakly
detected in the far-IR continuum (Capak et al.\ 2015; Willott et
al.\ 2015), but at lower levels than seen in similar galaxies at lower
redshifts using well known lower-redshift $z\sim0$-2 IRX-$\beta$
relations.

While current results generally suggest much lower dust emission from
$z>3$ galaxies than expected based on $z=0$-2 IRX-$\beta$ relations,
the number of $z>3$ sources where such constraints are available
remains modest.  This is especially the case when one only considers
sources which can be confidently placed in $z>3$ samples and which
have accurately-measured $UV$ continuum slopes, stellar masses, or
SFRs.  As such, it is clearly helpful to obtain deep continuum
observations with ALMA over fields with substantial amounts of legacy
observations at other wavelengths from {\it HST}, {\it Spitzer}, {\it
  Herschel}, and ground-based observatories.

Fortunately, with our recent 20-hour, 1.2-mm ALMA program
(2013.1.00718.S: Paper I and II in the ALMA Spectroscopic Survey in
the HUDF [ASPECS] series [Walter et al.\ 2016 and Aravena et
  al.\ 2016a]), we were able to acquire very deep continuum
observations over a 1 arcmin$^2$ region of the sky with high-quality
multi-wavelength observations.  That region was the {\it Hubble} Ultra
Deep Field (HUDF: Beckwith et al.\ 2006; Illingworth et al.\ 2013; Xue
et al.\ 2011), containing the most sensitive ultraviolet, optical,
near-infrared, X-ray, and radio observations available anywhere on the
sky (Bouwens et al.\ 2011; Ellis et al.\ 2013; Illingworth et
al.\ 2013; Teplitz et al.\ 2013; Xue et al.\ 2011).  Our new
observations (12.7$\mu$Jy/beam) were sufficiently deep to probe to an
almost-constant dust-enshrouded SFR of 4 $M_{\odot}$/yr at $2\sigma$
from $z\sim2$ to $z\sim10$ over the field [for $T_d$ $\sim$ 35 K],
allowing us to obtain a census of such star formation over a
substantial volume ($\sim$2$\times$10$^4$ Mpc$^3$) in the early
universe.

With these deep ALMA observations, we have the capacity not only to
detect dust emission from individual sources to almost unprecedented
limits, but also to systematically measure how the dust-enshrouded
SFRs in galaxies depend on redshift, stellar mass, and UV-continuum
slope $\beta$.  In an earlier paper in this series (paper II: Aravena
et al.\ 2016a), we identified all those sources that showed
significant individual detections ($>$3.5$\sigma$) in our 1 arcmin$^2$
mosaic and briefly examined the characteristics of the detected
sources, noting that the mean redshift of those sources was
approximately $z\sim1.5$.  We also considered stacks of various
sources over this field, subdividing these samples by redshift,
stellar mass, and SFR.  We demonstrated that by combining the
individual detections in ASPECS with the stacked measures we could
match the measured cosmic background at 1.2$\,$mm.

The purpose of this paper (paper VI in the ASPECS series) is to focus
in particular on the infrared excess ($L_{IR}/L_{UV}$) of $z=2$-10
galaxies over our 1 arcmin$^2$ field and to quantify the dependence of
this excess on stellar mass and $UV$-continuum slope $\beta$.  Over
the 1 arcmin$^2$ ASPECS field, we have sufficient $z=2$-10 sources,
i.e., $\sim$330 in total, to attempt a first exploration of the
IRX-$\beta$ and IRX-stellar mass relation at $z>2$ for normal
galaxies.  We can also investigate quantitatively whether the
IRX-$\beta$ relationship shows a dependence on stellar mass and how
strong that dependence is (if it exists).  Previous work (e.g., Baker
et al.\ 2001; Reddy et al.\ 2006; Siana et al.\ 2008, 2009;
{\'A}lvarez-M{\'a}rquez et al.\ 2016) have presented seemingly
compelling evidence for such a dependence, with higher and lower mass
galaxies showing a M99 and SMC IRX-$\beta$
relationship, respectively.\footnote{Reddy et al.\ 2006 frame the
  dependence of the IRX-$\beta$ relation in terms of the stellar
  population age of a galaxy.  Sources with ages $<$100 Myr and $>$100
  Myr were found to show an SMC and M99 IRX-$\beta$ relations,
  respectively.  In the context of the Reddy et al.\ (2006) samples,
  stellar population age is functionally equivalent to stellar mass.}
However, most of these studies have not extended into the same mass
regime we explore with the present data set and have not extended out
to $z\geq3$.

The outline of the paper is as follows.  We begin with a description
of the ASPECS 1.2$\,$mm data set we employ for this study, our
procedure for constructing $z=2$-10 samples from the HUDF data, and
finally our derivation of stellar population parameters for individual
sources (\S2).  In \S3, we discuss the number of $z=2$-10 galaxies we
would expect to individually detect based on $z=0$-2 results, compare
that with what we find, and then finally measure the stacked signal
from the observations subdividing the samples according to stellar
mass, $UV$-continuum slope $\beta$, and apparent magnitude.  In \S4,
we discuss our results and their likely implications in \S5.  Finally,
\S6 summarizes our results and the most important conclusions.

We refer to the {\it HST} F225W, F275W, F336W, F435W, F606W, F600LP,
F775W, F814W, F850LP, F105W, F125W, F140W, and F160W bands as
$UV_{225}$, $UV_{275}$, $U_{336}$, $B_{435}$, $V_{606}$, $V_{600}$,
$i_{775}$, $I_{814}$, $z_{850}$, $Y_{105}$, $J_{125}$, $JH_{140}$, and
$H_{160}$, respectively, for simplicity.  For consistency with
previous work, we find it convenient to quote results in terms of the
luminosity $L_{z=3}^{*}$ Steidel et al.\ (1999) derived at $z\sim3$,
i.e., $M_{1700,AB}=-21.07$.  Throughout the paper we assume a standard
``concordance'' cosmology with $H_0=70$ km s$^{-1}$ Mpc$^{-1}$,
$\Omega_{\rm m}=0.3$ and $\Omega_{\Lambda}=0.7$, which are in good
agreement with recent cosmological constraints (Planck Collaboration
et al.\ 2015).  Stellar masses and obscured SFRs are quoted assuming a
Chabrier (2003) IMF while the SFR density is presented adopting a
Salpeter (1955) IMF.  Magnitudes are in the AB system (Oke \& Gunn
1983).

\section{Observations and Sample}

\subsection{Band-6 Data Set and Flux Measurements}

The principal data used are the ALMA observations from the
2013.1.00718.S program (PI: Aravena) over the HUDF.  Those
observations were obtained through a full frequency scan in band 6
(212 $-$ 272 GHz) with ALMA in its most compact configuration.  The
observations are distributed over 7 pointings and cover an approximate
area of $\sim$1 arcmin$^2$ to near uniform depth.  As described in
paper II (Aravena et al.\ 2016a), we collapsed our spectral data cube
along the frequency axis in the $uv$-plane, inverting the visibilities
using the CASA task CLEAN using natural weighting and mosaic mode, for
producing the continuum image.  The peak sensitivity we measure in
these continuum observations is 12.7$\mu$Jy ($1\sigma$) per primary
beam.  Our observations fall within the region of the HUDF possessing
the deepest UV, optical, and near-infrared observations (see
Illingworth et al.\ 2013; Teplitz et al.\ 2013).

The sensitivity of our ALMA observations allows us to provide useful
individual constraints on the far-IR dust emission from normal
sub-L$^*$ galaxies.  If we adopt a modified blackbody form for the SED
shape with dust temperature of 35 K and a dust emissivity power-law
spectral index of $\beta_d = 1.6$ (Eales et al.\ 1989; Klaas et
al.\ 1997), and account for the impact of the CMB (e.g., da Cunha et
al.\ 2013: \S3.1.1), we estimate that we should be able to tentatively
detect at $2\sigma$ any star-forming galaxy at $z>3$ with an IR
luminosity (8-1000$\mu$m rest-frame) in excess of 4$\times10^{10}$
$L_{\odot}$ (see Table~\ref{tab:tempdep} and
Figure~\ref{fig:showsed36}).  For comparison, the characteristic
luminosity of galaxies in the rest-frame $UV$ is approximately equal
to $4\times10^{10}$ $L_{\odot}$ from $z\sim3$ to $z\sim8$.  The
implication is that the typical $L^*$ galaxy should be tentatively
detected at $\gtrsim$$2\sigma$ in our data set if it was outputing
equal amounts of energy in the far-IR and rest-frame $UV$.  

We consider tentative $2\sigma$ detections in our examination of our
ASPECS field, instead of the usual 3$\sigma$ or 3.5$\sigma$ limit, to
push as faint as possible in looking for evidence of obscured star
formation.  We can use this aggressive limit because of the relatively
modest number of $z=2$-10 sources over ASPECS and the availability of
sensitive MIPS $24\mu$m observations and photometrically-inferred
physical properties to evaluate any tentative detections.

\begin{deluxetable}{cccc}
\tablewidth{0cm}
\tablecolumns{4}
\tabletypesize{\footnotesize}
\tablecaption{Number of $UV$-selected $z\sim2$, $z\sim3$, $z\sim4$, $z\sim5$, $z\sim6$, $z\sim7$, $z\sim8$, $z\sim9$, and $z\sim10$ Galaxies Located within our deep ALMA HUDF pointing\label{tab:sample}}
\tablehead{\colhead{} & \colhead{} & \colhead{\# of} & \colhead{}\\
\colhead{Redshift} & \colhead{Selection Criterion} & \colhead{Sources} & \colhead{Ref\tablenotemark{a}}}
\startdata
$z\sim2$ & $UV_{275}$-dropout or &  & \\
         & $1.5<z_{phot}<2.5$ & 81 & This Work\\
$z\sim3$ & $U_{336}$-dropout or &  & \\
         & $2.5<z_{phot}<3.5$ & 81 & This Work\\
$z\sim4$ & $B_{435}$-dropout or &  & \\
         & $3.5<z_{phot}<4.5$ & 80 & B15/This Work\\
$z\sim5$ & $V_{606}$-dropout & 34 & B15\\
$z\sim6$ & $i_{775}$-dropout & 30 & B15\\
$z\sim7$ & $z_{850}$-dropout or \\
         & $6.5<z_{phot}<7.5$ & 16 & B15/This Work\\
$z\sim8$ & $Y_{105}$-dropout & 6 & B15\\
$z\sim9$ & $Y_{105}$-dropout & 1 & B16\\
$z\sim10$ & $J_{125}$-dropout & 1 & B16\\
\multicolumn{2}{c}{Total} & 330
\enddata
\tablenotetext{a}{References: B15 = Bouwens et al.\ (2015), B16 = Bouwens et al. (2016), in prep.}
\end{deluxetable}

To ensure accurate far-IR flux measurements for sources over our HUDF
mosaic, care was taken in determining the offset between the nominal
sky coordinates for sources in our deep ALMA continuum observations
and the positions in the ultraviolet, optical, and near-IR
observations using the 6 best-continuum-detected sources over the HUDF
(Aravena et al.\ 2016a).  The positional offset between the images was
found to be such that sources in our ALMA continuum image were
positioned $\sim$0.3'' to the south of sources in the {\it HST}
mosaic, with $\sim$0.2$''$ source-to-source scatter in the derived
offset.  Such source-to-source UV-to-far-IR offsets are not surprising
for bright sources, but are expected to be smaller for most of the
fainter sources we are stacking, as the results we present in \S3.2
indicate.  Overall, on the basis of the source-to-source scatter, we
estimate that we can register the {\it HST} and ALMA mosaics to better
than 0.1$''$ on average.

Flux measurements themselves were made using the nominal flux at the
position of the source in the continuum map divided by the primary
beam.  We checked those flux measurements against those we derive
after convolving the maps by the primary beam and looking at the flux
at source center.  For this latter procedure, we found that we
recovered a flux that was less than 5\% higher than using the flux at
the position of the source.

For a more detailed summary of the ASPECS data set and the basic
results, we refer the reader to paper I and II in this series (Walter
et al.\ 2016; Aravena et al.\ 2016a).  Paper V in this series (Aravena
et al.\ 2016b) provides a comprehensive discussion of the candidate
[CII]158$\mu$m lines identified in the band-6 data.

\begin{deluxetable*}{ccccccccccccc}
\tablewidth{0cm}
\tablecolumns{13}
\tabletypesize{\footnotesize}
\tablecaption{$z\gtrsim 2$ $UV$-selected sources expected to show tentative $2\sigma$ detections adopting the M99 IRX-$\beta$ relationship and assuming a 35K modified blackbody SED and $\beta_d=1.6$\label{tab:catalog}}
\tablehead{
\colhead{} & \colhead{} & \colhead{} & \colhead{} & \colhead{} & \colhead{} & \colhead{} & \multicolumn{3}{c}{Predicted} & \colhead{Measured} & \colhead{Inferred}\\
\colhead{} & \colhead{} & \colhead{} & \colhead{$m_{UV,0}$} & \colhead{} & \colhead{$\log_{10}$} & \colhead{} & \multicolumn{3}{c}{$f_{1.2mm}$ [$\mu$Jy]} & \colhead{$f_{1.2mm}$} & \colhead{$[L_{IR}/$}\\
\colhead{ID\tablenotemark{a}} & \colhead{R.A.} & \colhead{DEC} & \colhead{[mag]} & \colhead{$z_{ph}$} & \colhead{$M/M_{\odot}$} & \colhead{$\beta$\tablenotemark{b}} & \colhead{Calz\tablenotemark{c}} & \colhead{SMC\tablenotemark{c}} & \colhead{Mass\tablenotemark{c,d}} & \colhead{[$\mu$Jy]} & \colhead{$10^{10}\,$$L_{\odot}$]}}
\startdata
\multicolumn{13}{c}{$z\sim2$-3 Sample}\\
XDFU-2397246112(C2) & 03:32:39.72 & $-$27:46:11.2 & 24.4 & 1.55\tablenotemark{*} & 11.21 & 0.3$\pm$0.1 & 1426 & 99 & 946\tablenotemark{$\ddagger$} & 261$\pm$25 & 50$\pm$5\\
XDFU-2373546453(C5) & 03:32:37.35 & $-$27:46:45.3 & 23.7 & 1.85\tablenotemark{*} & 10.52 & $-$0.5$\pm$0.1 & 1028 & 121 & 552\tablenotemark{$\ddagger$} & 71$\pm$14 & 14$\pm$3\\
XDFU-2393346236 & 03:32:39.33 & $-$27:46:23.6 & 25.5 & 2.59\tablenotemark{*} & 10.18 & $-$0.5$\pm$0.1 & 369 & 44 & 93\tablenotemark{$\ddagger$} & $-$12$\pm$13 & $-$2$\pm$2\\
XDFU-2370746171 & 03:32:37.07 & $-$27:46:17.1 & 23.7 & 2.24\tablenotemark{*} & 10.09 & $-$1.2$\pm$0.1 & 350 & 67 & 306\tablenotemark{$\ddagger$} & 34$\pm$14 & 6$\pm$2\\
XDFU-2358146436 & 03:32:35.81 & $-$27:46:43.6 & 24.6 & 1.90\tablenotemark{*} & 9.98 & $-$0.9$\pm$0.1 & 194 & 31 & 75 & 16$\pm$52 & 3$\pm$10\\
XDFU-2356746283 & 03:32:35.67 & $-$27:46:28.3 & 25.2 & 3.17 & 9.93 & $-$1.3$\pm$0.1 & 153 & 30 & 108\tablenotemark{$\ddagger$} & $-$13$\pm$21 & $-$2$\pm$4\\
XDFU-2385446340(C1) & 03:32:38.54 & $-$27:46:34.0 & 24.3 & 2.54\tablenotemark{*} & 9.9 & $-$1.2$\pm$0.1 & 242 & 47 & 143\tablenotemark{$\ddagger$} & 571$\pm$14 & 97$\pm$2\\
XDFU-2388246143 & 03:32:38.82 & $-$27:46:14.3 & 26.3 & 3.38 & 9.85 & $-$0.5$\pm$0.1 & 321 & 38 & 37\tablenotemark{$\ddagger$} & 10$\pm$14 & 2$\pm$2\\
XDFU-2387446541 & 03:32:38.74 & $-$27:46:54.1 & 25.7 & 2.74 & 9.85 & $-$1.2$\pm$0.1 & 81 & 15 & 40 & 18$\pm$26 & 3$\pm$4\\
XDFU-2365446123 & 03:32:36.54 & $-$27:46:12.3 & 24.1 & 1.87\tablenotemark{*} & 9.77 & $-$1.5$\pm$0.1 & 85 & 19 & 68\tablenotemark{$\ddagger$} & 38$\pm$16 & 7$\pm$3\\
XDFU-2384246348 & 03:32:38.42 & $-$27:46:34.8 & 23.8 & 2.70 & 9.75 & $-$2.0$\pm$0.1 & 46 & 13 & 181\tablenotemark{$\ddagger$} & 36$\pm$14 & 6$\pm$2\\
XDFU-2369146023 & 03:32:36.91 & $-$27:46:02.3 & 24.1 & 2.33\tablenotemark{*} & 9.47 & $-$1.8$\pm$0.1 & 65 & 17 & 56\tablenotemark{$\ddagger$} & 13$\pm$16 & 3$\pm$3\\
XDFU-2395845544 & 03:32:39.58 & $-$27:45:54.4 & 24.4 & 3.21\tablenotemark{*} & 9.46 & $-$1.5$\pm$0.1 & 215 & 47 & 78 & 14$\pm$64 & 2$\pm$11\\
XDFU-2369146348 & 03:32:36.91 & $-$27:46:34.8 & 24.7 & 1.76 & 9.46 & $-$1.3$\pm$0.1 & 75 & 15 & 17 & 9$\pm$14 & 2$\pm$3\\
XDFU-2370846470 & 03:32:37.08 & $-$27:46:47.0 & 24.4 & 1.85\tablenotemark{*} & 9.44 & $-$1.3$\pm$0.1 & 102 & 20 & 23 & 0$\pm$15 & 0$\pm$3\\
XDFU-2363346155 & 03:32:36.33 & $-$27:46:15.5 & 25.3 & 2.34 & 9.32 & $-$1.6$\pm$0.1 & 35 & 8 & 13 & 1$\pm$16 & 0$\pm$3\\
XDFU-2366846484 & 03:32:36.68 & $-$27:46:48.4 & 25.3 & 1.88\tablenotemark{*} & 9.27 & $-$1.0$\pm$0.1 & 84 & 14 & 8 & $-$10$\pm$19 & $-$2$\pm$4\\
XDFU-2366946210 & 03:32:36.69 & $-$27:46:21.0 & 24.9 & 1.96 & 9.26 & $-$1.7$\pm$0.1 & 28 & 7 & 11 & 2$\pm$13 & 0$\pm$3\\
XDFU-2382946284 & 03:32:38.29 & $-$27:46:28.4 & 26.2 & 1.76 & 9.22 & $-$1.0$\pm$0.1 & 32 & 5 & 3 & $-$2$\pm$14 & $-$0$\pm$3\\
XDFU-2378846451 & 03:32:37.88 & $-$27:46:45.1 & 26.4 & 1.89 & 8.9 & $-$0.9$\pm$0.1 & 37 & 6 & 1 & 3$\pm$15 & 0$\pm$3\\
XDFU-2372446294 & 03:32:37.24 & $-$27:46:29.4 & 27.2 & 3.25 & 8.77 & $-$1.2$\pm$0.1 & 28 & 5 & 1 & $-$34$\pm$14 & $-$6$\pm$2\\
XDFU-2379146261 & 03:32:37.91 & $-$27:46:26.1 & 27.1 & 2.48 & 8.41 & $-$0.4$\pm$0.1 & 85 & 10 & 0 & 0$\pm$13 & 0$\pm$2\\
XDFU-2379046328 & 03:32:37.90 & $-$27:46:32.8 & 26.8 & 3.38 & 8.25 & $-$1.4$\pm$0.1 & 30 & 6 & 1 & $-$3$\pm$14 & $-$1$\pm$2\\\\
\multicolumn{13}{c}{$z\sim4$ Sample}\\
XDFB-2394046224 & 03:32:39.40 & $-$27:46:22.4 & 25.5 & 2.94\tablenotemark{$\dagger$} & 9.76 & $-$1.2$\pm$0.1 & 119 & 22 & 45\tablenotemark{$\ddagger$} & $-$22$\pm$13 & $-$3$\pm$2\\
XDFB-2368245580 & 03:32:36.82 & $-$27:45:58.0 & 24.4 & 3.87\tablenotemark{*} & 9.45 & $-$1.5$\pm$0.1 & 291 & 64 & 105\tablenotemark{$\ddagger$} & $-$6$\pm$22 & $-$1$\pm$3\\
XDFB-2375446199 & 03:32:37.54 & $-$27:46:19.9 & 25.9 & 4.21 & 9.14 & $-$1.3$\pm$0.1 & 139 & 27 & 16 & 4$\pm$14 & 1$\pm$2\\
XDFB-2394246267 & 03:32:39.42 & $-$27:46:26.7 & 27.1 & 4.99 & 8.73 & $-$0.7$\pm$0.2 & 200 & 28 & 3 & $-$8$\pm$13 & $-$1$\pm$2\\
XDFB-2381646267\tablenotemark{e} & 03:32:38.16 & $-$27:46:26.7 & 28.5 & 4.16 & 8.57 & $-$0.6$\pm$0.4 & 50 & 7 & 0 & $-$21$\pm$13 & $-$3$\pm$2\\\\
\multicolumn{13}{c}{$z\sim5$ Sample}\\
XDFV-2372946175 & 03:32:37.29 & $-$27:46:17.5 & 28.1 & 5.00 & 8.92 & $-$1.1$\pm$0.3 & 38 & 7 & 2 & $-$1$\pm$14 & $-$0$\pm$2\\\\
\multicolumn{13}{c}{$z\sim6$ Sample}\\
GSDI-2374046045 & 03:32:37.40 & $-$27:46:04.5 & 26.7 & 5.85 & 9.52 & $-$1.5$\pm$0.8 & 71 & 16 & 34\tablenotemark{$\ddagger$} & 4$\pm$14 & 1$\pm$2\\
XDFI-2374646327 & 03:32:37.46 & $-$27:46:32.7 & 26.4 & 6.49 & 9.35 & $-$1.5$\pm$0.2 & 111 & 25 & 34\tablenotemark{$\ddagger$} & 12$\pm$14 & 2$\pm$2\\
XDFI-2364964171 & 03:32:36.49 & $-$27:46:41.71 & 25.5 & 6.12 & 8.97 & $-$2.0$\pm$0.2 & 50 & 14 & 31 & $-$11$\pm$18 & $-$2$\pm$2\\
GSDI-2382846172 & 03:32:38.28 & $-$27:46:17.2 & 26.2 & 6.12 & 8.66 & $-$1.9$\pm$0.3 & 34 & 9 & 8 & $-$4$\pm$14 & $-$1$\pm$2\\
XDFI-2378346180\tablenotemark{e} & 03:32:37.83 & $-$27:46:18.0 & 29.3 & 6.20 & 8.54 & $-$0.3$\pm$0.6 & 87 & 9 & 0 & 1$\pm$14 & 0$\pm$2\\\\
\multicolumn{13}{c}{$z\sim7$ Sample}\\
XDFZ-2381446048\tablenotemark{e} & 03:32:38.14 & $-$27:46:04.8 & 29.5 & 6.66 & 7.98 & 0.2$\pm$1.7 & 196 & 14 & 0 & $-$11$\pm$15 & $-$2$\pm$2\\
\enddata
\tablenotetext{*}{Spectroscopic Redshift from 3D-HST (Momcheva et al.\ 2015).}
\tablenotetext{$\dagger$}{The $B_{435}$-dropout color selection
  criteria from Bouwens et al.\ (2015) identifies galaxies with
  photometric redshifts as low as $z\sim3$.  We retain this source in
  our $z\sim4$ sample consistent with the Bouwens et al\ (2015)
  $B_{435}$-dropout selection function.}
\tablenotetext{$\ddagger$}{Tentative $\gtrsim2\sigma$ detection of source is expected.}
\tablenotetext{a}{Source ID from Bouwens et al.\ 2015.  Otherwise
  selected from either a new catalog constructed here or the Rafelski
  et al.\ (2015) catalog based on the {\it HST} WFC3/UVIS, ACS, and WFC3/IR
  observations over the HUDF.  C1, C2, and C5 correspond to the continuum
  detections identified in our blind search of our ALMA 1.2$\,$mm
  observations (paper II from this series: Aravena et al.\ 2016a).}
\tablenotetext{b}{$UV$-continuum slope $\beta$ estimated by fitting
  the $UV$-continuum fluxes to a power-law (Castellano et al.\ 2012;
  Bouwens et al.\ 2012; Rogers et al.\ 2013).  }
\tablenotetext{c}{Assuming a standard modified blackbody SED with dust
  temperature of 35 K and accounting for the impact of the CMB on the
  measured flux (da Cunha et al.\ 2013).}
\tablenotetext{d}{Assuming the consensus $z\sim2$-3 relationship between 
the infrared excess and the inferred stellar mass of the galaxy (Appendix A).}
\tablenotetext{e}{These sources are nominally expected to be detected
  in our ALMA mosaic based on their very red measured $\beta$'s.
  However, the $\beta$ measurement for these sources are quite
  uncertain.  It is anticipated that a few of the reddest $z=6$-8
  galaxies over our small field would have these colors due to the
  impact of noise.}
\end{deluxetable*}

\subsection{LBG samples}

The high-redshift star-forming galaxies that we analyzed in this study
were selected for this study using the best existing {\it HST}
observations over the HUDF.

\begin{figure*}
\epsscale{1.15} \plotone{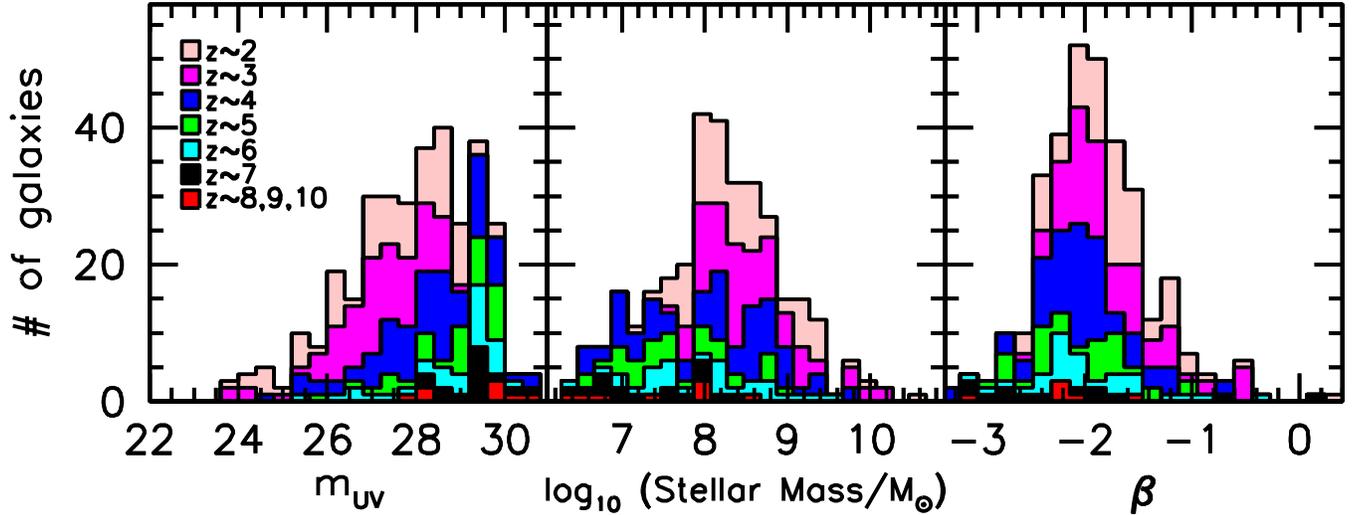}
\caption{Cumulative histograms showing the composition of the HUDF
  $z\sim2$, $z\sim3$, $z\sim4$, $z\sim5$, $z\sim6$, $z\sim7$, and
  $z\sim8$-10 samples we consider here (\textit{pink, magenta, blue,
    green, cyan, black, and red shaded histograms, respectively}) as a
  function of apparent magnitude (measured at wavelengths probing the
  $UV$ continuum), stellar mass, and $UV$-continuum slope $\beta$
  (\textit{left, central, and right panels, respectively}).  The deep
  far-IR continuum observations (12.7$\mu$Jy/beam RMS) from ASPECS
  make it possible for us to set the best constraints on the average
  dust emission from ($m_{UV}>25$, $<$10$^{10}$ $M_{\odot}$) galaxies
  thus far.  \label{fig:sample}}
\end{figure*}

We briefly describe the $z=1.5$-3.5 samples that we constructed over
the HUDF using the available WFC3/UVIS, ACS/WFC, and WFC3/IR
observations.  For {\it HST} optical ACS/WFC and near-infrared WFC3/IR
observations, we make use of the XDF reductions (Illingworth et
al.\ 2013), which incorporated all ACS+WFC3/IR data available over the
HUDF in 2013.  The XDF reductions are $\sim$0.1-0.2 mag deeper than
original Beckwith et al.\ (2006) reductions at optical wavelengths and
also provide coverage in the F814W band.  The WFC3/IR reductions made
available as part of the XDF release include all data from the
original HUDF09 (Bouwens et al.\ 2011), CANDELS (Grogin et al.\ 2011;
Koekemoer et al.\ 2011), and the HUDF12 (Ellis et al.\ 2013) programs.

In the process of assembling our $z=1.5$-3.5 samples, we derived our
initial source catalogs and performed photometry on sources using our
own modified version of the SExtractor (Bertin \& Arnouts 1996)
software.  Source detection was performed on the square-root of
$\chi^2$ image (Szalay et al.\ 1999: similar to a coadded image)
constructed from the $V_{606}$, $i_{775}$, $Y_{105}$, $J_{125}$,
$JH_{140}$, and $H_{160}$ images.  After PSF-correcting fluxes to
match the $H_{160}$-band image, color measurements were made in
Kron-style (1980) scalable apertures with a Kron factor of 1.6.
``Total magnitude'' fluxes were derived from the smaller-scalable
apertures by (1) correcting up the flux to account for the additional
flux seen in a larger-scalable aperture (Kron factor of 2.5) seen on
the square root of $\chi^2$ image and (2) correcting for the flux
outside these larger scalable apertures and on the wings of the PSF
using tabulations of the encircled energy (Dressel 2012).

Intermediate redshift $z\sim2$-3 galaxies were selected by applying
simple two-color criteria using the Lyman-Break Galaxy (LBG) strategy:
\begin{eqnarray*}
(UV_{275}-U_{336}>1)\wedge (U_{336}-B_{435}<1)\wedge \\
(V_{606}-Y_{105}<0.7)\wedge (S/N(UV_{225})<1.5)
\end{eqnarray*}
for $z\sim2$ galaxies and
\begin{eqnarray*}
(U_{336}-B_{435}>1)\wedge(B_{435}-V_{606}<1.2)\wedge \\
(i_{775}-Y_{105}<0.7)\wedge(\chi_{UV_{225},UV_{275}}^2<2)
\end{eqnarray*}
for $z\sim3$ galaxies, where $\wedge$, $\vee$, and S/N represent the
logical \textbf{AND}, \textbf{OR} symbols, and signal-to-noise in our
smaller scalable apertures, respectively.  We define
$\chi_{UV_{225},UV_{275}} ^2$ as $\Sigma_{i} \textrm{SGN}(f_{i})
(f_{i}/\sigma_{i})^2$ where $f_{i}$ is the flux in band $UV_{225}$ and
$UV_{275}$ in a small-scalable aperture, $\sigma_i$ is the uncertainty
in this flux, and SGN($f_{i}$) is equal to 1 if $f_{i}>0$ and $-1$ if
$f_{i}<0$.  These criteria are similar to the two color criteria
previously utilized in Oesch et al.\ (2010) and Hathi et al.\ (2010).

Our $z=4$-8 samples were drawn from the Bouwens et al.\ (2015) samples
and include all $z=3.5$-8.5 galaxies located over the 1 arcmin$^2$
ASPECS region.  The Bouwens et al.\ (2015) samples were based on the
deep optical ACS and WFC3/IR observations within the HUDF.  $z=4$-8
samples were constructed by applying Lyman-break-like color criteria
to the XDF reduction (Illingworth et al.\ 2013) of the Hubble Ultra
Deep Field.

The Bouwens et al.\ (2016, in prep) $z=9$-10 samples were constructed
using a $Y_{105}$ or $J_{125}$-dropout Lyman-break color criteria to
the available {\it HST} data and then splitting the selected sources
into $z\sim9$ and $z\sim10$ subsamples.  Two sources from that
$z=9$-10 sample lie within the ASPECS region (see also Bouwens et
al.\ 2011; Ellis et al.\ 2013; Oesch et al.\ 2013).

To maximize the total number of star-forming galaxies at $z\sim2$-10
considered in this study, we also applied the EAZY photometric
redshift code to the {\it HST} WFC3/UVIS, ACS, and WFC3/IR photometric
catalogs we had available over our deep ALMA field and included all
sources with a best-fit redshift solution between $z\sim1.5$ and
$z\sim8.5$, which were not in our Lyman-break catalogs, and which
utilized star-forming or dusty SED templates to reproduce the observed
SED.  We also made use of the photometric catalog of Rafelski et
al.\ (2015) and included those sources in our samples, if not present
in our primary two selections.

The photometric-redshift-selected star-forming galaxies added 64, 31,
2, and 1 $z\sim2$, $z\sim3$, $z\sim4$, and $z\sim7$ galaxies to our
study, respectively.  Sources in our photometric-redshift selections
showed almost identical distributions of properties to our LBG
selections at $z\sim2$-3 (where our photometric-redshift selections
add sources), with a median $\beta$ and stellar mass of $-1.84$ and
$10^{8.41}$ $M_{\odot}$ for the photometric-redshift selections
vs. $-1.82$ and $10^{8.37}$ $M_{\odot}$ for the $z\sim2$-3 LBG
selections.  9\% (7/79) and 3\% (3/96) of the sources in our LBG and
photometric-redshift selections, respectively, have measured $\beta$'s
redder than $-1$.

Eighty-one $z\sim2$ and 81 $z\sim3$ sources in total were identified
using our dropout + photometric-redshift criteria over the 1
arcmin$^2$ region of the HUDF where we have deep ALMA observations.
Our higher redshift $z\sim4$, $z\sim5$, $z\sim6$, $z\sim7$, $z\sim8$,
$z\sim9$, and $z\sim10$ samples (Bouwens et al.\ 2015; Bouwens et
al.\ 2016, in prep) contain 80, 34, 30, 16, 6, 1, and 1 sources,
respectively (Table~\ref{tab:sample}) over this same region.  The
expected contamination rate in these color-selected samples by
lower-redshift galaxies (or stars) is estimated to be on the order of
3-8\% (e.g., Bouwens et al.\ 2015).  In terms of apparent magnitude in
the $UV$-continuum, these sources extend from 21.7 mag to 30.8 mag
(Figure~\ref{fig:sample}: \textit{left panel}).

\begin{figure*}
\epsscale{1.00}
\plotone{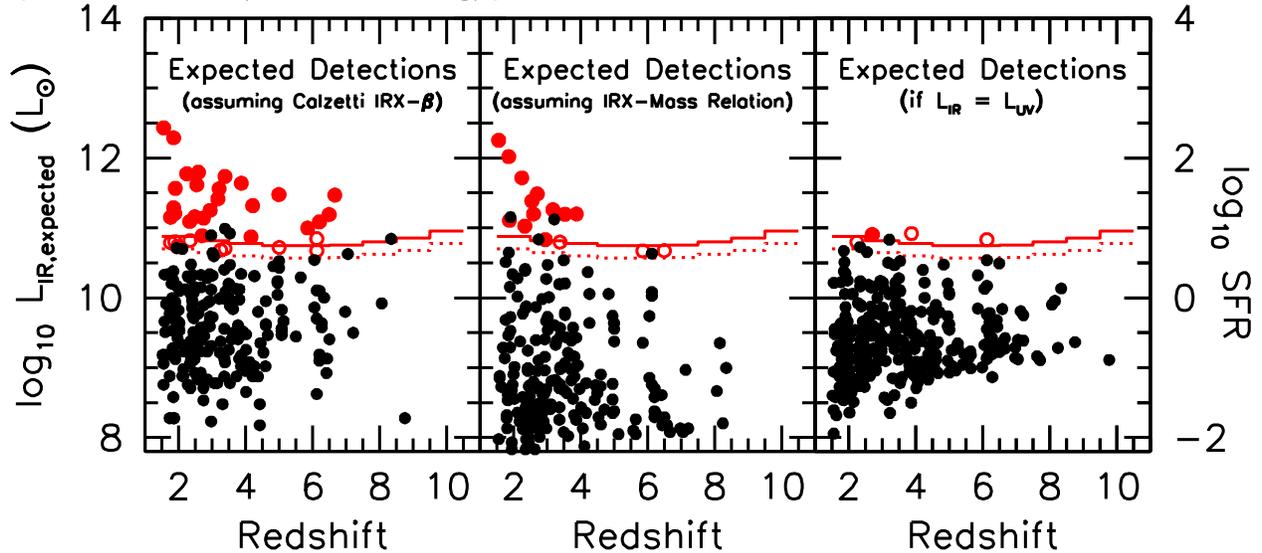}
\caption{Expected IR luminosities (in $L_{\odot}$) versus photometric
  redshift of $z=2$-10 galaxies (solid circles) within the 1
  arcmin$^2$ ASPECS region.  Expected IR luminosities are based on (1)
  the M99 IRX-$\beta$ relationship (\textit{left panel}), (2) the
  approximate $z\sim2$-3 IRX-stellar mass relationship (\textit{center
    panel}: see Appendix A), or (3) assuming $L_{IR}=L_{UV}$
  (\textit{right panel}).  The solid and dotted red lines indicate the
  $3\sigma$ and $2\sigma$ limiting luminosities, respectively, to
  which we can probe as a function of redshift in the deepest regions
  within our 1 arcmin$^2$ field (assuming that the SED is well
  represented by a 35 K-modified blackbody).  The solid and open red
  circles correspond to sources where $3\sigma$ and $2\sigma$
  detections are expected, respectively, adopting the assumptions from
  a given panel, while the black circles indicate sources where a
  tentative $2\sigma$ detection is not expected.  Black sources can
  appear above the red lines if these sources fall in regions of
  ASPECS where the sensitivities are lower than the
  maximum.\label{fig:probe}}
\end{figure*}

\subsection{Estimates of the Stellar Mass for Individual Sources in our $z=2$-10 Sample}

We provide a brief description of our estimates of the stellar mass
for $z=2$-10 sources over the HUDF.  As in other work (e.g., Sawicki
et al.\ 1998; Brinchmann et al.\ 2000; Papovich et al.\ 2001;
Labb{\'e} et al.\ 2005; Gonzalez et al.\ 2014), we estimate stellar
masses for individual sources in our samples by modeling the observed
photometry using stellar population libraries and considering variable
(or fixed) star formation histories, metallicities, and dust content.

For convenience, we make use of the publicly-available code
\textsc{FAST} (Kriek et al.\ 2009) to perform this fitting.  We assume
a Chabrier (2003) IMF, a metallicity of 0.2 $Z_{\odot}$, an
approximately constant star formation rate in modeling the star
formation history while performing the fits (keeping $\tau$ parameter
for a $e^{-t/\tau}$ star formation history equal to 100 Gyr), and
allowing the dust extinction in the rest-frame V to range from zero to
2 mag.  Our fixing the fiducial metallicity to 0.2 $Z_{\odot}$ is well
motivated based on studies of the metallicity of individual $z\sim2$-4
galaxies (Pettini et al.\ 2000) or as predicted from cosmological
hydrodynamical simulations (Finlator et al.\ 2011; Wise et al.\ 2012).
While the current choice of parameters can have a sizeable impact on
inferred quantities like the age of a stellar population (changing by
$>$0.3-0.5 dex), these choices typically do not have a major impact
($\gtrsim$0.2 dex) on the inferred stellar masses.

In deriving the stellar masses for individual sources, we made use of
flux measurements from 11 {\it HST} bands ($UV_{225}$, $UV_{275}$,
$U_{336}$, $B_{435}$, $V_{606}$, $i_{775}$, $z_{850}$, $Y_{105}$,
$J_{125}$, $JH_{140}$, $H_{160}$), 1 band in the near-IR from the
ground ($K_s$), and 4 {\it Spitzer}/IRAC bands (3.6$\mu$m, 4.5$\mu$m,
5.8$\mu$m, and 8.0$\mu$m).  The {\it HST} photometry we use for
estimating stellar masses was derived applying the same procedure as
used for selecting our $z\sim2$-3 LBG samples (see \S2.2).

Our {\it Spitzer}/IRAC flux measurements were derived for individual
sources from $\sim$100-200-hour stacks of the IRAC observations over
the HUDF (Labb{\'e} et al.\ 2015) from the IUDF program (PI:
Labb{\'e}) and Oesch et al.\ (2013c) {\it Spitzer}/IRAC program.  As
has become standard procedure (e.g., Shapley et al.\ 2005; Labb{\'e}
et al.\ 2005, 2015; Grazian et al.\ 2006; Laidler et al.\ 2007; Merlin
et al.\ 2015), we use the {\it HST} observations as a template to model the
fluxes of sources in the {\it Spitzer}/IRAC observations and thus
perform photometry below the nominal confusion limit.  The model flux
from neighboring sources is subtracted before attempting to measure
fluxes for the sources of interest.  Source photometry is performed in
1.8$''$-diameter circular apertures for the {\it Spitzer}/IRAC
3.6$\mu$m and 4.5$\mu$m bands and 2.0$''$-diameter circular apertures
for the 5.8$\mu$m and 8.0$\mu$m bands.  The observed fluxes are
corrected to total based on the inferred growth curve for sources
after PSF-correction to the {\it Spitzer}/IRAC PSF.  We utilize a
similar procedure to derive fluxes for sources based on the deep
ground-based $K$-band observations available from VLT/HAWK-I,
VLT/ISAAC, and PANIC observations (Fontana et al.\ 2014) over the HUDF
($5\sigma$ depths of 26.5 mag).

A modest correction is made to the IRAC 3.6$\mu$m and 4.5$\mu$m
photometry to account for the impact on nebular emission lines on the
observed IRAC fluxes, decreasing the brightness of the 3.6$\mu$m and
4.5$\mu$m band fluxes by 0.32 mag to account for the presence of
H$\alpha$ and 0.4 mag to account for the presence of [OIII]+H$\beta$
emission where present.  These corrections are well motivated based on
observations of $z\sim4$-8 galaxies (Stark et al.\ 2013; Labb{\'e} et
al.\ 2013; Marmol-Queralto et al.\ 2016; Smit et al.\ 2014, 2016;
Rasappu et al.\ 2016) and lower the median inferred stellar mass for
$z>3.8$ galaxies in our sample by $\sim$0.1 dex.

The stellar masses we estimate for the sources for the
highest-redshift sources in our selection, $z>5$, are not as well
constrained as at lower redshifts where our sensitive photometry
extend to rest-frame $1\mu$m.  To guard against noise in the modeling
process scattering lower-mass galaxies into higher-mass bins, we also
model the photometry of galaxies forcing the dust extinction to be
zero in fitting the observed SEDs with FAST.  For sources where the
stellar-mass estimates exceed the dust-free stellar-mass estimates by
more than 0.9 dex and the photometric evidence for a particularly
dusty SED was weak (applicable to only 6 sources from our total sample
of 330 sources), we made use of the dust-free stellar mass estimates
instead.

We also estimated stellar masses for our sources by using the
\textsc{MAGPHYS} software (da Cunha et al.\ 2008) to model the
photometry for the 330 $z=2$-10 sources that make up our samples.  For
sources with redshifts $z<3.8$, the stellar masses we estimated were
in excellent agreement with our fiducial results, with the median and
mean stellar mass derived by \textsc{MAGPHYS} being 0.02 dex and 0.04
dex lower, respectively.  This points toward no major systematic
biases in the results from the present study -- which rely on
\textsc{FAST}-estimated masses -- and the other papers in the ASPECS
series -- where the reliance is on \textsc{MAGPHYS}-estimated masses.

The middle panel of Figure~\ref{fig:sample} illustrates the effective
range in stellar mass probed by our $z=2$-10 sample.  Most sources
from our HUDF $z=2$-10 sample have stellar masses in the range
$10^{7.5}$ $M_{\odot}$ to $10^{9.5}$ $M_{\odot}$.  The most massive
sources probed by our program extend from $10^{10}$ to $10^{11.2}$
$M_{\odot}$.  Beyond the stellar mass itself, Figure~\ref{fig:sample}
also illustrates the range in $UV$-continuum slope $\beta$ probed by
our samples (see \S3.1 for details on how $\beta$ is derived).  Since
the measured $\beta$ has been demonstrated to be quite effective in
estimating the infrared excess for lower-redshift $UV$-selected
samples (e.g., M99; Reddy et al.\ 2006; Daddi et al.\ 2007), it is
useful for us to probe a broad range in $\beta$.  As can be seen from
Figure~\ref{fig:sample}, our samples probe the range $\beta\sim-1.5$
to $\sim-2.5$ quite effectively.

\begin{deluxetable*}{cccccccccccc}
\tablewidth{0cm}
\tablecolumns{12}
\tabletypesize{\footnotesize}
\tablecaption{$z\gtrsim 2$ $UV$-selected galaxies showing tentative 2$\sigma$ detections in our deep ALMA continuum observations\tablenotemark{a}\label{tab:detect}}
\tablehead{
\colhead{} & \colhead{} & \colhead{} & \colhead{} & \colhead{} & \colhead{} & \colhead{} & \multicolumn{3}{c}{Predicted} & \colhead{Measured} & \colhead{Inferred}\\
\colhead{} & \colhead{} & \colhead{} & \colhead{$m_{UV,0}$} & \colhead{} & \colhead{$\log_{10}$} & \colhead{} & \multicolumn{3}{c}{$f_{1.2mm}$ [$\mu$Jy]} & \colhead{$f_{1.2mm}$} & \colhead{$[L_{IR}/$}\\
\colhead{ID} & \colhead{R.A.} & \colhead{DEC} & \colhead{[mag]} & \colhead{$z_{ph}$} & \colhead{$M/M_{\odot}$} & \colhead{$\beta$} & \colhead{Calz} & \colhead{SMC} & \colhead{Mass} & \colhead{[$\mu$Jy]} & \colhead{$10^{10}\,$$L_{\odot}$]}}
\startdata
\multicolumn{12}{c}{Tentative $>$2$\sigma$ Detections (Most Credible)\tablenotemark{$\dagger$}}\\
XDFU-2397246112(C2) & 03:32:39.72 & $-$27:46:11.2 & 24.4 & 1.55\tablenotemark{*} & 11.21 & 0.3$\pm$0.1 & 1426 & 99 & 946 & 261$\pm$25 & 50$\pm$5\\
XDFU-2373546453(C5) & 03:32:37.35 & $-$27:46:45.3 & 23.7 & 1.85\tablenotemark{*} & 10.52 & $-$0.5$\pm$0.1 & 1028 & 121 & 552 & 71$\pm$14 & 14$\pm$3\\
XDFU-2370746171 & 03:32:37.07 & $-$27:46:17.1 & 23.7 & 2.24\tablenotemark{*} & 10.09 & $-$1.2$\pm$0.1 & 350 & 67 & 306 & 34$\pm$14 & 6$\pm$2\\
XDFU-2385446340(C1) & 03:32:38.54 & $-$27:46:34.0 & 24.3 & 2.54\tablenotemark{*} & 9.90 & $-$1.2$\pm$0.1 & 242 & 47 & 143 & 571$\pm$14 & 97$\pm$2\\
XDFU-2365446123 & 03:32:36.54 & $-$27:46:12.3 & 24.1 & 1.87\tablenotemark{*} & 9.77 & $-$1.5$\pm$0.1 & 85 & 19 & 68 & 38$\pm$16 & 7$\pm$3\\
XDFU-2384246348 & 03:32:38.42 & $-$27:46:34.8 & 23.8 & 2.70 & 9.75 & $-$2.0$\pm$0.1 & 46 & 13 & 181 & 36$\pm$14 & 6$\pm$2\\\\
\multicolumn{12}{c}{Not Especially Credible $>$2$\sigma$ Detections\tablenotemark{$\ddagger$}}\\
XDFU-2403146258 & 03:32:40.31 & $-$27:46:25.8 & 27.6 & 1.55 & 8.21 & $-$1.9$\pm$0.1 & 1 & 0 & 0 & 54$\pm$26 & 10$\pm$5\\
XDFB-2355846304 & 03:32:35.58 & $-$27:46:30.4 & 29.9 & 4.06 & 8.15 & $-$1.9$\pm$1.0 & 1 & 0 & 0 & 71$\pm$26 & 11$\pm$4\\\\
\multicolumn{12}{c}{$>$2$\sigma$ ``Detections'' in the Negative Continuum Image}\\
XDFU-2372446294 & 03:32:37.24 & $-$27:46:29.4 & 27.2 & 3.25 & 8.77 & $-$1.2$\pm$0.1 & 28 & 5 & 1 & $-$34$\pm$14 & $-$6$\pm$2\\
XDFU-2390646560 & 03:32:39.06 & $-$27:46:56.0 & 26.7 & 2.42 & 8.75 & $-$1.5$\pm$0.1 & 12 & 3 & 1 & $-$82$\pm$38 & $-$14$\pm$7\\
XDFU-2390446358 & 03:32:39.04 & $-$27:46:35.8 & 28.1 & 3.25 & 8.62 & $-$2.0$\pm$0.2 & 1 & 0 & 0 & $-$33$\pm$14 & $-$6$\pm$2\\
XDFU-2375346041 & 03:32:37.53 & $-$27:46:04.1 & 26.4 & 1.96 & 8.26 & $-$2.1$\pm$0.1 & 1 & 0 & 0 & $-$32$\pm$14 & $-$6$\pm$3\\
XDFU-2369446426 & 03:32:36.94 & $-$27:46:42.6 & 28.6 & 1.60 & 8.08 & $-$1.3$\pm$0.2 & 2 & 0 & 0 & $-$32$\pm$14 & $-$6$\pm$3\\
XDFB-2401746314 & 03:32:40.17 & $-$27:46:31.4 & 28.3 & 4.01 & 8.21 & $-$2.0$\pm$0.3 & 1 & 0 & 0 & $-$56$\pm$26 & $-$8$\pm$4\\
XDFV-2385645553 & 03:32:38.56 & $-$27:45:55.3 & 29.8 & 5.07 & 6.99 & $-$1.6$\pm$0.9 & 3 & 1 & 0 & $-$72$\pm$26 & $-$10$\pm$4\\
XDFZ-2375446018 & 03:32:37.54 & $-$27:46:01.8 & 29.3 & 7.05 & 7.85 & $-$2.3$\pm$1.5 & 0 & 0 & 0 & $-$40$\pm$16 & $-$6$\pm$2
\enddata
\tablenotetext{*}{Spectroscopic Redshift from 3D-HST (Momcheva et
  al.\ 2015).}  
\tablenotetext{$\dagger$}{The reality of each of these tentatively
  detected sources is supported by their not being any comparable
  detections of $>10^{9.75}$ $M_{\odot}$ sources in the negative
  continuum images and each of these sources also showing a detection
  in the MIPS $24\mu$m observations (see Table~\ref{tab:mipsalma} from
  Appendix B).}
\tablenotetext{$\ddagger$}{While fewer low-mass ($<10^{9}$
  $M_{\odot}$) galaxies are tentatively detected at $>2\sigma$ in the
  positive continuum image than in the negative continuum image, this
  appears not to be statistically significant.  Use of spatially
  offset positions (by 0.3$''$) to measure the flux in sources
  typically result in an essentially equal number of tentative
  $>2\sigma$ detections in the positive and negative continuum
  images.}
\tablenotetext{a}{Columns in this table are essentially identical to
  those in Table~\ref{tab:catalog}.}
\end{deluxetable*}

\begin{figure}
\epsscale{1.15}
\plotone{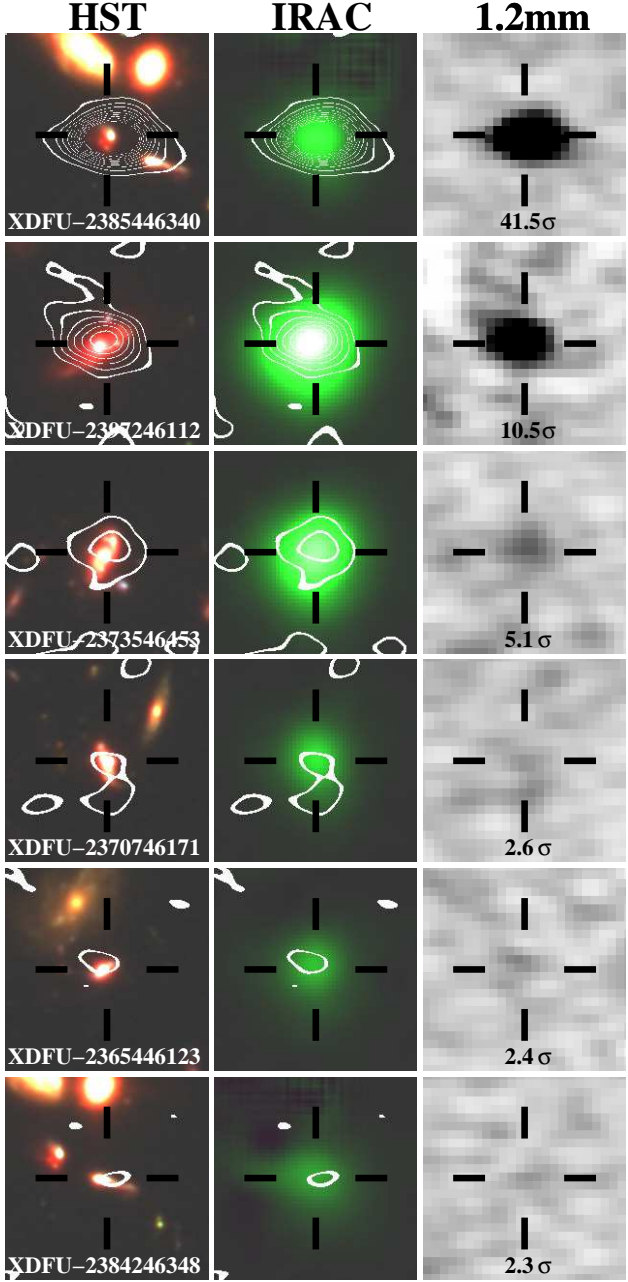}
\caption{{\it HST} $B_{435}i_{775}H_{160}$ (\textit{left}), IRAC
  3.6$\mu$m (\textit{middle}), and 1.2$\,$mm ALMA-continuum images
  (\textit{right}) for six $z\sim2$-3 galaxies that we detect
  ($>$3.5$\sigma$) or tentatively detect ($>$2$\sigma$) in our 1
  arcmin$^2$ deep ALMA map over the HUDF.  The size of the stamps is
  7.2''$\times$7.2''.  The position of our 1.2$\,$mm-continuum
  detections relative to the position of sources in our {\it HST} or
  {\it Spitzer}/IRAC images are illustrated in the left and center
  stamps with the $2\sigma$, $4\sigma$, $6\sigma$, $8\sigma$, and
  $10\sigma$ contours (\textit{white lines}).  Light from neighboring
  sources on the IRAC images have been removed for clarity.
  Significantly enough, these sources are among the 13 $z=2$-10
  candidates from that 1 arcmin$^2$ region with the highest stellar
  mass estimates.  All six have estimated stellar masses $\geq$
  $10^{9.75}$ $M_{\odot}$.  Given evidence that dust emission from
  star-forming galaxies correlates with stellar mass in many studies
  (e.g., Pannella et al.\ 2009, 2015; Reddy et al.\ 2010; Whitaker et
  al.\ 2014; {\'A}lvarez-M{\'a}rquez et al.\ 2016), these sources are
  amongst the 6 sources most likely to show dust emission from our
  entire $z=2$-10 LBG selection.  The fact that each of them shows
  ALMA flux at $\geq$2.3$\sigma$ seems to confirm that stellar mass is
  an especially useful predictor of dust emission for normal
  star-forming galaxies at $z\gtrsim 2$.  Each of these sources also
  shows evidence for being detected ($\gtrsim2\sigma$) in the MIPS
  $24\mu$m observations (Table~\ref{tab:mipsalma} from Appendix
  B).\label{fig:ddetect}}
\end{figure}

\begin{figure}
\epsscale{1.15}
\plotone{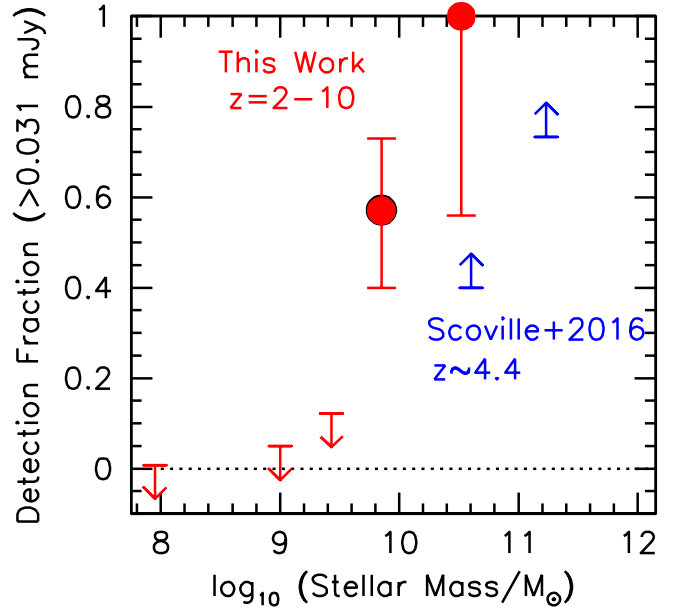}
\caption{Fraction of tentatively detected ($>$2$\sigma$) $z=2$-10
  galaxies in our ALMA $1.2\,$mm continuum observations versus the
  inferred stellar mass (\textit{solid red circle}).  Errors and upper
  limits are $1\sigma$.  Only the 172 $z=2$-10 galaxies where our
  $2\sigma$ continuum sensitivity is highest ($<$34$\mu$Jy) are
  included in this determination.  The blue upperward arrows are from
  Scoville et al.\ (2016) and indicate lower limits on the detected
  fraction (i.e., at $>$0.031 mJy) based on the results of that study.
  Stellar mass appears to be a very good predictor of dust emission in
  $z=2$-10 galaxies, with 5 of the 8 $>10^{9.75}$ $M_{\odot}$ galaxies
  probed at the requisite sensitivity being detected at $>$2$\sigma$
  (and several other $>10^{9.75}$ $M_{\odot}$ galaxies probed by our
  field show measured 1.2$\,$mm fluxes consistent with the other
  measurements).\label{fig:detsm}}
\end{figure}

\begin{figure*}
\epsscale{0.95}
\plotone{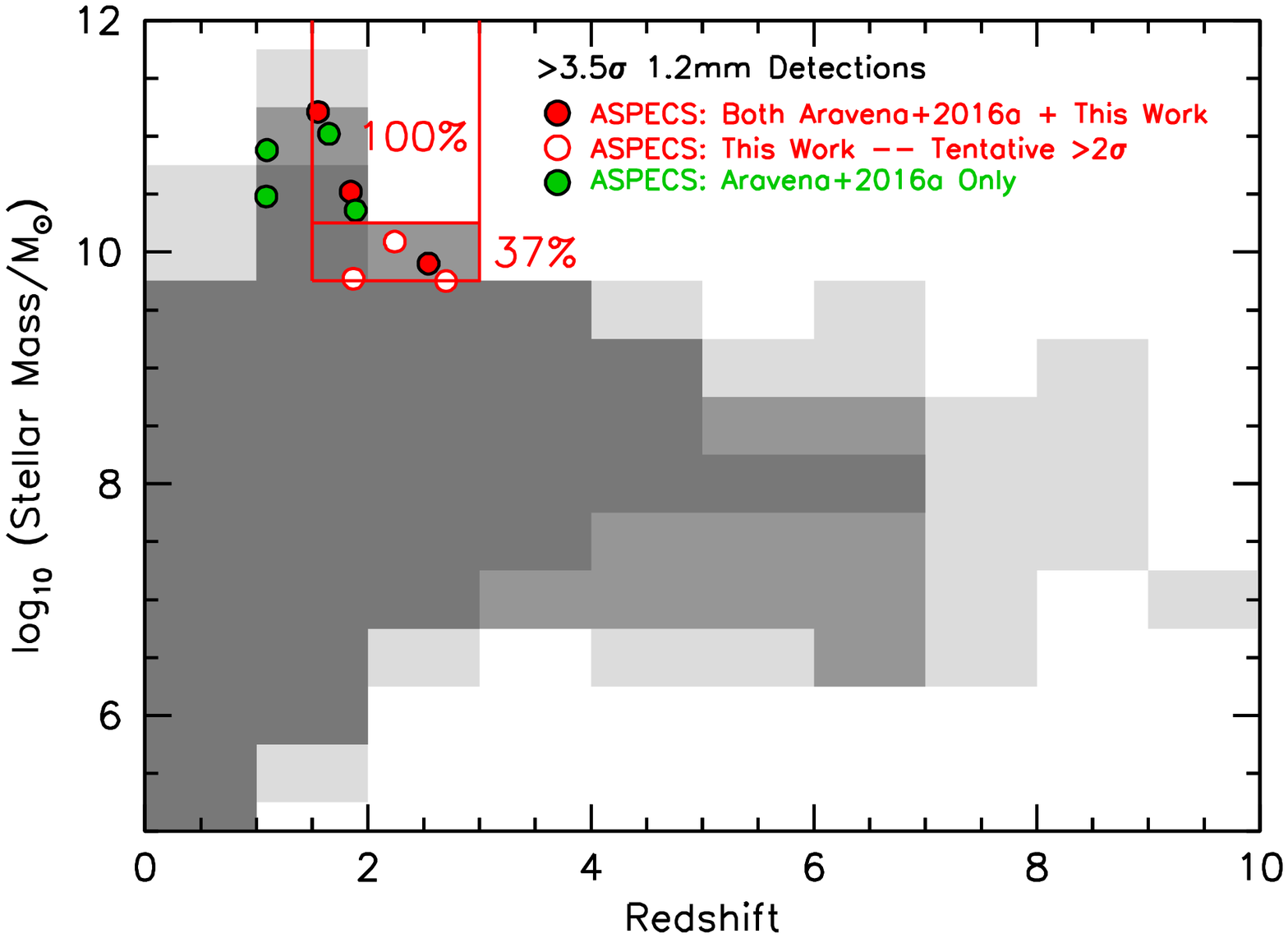}
\caption{Stellar mass vs. redshift range covered by sources identified
  over the $\sim$1 arcmin$^2$ ASPECS region, with light, medium dark,
  and dark gray indicating regions where $\geq$1, $\geq$3, and $\geq$7
  sources are found (where each region subtends $\Delta M\sim 0.5$ dex
  $\times$ $\Delta z\sim1$).  Large filled red circles indicate those
  sources where continuum detections ($>$3.5$\sigma$) are found in
  both present study + paper II of the ASPECS series (Aravena et
  al.\ 2016a).  Green circles indicate sources that are only found in
  paper II of ASPECS (Aravena et al.\ 2016a).  Open red circles
  indicate those galaxies which show tentative $>2\sigma$ detections
  in ASPECS.  100\% and 37\% of the star-forming $z=1.5$-3.0 sources
  in the stellar mass range $\log_{10} M/M_{\odot} > 10.25$ and 9.75-10.25 (\textit{indicated with the red boxes}), respectively,
  show detections in the ALMA continuum 1.2$\,$mm data.  It is clear
  that stellar mass is an especially useful predictor of IR luminosity
  over a wide range in redshift.  Inspiration for this figure came in
  part from Figure 6 of Dunlop et al.\ (2016).\label{fig:massz}}
\end{figure*}

\section{Results}

\subsection{Expected Detections in the Far-IR Continuum}

\subsubsection{Expectations Using the $z\sim0$ IRX-$\beta$ Relations}

We commence our analysis of our ALMA HUDF observations by first asking
ourselves which sources we might expect to detect, given various
results at lower redshift.  Such an exercise will help us interpret
the results which follow and also to evaluate whether or not the
number of sources we detect and rest-frame far-IR flux density we
measure for $z>2$ galaxies are similar to that found for galaxies at
$z\sim0$.

We adopt as our $z\sim0$ baseline the now canonical IRX-$\beta$
relationship of M99 where a connection was found between the infrared
excess (IRX) of galaxies and the spectral slope of the $UV$ continuum:
\begin{equation}
M99: IRX_{M99} = 1.75 (10^{0.4(1.99(\beta+2.23))} - 1)
\label{eq:m99}
\end{equation}
The factor of 1.75 in the above relationship is needed to express the
M99 relation in terms of the IR luminosity, rather than the far-IR
luminosity utilized by M99.  See discussion in \S5.1 of Reddy et
al.\ (2006).  This relationship implicitly includes the slope of the
Calzetti et al.\ (2000) dust law.  Despite modest scatter ($\sim$0.3
dex), redder galaxies were systematically found to show higher
infrared excesses than blue galaxies.  For simplicity, the $B =
BC(1600)_*/BC(FIR)$ factor from M99 is taken to equal one, consistent
with the measurements made in that study.

Importantly, the M99 IRX-$\beta$ relationship was shown to have a
basic utility that went beyond the $z\sim0$ universe for $UV$-selected
samples.  A series of intermediate-redshift studies (Reddy et
al.\ 2006, 2010; Daddi et al.\ 2007; Pannella et al.\ 2009) found this
relationship to be approximately valid when comparing the observed IR
luminosities of galaxies to the predictions from the M99 IRX-$\beta$
relationship.

As an alternative baseline, we also consider the expectations adopting
the so-called Small Magellanic Cloud (SMC) IRX-$\beta$ relationship,
where a connection is again assumed between the infrared excess of a
galaxy and its spectral slope in the $UV$-continuum.  However, since
the SMC dust curve is steeper in the near-$UV$ than dust laws like
Calzetti et al.\ (2000), a small optical depth in dust extinction can
have a large impact on the observed color of a galaxy in the $UV$
continuum.  The infrared excess given an SMC extinction can be
expressed as follows:
\begin{equation}
SMC: IRX_{SMC} = 10^{0.4(1.1(\beta+2.23))} - 1
\label{eq:smc}
\end{equation}
This relationship is derived based on the observational results of
Lequeux et al.\ (1982), Prevot et al.\ (1984), and Bouchet et
al.\ (1985: see also Pei 1992; Pettini et al.\ 1998; Smit et
al.\ 2016).  

For each of the $z=2$-10 sources in our ALMA field, we fit the {\it
  HST} photometry in various bands probing the $UV$-continuum to a
power-law $f_{1600}(\lambda/1600\AA)^{\beta}$ to derive a mean flux at
$\sim$1600$\AA$ and also a spectral slope $\beta$.  We derive a
nominal luminosity for the source in the rest-frame $UV$ by
multiplying the flux density of the source at 1600\AA$\,\,$by the
frequency at that wavelength ($\nu_{1600} f_{1600}$) and convert that
to an expected IR luminosity for the source (considered to extend from
8$\mu$m to 1000$\mu$m).\footnote{In performing these fits, we fix this
  slope to $-2.2$ for our two $z=9$-10 candidate galaxies given the
  lack of sufficiently deep long-wavelength data to constrain the
  $UV$-continuum slopes $\beta$.  The $\beta$ value we utilize here is
  motivated by the results of Bouwens et al.\ (2012), Finkelstein et
  al.\ (2012), Bouwens et al.\ (2014), Kurczynski et al.\ (2014), and
  Wilkins et al.\ (2016a) who find evidence for bluer slopes for
  higher redshift and generally fainter, lower-mass galaxies.}

\begin{figure*}
\epsscale{1.15}
\plotone{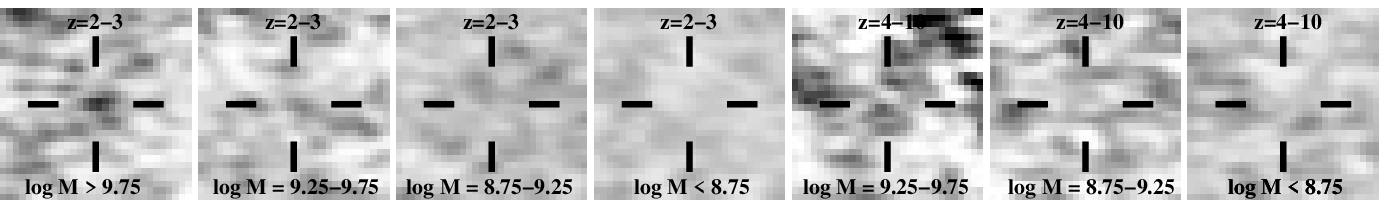}
\caption{Stacked 1.2$\,$mm-continuum images (9''$\times$9'') for all
  candidate $z=2$-3 and $z=4$-10 galaxies falling in four different
  ranges of stellar mass ($<10^{8.75}$ $M_{\odot}$, $10^{8.75}$ to
  $10^{9.25}$ $M_{\odot}$, 10$^{9.25}$ to $10^{9.75}$ $M_{\odot}$, and
  $>10^{9.75}$ $M_{\odot}$).  In the stacks, sources are weighted
  according to the square of their $UV$ flux and the inverse square of
  the noise.  The 3 sources from this analysis individually detected
  at $>$4$\sigma$ are not included in the stack results shown in this
  figure.\label{fig:massstack}}
\end{figure*}

The equivalent flux at an observed wavelength of 1.2$\,$mm is then
computed adopting a modified blackbody form with a dust temperature of
35 K and a dust emissivity power-law spectral index of $\beta_d =
1.6$.  A value of $T_d = 35$ K is intermediate between the
temperatures found for main sequence galaxies by Elbaz et al.\ (2011)
and Genzel et al.\ (2015), i.e., $\sim$30 K at $z\sim2$, and the
$\sim$37-38 K temperatures found for stacked sources in other studies
(Coppin et al.\ 2015).

Since the flux density we would measure with ALMA is reduced somewhat
by the effective temperature of the CMB at $z\sim2$-10, we multiply
the predicted flux (before consideration of CMB effects) by $C_{\nu}$
\begin{equation}
C_{\nu} = \left[ 1 - \frac{B_{\nu} (T_{CMB} (z))}{B_{\nu} (35 K)} \right]
\end{equation}
to compute the expected signal.  This treatment follows prescriptions
given in da Cunha et al.\ (2013).

Performing this exercise over all 330 $z=2$-10 galaxies with coverage
from our ALMA mosaic, we calculated expected fluxes for these sources
at 1.2$\,$mm assuming that sources follow the M99 and SMC IRX-$\beta$
relations.  These calculations suggested that 35 and 26 of these
galaxies should be detected at $\geq 2\sigma$ and $\geq3\sigma$
significance, respectively, in our observations if the M99 IRX-$\beta$
relation applied, while 8 and 5 of these sources would be detected at
$\geq 2\sigma$ and $\geq 3\sigma$ significance, respectively, if the
SMC IRX-$\beta$ relationship applied to $z=2$-10 galaxies in our
samples.

\begin{figure}
\epsscale{1.15} \plotone{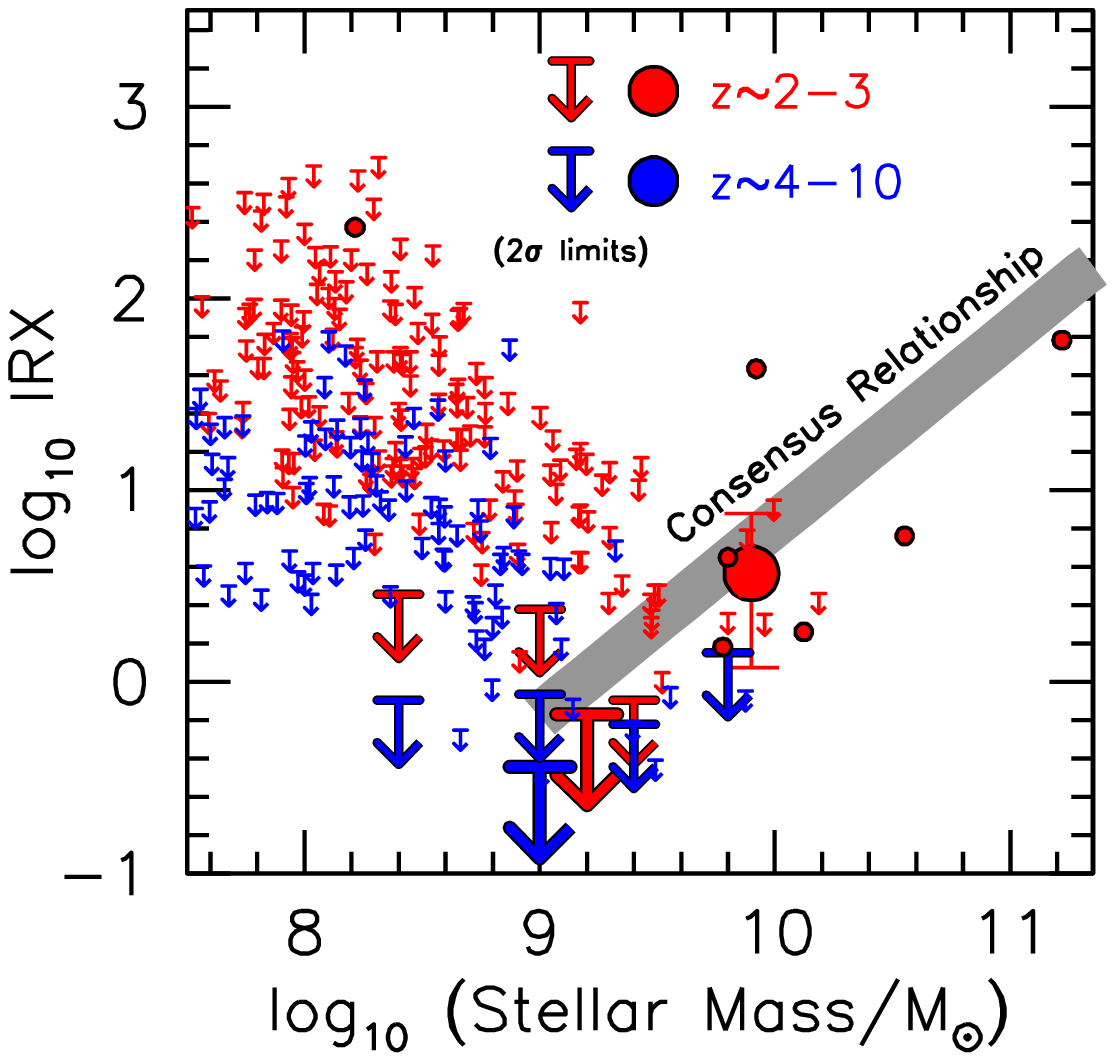}
\caption{Constraints on the infrared excess of $z=2$-3 and $z=4$-10
  galaxies (\textit{large red and blue circles and downward arrows,
    respectively}) obtained by stacking the ALMA 1.2$\,$mm
  observations available for many individual sources over the 1
  arcmin$^2$ ASPECS region (excluding the one source XDFU-2397246112
  with an AGN, but not excluding any other sources).  The small filled
  circles and downward arrows are for sources with a positive
  $2\sigma$ measurement of IRX and $2\sigma$ upper limit on IRX,
  respectively.  Upper limits and errorbars are $2\sigma$ and
  $1\sigma$, respectively.  A redshift-independent dust temperature of
  35 K is assumed in deriving these results.  The presented upper
  limits on our $z=4$-10 sample would be 0.4 dex higher if the dust
  temperature in higher-redshift galaxies is significantly higher than
  at $z\sim1.5$ (i.e., 44-50 K as suggested by the results of Magdis
  et al.\ 2012 and Bethermin et al.\ 2015).  The thick-shaded grey
  line shows the consensus dependence of IRX on galaxy stellar mass we
  derive for $z\sim2$-3 galaxies (Appendix A) from the literature
  (Reddy et al.\ 2010; Whitaker et al.\ 2014; {\'A}lvarez-M{\'a}rquez
  et al.\ 2016).  The ALMA stack results suggest that only galaxies
  with stellar masses in excess of $\gtrsim$$10^{9.75}$ $M_{\odot}$
  tend to output a significant ($\gg$50\%) fraction of their energy at
  far-infrared wavelengths.\label{fig:irxsm}}
\end{figure}

\begin{figure*}
\epsscale{1.15}
\plotone{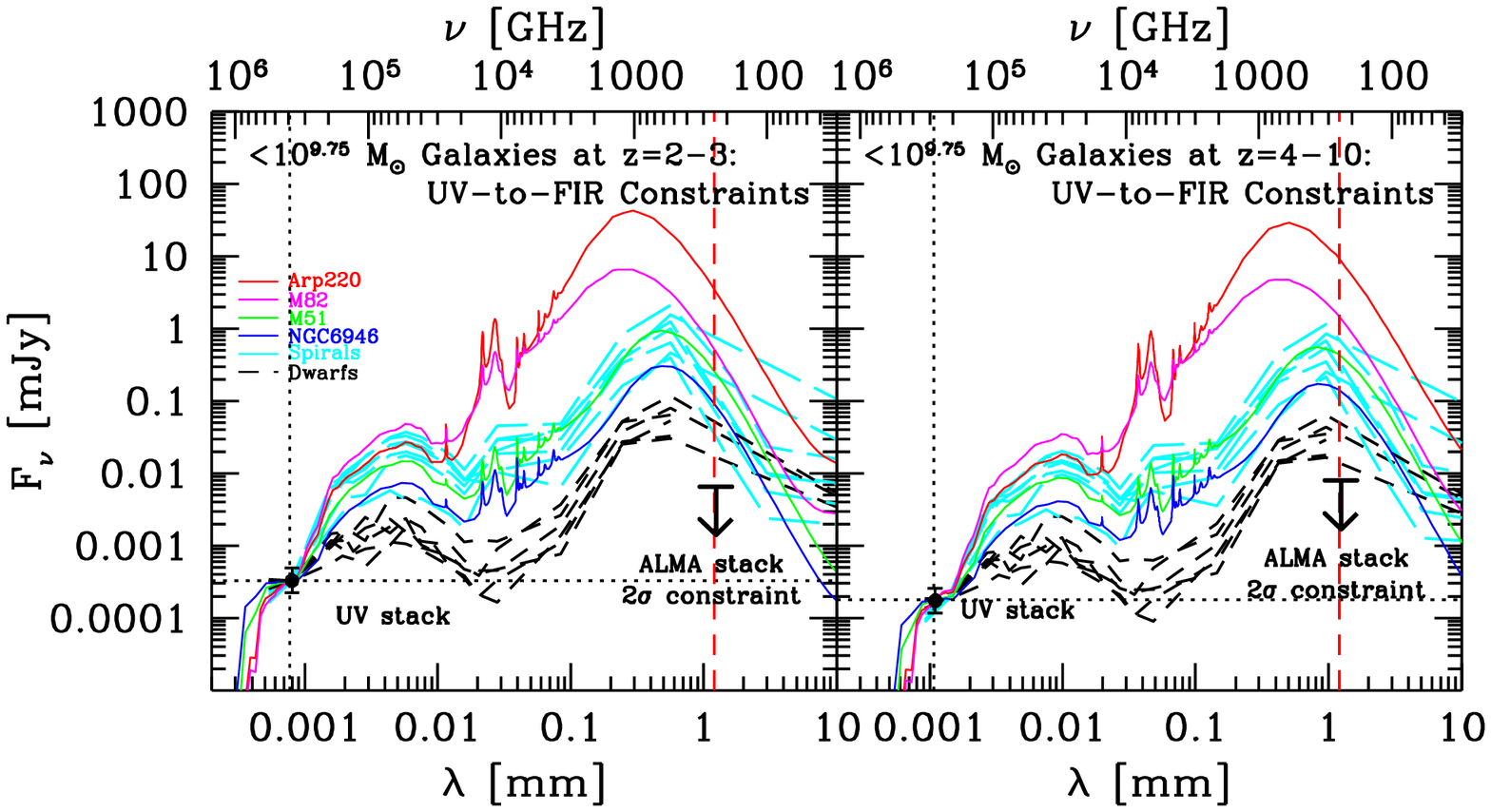}
\caption{Illustration of the general constraints we can set on the
  overall shape of the near-UV + far-infrared spectral energy
  distribution for lower-mass ($<10^{9.75}$ $M_{\odot}$) galaxies at
  $z=2$-3 (\textit{left panel}) and $z=4$-10 (\textit{right panel})
  based on the results we obtain by stacking our deep 1.2$\,$mm ALMA
  HUDF observations at the positions of candidate $z=2$-10 galaxies.
  The flux points at $\sim$0.8 and 1.1$\mu$m shows the UV flux for our
  lower-mass stack of $z=2$-3 and $z=4$-10 candidates, respectively,
  while the large upper limit shows the $2\sigma$ upper limit we can
  set on the flux at $1.2\,$mm (which is just a factor of 20 and 44
  higher, respectively, than the weighted flux in the $UV$ continuum
  that contributes to the stack results shown here: see
  Table~\ref{tab:irxsm} from Appendix D).  For reference, we have
  included model SED templates from Silva et al.\ (1998), Dale et
  al.\ (2007), and some lower-metallicity galaxies studied in the
  KINGFISH project (Dale et al.\ 2012) redshifted to $z\sim4$.  Our
  ALMA observations suggest that faint high-redshift galaxies show
  less dust emission at 1.2$\,$mm than any of these SED
  templates.\label{fig:showsed}}
\end{figure*}

To illustrate these expectations for our $z\geq2$ study, we present
the predicted IR luminosities for our $z\sim2$-10 sample versus
redshift in the left panel of Figure~\ref{fig:probe}.  The red solid
and open circles indicate those sources for which a $3\sigma$ and
$2\sigma$ detection, respectively, is expected in our 1.2$\,$mm continuum
observations while the solid black circles indicate those sources for
which a detection is not expected.  The solid and dotted red line
indicates the lowest IR luminosities that we would detect sources at
$3\sigma$ and $2\sigma$, respectively, over the $\sim$1 arcmin$^2$
ASPECS region.

To help guide the discussion which follows, we provide a complete list
of the sources with expected detections in Table~\ref{tab:catalog}.
Comparisons of the actual flux measurements with estimates based upon
various $z\sim0$ IRX-$\beta$ relations provides us with a quantitative
sense of how much these relations have evolved from $z\geq3$, while
also illustrating the source-to-source scatter.

M99 found that individual sources exhibited a 0.3 dex scatter in
$L_{FIR}$ around the IRX-$\beta$ relationship preferred in that study.
If we include a similar scatter in predicting $L_{IR}$ for individual
sources, we predict 36.9 $2\sigma$ and 28.4 $3\sigma$ detections
instead of 35 and 26, respectively.

The $UV$-continuum slopes $\beta$ we use in setting these expectations
are not known precisely, especially for the faintest sources in our
$z=6$-8 samples.  In particular, if a source is measured to have an
especially red $\beta$ due to the impact of noise, we would predict
its detection in the ASPECS data even if this source is actually
intrinsically blue.  The impact of the scatter is asymmetric since
faint blue sources -- with $\beta$'s in the range of $\sim-2$ to
$\sim-2.3$ (e.g., Wilkins et al.\ 2011; Dunlop et al.\ 2013;
Kurczynski et al.\ 2014) -- are already predicted to show essentially
no dust emission and so the expected emission can only be larger when
adding noise to the photometry of faint sources.

To determine the impact that this would have on the expected number of
detected sources, we perturbed the measured $\beta$'s for individual
sources by the estimated uncertainty, and we calculated the total
number of sources we would expect to find.  Repeating this exercise
multiple times, we find that this would boost the expected number of
detections by $\sim$3.8 sources to 38.8 in total.  This simulation
result suggests that noise in the HST photometry does boost the
expected numbers above what it would be in the noise-free case (by
$\sim$11\%).  If we suppose that a similar correction applies to our
nominal expectations for tentative detections (35 sources), $\sim$31.6
may be a better estimate for the expected number of tentative
detections of $z\sim2$-10 galaxies in ASPECS.

\subsubsection{Expectations Using the $z\sim2$ IRX-Stellar Mass 
Relation and Assuming $L_{IR}=L_{UV}$}

Alternatively, we (1) can use the inferred stellar masses of
$z\sim2$-10 galaxies to estimate their IR luminosities or (2) assume
that the IR luminosities of galaxies matches their luminosities in the
rest-frame UV.  Previous work at $z\sim0$-3 (e.g. Pannella et
al.\ 2009) has demonstrated that the infrared excess of galaxies
exhibits a strong correlation with the stellar mass, and many
different authors (Reddy et al.\ 2010; Whitaker et al.\ 2014; Pannella
et al.\ 2015) recover approximately the same relationship over a wide
range in redshift, i.e., $z\sim0$-3.

As an alternate demonstration of the utility of our ALMA observations,
we show in the center and right panel to Figure~\ref{fig:probe} the
expected detections in our data (\textit{red solid circles}) if we
assume that the luminosities of sources in the IR are either (1)
dictated by the observed relationship between IRX and stellar mass at
$z\sim2$ or (2) equal to their luminosity in the rest-frame $UV$,
respectively.  We adopt the IRX-stellar mass relation presented in
Figure~\ref{fig:lit} from Appendix A, which shows the approximate
consensus relationship at $z\sim2$-3 from three separate studies
(Reddy et al.\ 2010; Whitaker et al.\ 2014; {\'A}lvarez-M{\'a}rquez et
al.\ 2016).

Fifteen and four tentative $>$2$\sigma$ detections are expected,
respectively, for those two cases.

\subsubsection{Impact of the Dust Temperature}

We consider the impact that the assumed dust temperature has on these
results.  If, for example, the mean dust temperature were equal to 30
K as found by Elbaz et al.\ (2011), the expected number of detections
would increase quite significantly.  The totals would be 40 and 13
adopting the M99 IRX-$\beta$ and SMC IRX-$\beta$ relationships,
respectively, while 20 detections would be expected based on the
consensus $z\sim2$-3 IRX-stellar mass relation.

Perhaps, even more importantly, we should consider the possibility
that the dust temperature may increase quite substantially as we move
out to higher redshift.  A variety of work (e.g., Magdis et al.\ 2012;
Bethermin et al.\ 2015) have found considerable evidence for such an
evolution in the typical dust temperature from $z\sim1.5$ to $z\sim4$,
in terms of the mean intensity in the radiation field $<U>$ found to
evolve as $(1+z)^{1.8\pm0.4}$.  As $<U>$ $\propto T^{4+\beta}$, the
temperature can be inferred to evolve as
$(1+z)^{(1.8\pm0.4)/(4+\beta)}\sim (1+z)^{0.32}$, such that for a mean
dust temperature of 35 K at $z\sim1.5$, the implied dust temperature
at $z\sim4$ and $z\sim6$ is equal to 44 K and 49 K, respectively.
Such temperatures are very similar to the 40$\pm$2 K found for a
massive sample of $z\sim4$ galaxies by Schreiber et al.\ (2016) and
the 40-50 K found by Sklias et al.\ (2014) for typical $z\sim2$-3
sources from the Herschel Lens Survey (Egami et al.\ 2010).  We remark
that one might naturally expect an evolution in dust temperature given
the observed evolution in the SFR surface densities observed in
galaxies with cosmic time (e.g., Shibuya et al.\ 2015) and the
correlation of dust temperature with SFR surface density (e.g., Elbaz
et al.\ 2011).

Assuming that the dust temperatures monotonically increase towards
high redshift, as (35 K)$((1+z)/2.5)^{0.32}$, we predict that we
should tentatively detect 20 sources using the M99 IRX-$\beta$
relationship, 3 sources using the SMC IRX-$\beta$ relationship, 11
sources using the consensus $z\sim2$-3 IRX-stellar mass relationship,
and 0 sources assuming $L_{IR} = L_{UV}$.

\subsection{Continuum Detections of Individual Sources at 1.2$\,$mm}

Here we look for possible individual detections of $z=2$-10 galaxies
over our deep ALMA continuum map at 1.2$\,$mm.  As results from the
previous section illustrate, we could reasonably expect the number of
detections to be modest if various IRX-$\beta$ relations from the
$z\sim0$ universe serve as a useful guide.

Table~\ref{tab:detect} provides a summary of the properties of the
$z\gtrsim2$ sources from our catalog of 330 $z=2$-10 sources that are
nominally tentatively detected at $\gtrsim2\sigma$ in our data.  The
measured flux density for the detected sources was derived taking the
value in our 1.2$\,$mm-continuum image at the nominal optical position
of each source in our LBG samples (after correcting for the 0.3$''$
positional offset between the ALMA and optical maps).  We verified
that we would retain all of our most significantly detected sources
from this table, if we derived flux densities for sources using other
methods (e.g., by scaling the normalization of the primary beam to fit
the pixels in a $3''\times3''$ aperture centered on a source).

Only 2 sources from the entire catalog are detected at $\gg$3$\sigma$
significance.  They are the $z=2.54$ source XDFU-2385446340, where the
detection significance is indeed very high, i.e., 41$\sigma$, with
$f_{1.2mm}=571\pm14\mu$Jy, and the $z=1.55$ source XDFU-2397246112,
where the detection significance is $10\sigma$, with
$f_{1.2mm}=261\pm25\mu$Jy.  In paper IV in this series (Decarli et
al.\ 2016a), we discuss the far-IR SED and molecular gas properties of
both sources in more detail.  Based on its X-ray flux in the deep
Chandra observations over the CDF South (Xue et al.\ 2011), the latter
source (XDFU-2397246112) is known to host an X-ray AGN.

Six other sources from our catalogs show convincing $>$3.5$\sigma$ or
tentative $>$2$\sigma$ detections in our ASPECS data.  However, 2 of
these detections appear to be noise spikes.  This can been seen by
looking for similar $>2\sigma$ detections in the negative continuum
image for sources with similar stellar masses (also presented in
Table~\ref{tab:detect}).  For the sources with the highest masses
(i.e., $>10^{9.75}$ $M_{\odot}$), only four positive $2\sigma$
detections are found and no $2\sigma$ ``detections'' in the negative
images.  The positive detections correspond to XDFU-2373546453,
XDFU-2370746171, XDFU-2365446123, and XDFU-2384246348 with
$5.1\sigma$, $2.6\sigma$, $2.4\sigma$, and $2.3\sigma$ detections,
respectively.  Given that there are only 13 sources in our highest
mass sample and six of them show at least a tentative $>$2.3$\sigma$
detection in our ALMA observations (expected only 1\% of the time
assuming gaussian noise), each of these detections is likely
real.\footnote{Our conclusions here differ significantly from what we
  would conclude based on a blind search for$>$2$\sigma$ detected
  sources across the entire 1 arcmin$^2$ mosaic (where the fidelity is
  only 25\% for sources with flux densities between 30 and 40$\mu$Jy:
  Aravena et al.\ 2016a).  With a blind search, one has many
  opportunities to find tentative $>2\sigma$-detected sources;
  however, for the present high-mass sample, one only has 13
  opportunities.}

We remark that each of these four sources is also detected in the MIPS
24$\mu$m observations at $\gtrsim$2$\sigma$ significance, providing
further support for our conclusions here (see Table~\ref{tab:mipsalma}
from Appendix B).  This also points towards MIPS 24$\mu$m data being a
valuable probe of the infrared excess to $z\sim 3$, given its
competitive sensitivity to long exposures with ALMA.

For sources with estimated stellar masses in the range $10^7$ to
$10^9$ $M_{\odot}$, tentative $2\sigma$ detections are seen in both
the positive and negative images.  The excess numbers in the negative
image appear not to be statistically significant, as small changes to
the positions where the flux measurements are made (by $\sim$0.1$''$)
typically result in essentially identical numbers of tentative
$2\sigma$ detections in the positive and negative continuum images.

\begin{deluxetable}{cccc}
\tablewidth{0cm}
\tablecolumns{4}
\tabletypesize{\footnotesize}
\tablecaption{Inferred IRX vs. Galaxy Stellar Mass and $\beta$ from ASPECS (assuming $T_d=35$ K and $\beta_d=1.6$)\tablenotemark{$\dagger$}\label{tab:irx}}
\tablehead{\colhead{} & \colhead{} & \colhead{\# of} & \colhead{}\\
\colhead{Stellar Mass} & \colhead{$\beta$} & \colhead{sources} & \colhead{IRX\tablenotemark{a}}}
\startdata
\multicolumn{4}{c}{$z=2$-3}\\
$>$10$^{9.75}$ $M_{\odot}$ & All & 11 & 3.80$_{-2.40}^{+3.61}$$\pm$0.19\\
$<$10$^{9.75}$ $M_{\odot}$ & All & 151 & 0.11$_{-0.42}^{+0.32}$$\pm$0.34\\\\
\multicolumn{4}{c}{$z=4$-10}\\
$>$10$^{9.75}$ $M_{\odot}$ & All & 2 & $-$0.49$_{-1.13}^{+0.69}$$\pm$0.71\\
$<$10$^{9.75}$ $M_{\odot}$ & All & 166 & 0.14$_{-0.14}^{+0.15}$$\pm$0.18\\\\
\multicolumn{4}{c}{$z=2$-3}\\
$>$10$^{9.75}$ $M_{\odot}$ & $-4<\beta<-1.75$ & 1 & 0.54$_{-0.00}^{+0.00}$$\pm$0.29\\
 & $-1.75<\beta<-1.25$ & 2 & 1.31$_{-0.94}^{+0.67}$$\pm$0.72\\
 & $\beta<-1.25$ & 8 & 6.79$_{-4.51}^{+5.38}$$\pm$0.26\\
$<$10$^{9.75}$ $M_{\odot}$ & $-4<\beta<-1.75$ & 89 & 0.19$_{-0.76}^{+0.40}$$\pm$0.44\\
 & $-1.75<\beta<-1.25$ & 49 & $-0.01$$_{-0.35}^{+0.39}$$\pm$0.58\\
 & $\beta<-1.25$ & 13 & $-0.14$$_{-3.64}^{+4.65}$$\pm$2.11\\\\
\multicolumn{4}{c}{$z=4$-10}\\
$<$10$^{9.75}$ $M_{\odot}$ & $-4<\beta<-1.75$ & 122 & 0.03$_{-0.15}^{+0.22}$$\pm$0.24\\
 & $-1.75<\beta<-1.25$ & 29 & 0.33$_{-0.16}^{+0.11}$$\pm$0.29\\
 & $\beta<-1.25$ & 11 & $-$1.03$_{-1.23}^{+0.29}$$\pm$1.46
\enddata
\tablenotetext{$\dagger$}{See Tables~\ref{tab:irxsm}-\ref{tab:irxmuv} from Appendix D for a more detailed presentation of the stack results summarized here.}
\tablenotetext{a}{Both the bootstrap and formal uncertainties are quoted on the result (presented first and second, respectively).}
\end{deluxetable}

To illustrate the significance of this apparent dependence on the
inferred stellar mass, we present in Figure~\ref{fig:detsm} the
fraction of detected $z=2$-10 galaxies versus mass, after correcting
the positive $>2\sigma$ detections in a mass bin for the $>2\sigma$
detection seen in the negative image.  For this figure, we only
consider those sources (172 out of 330) over the ASPECS field which
are the $1.2\,$mm-continuum senstivities are the highest, i.e., with
$1\sigma$ RMS noise $<$17$\mu$Jy.  63$_{-17}^{+14}$\% of the galaxies
with stellar mass estimates $>10^{9.75}$ $M_{\odot}$ are detected;
none of the sources with masses lower than $10^{9.75}$ $M_{\odot}$ are
detected.  Dust-continuum emission shows a clear connection with the
apparent stellar mass in galaxies -- which is similar to a few
prominent earlier predictions for the expected findings from a deep
far-IR continuum survey over the HUDF with ALMA (da Cunha et
al.\ 2013).

As a separate illustration of the predictive power of stellar mass in
estimating the approximate luminosity of galaxies in the IR, we
present in Figure~\ref{fig:massz} the range in stellar mass
vs. redshift probed by our HUDF sample and indicate the sources we
identify as detected in red (open and solid circles detected at
$>$3$\sigma$ and $2$-3$\sigma$, respectively) and as only found in a
blind search in green (ASPECS paper II: Aravena et al.\ 2016a).  The
sources we detected at $>$3$\sigma$ also appear in the Aravena et
al.\ (2016a) blind search.  It is clear that stellar mass is a good
predictor of which sources are IR luminous, for galaxies with $z>1.5$.
The stellar masses used for constructing Figure~\ref{fig:massz} are
taken from the 3D-HST catalogs (Skelton et al.\ 2014) if at $z<1.5$
(if available); otherwise, they are inferred as in \S2.3.  For the
most obscured systems, estimates of the redshift and stellar mass can
be quite uncertain (given degeneracies between dust and age and
challenge in locating spectral breaks), so some caution is needed in
interpreting this figure.

We now return to discussing continuum-detected sources in ASPECS.
1.2$\,$mm-continuum images of the 6 sources showing meaningful
detections are presented in Figure~\ref{fig:ddetect} together with
their {\it HST} and {\it Spitzer}/IRAC images.  We remark that the
detected sources in the present samples are much fainter than those
identified in many previous programs.  For example, the typical flux
measured by Scoville et al.\ (2016) for detected $z\sim4.4$ galaxies
in their very high-mass ($\gtrsim2\times10^{10}$ $M_{\odot}$) sample
is $\sim$200$\mu$Jy, which contrasts with the $\sim$35$\mu$Jy seen in
the 3 faintest $z\sim2$-3 galaxies tentatively detected here.  The
observed differences in the typical fluxes of detected sources is a
natural consequence of the relative sensitivities of the data sets,
i.e., 12.7$\mu$Jy/beam RMS for ASPECS vs. 65$\mu$Jy/beam RMS in the
Scoville et al.\ (2016) observations.

\subsection{Stacked constraints on the Infrared Excess}

In addition to looking at which $z=2$-10 galaxies over the HUDF we can
individually detect in our ALMA continuuum observations, we can gain
powerful constraints on dust emission from high-redshift galaxies by
stacking.  For this, we subdivide our samples in terms of various
physical properties and then do a weighted stack of the ALMA-continuum
observations at the positions of the candidates.

For sources included in the stack, we map the ALMA continuum maps onto
the same position and weight the contribution of each source to the
stack according to its expected 1.2$\,$mm continuum signal assuming
$L_{IR} \propto L_{UV}$ and according to the inverse square of the
noise (per beam).  We derive a flux from the stack based on a
convolution of the image stack (3.3$''$$\times$3.3$''$ aperture) with
the primary beam.  No spatial extension is assumed in the stacked
flux.

\begin{figure}
\epsscale{1.15}
\plotone{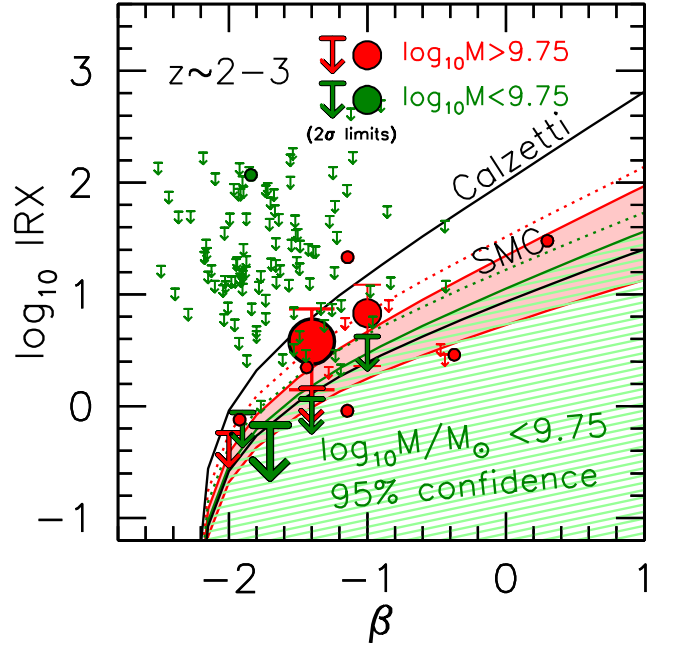}
\caption{Stacked constraints on the infrared excess in $z=2$-3
  galaxies versus the $UV$-continuum slope $\beta$.  These results are
  shown for higher- and lower-mass subsamples ($>10^{9.75}$
  $M_{\odot}$ and $<10^{9.75}$ $M_{\odot}$) of $z=2$-3 galaxies
  (\textit{red and green solid circles and downward arrows,
    respectively}) and were obtained by stacking the ALMA 1.2$\,$mm
  observations of individual sources over the ASPECS region (excluding
  the one source XDFU-2397246112 with an AGN, but not excluding any
  other sources).  Upper limits and errorbars are $2\sigma$ and
  $1\sigma$, respectively.  The smaller solid circles and downward
  arrows indicate a $>2\sigma$ measurement of the infrared excess and
  $2\sigma$ upper limits on this excess.  The purpose of the smaller
  points is to show the constraints that can be derived from
  individual sources before stacking.  The very large red circle and
  downward-pointing green arrow gives the value and $2\sigma$ upper
  limit, respectively, on IRX based on all $>$10$^{9.75}$ $M_{\odot}$
  and $<$10$^{9.75}$ $M_{\odot}$ $z=2$-3 sources from ASPECS plotted
  at the weighted value of $\beta$ contributing the most signal to
  this measurement of IRX.  The solid lines show the nominal
  IRX-$\beta$ relation one would derive based on the Calzetti and SMC
  dust laws.  The shaded red and light green regions indicate the 68\%
  and 95\% confidence intervals we can derive on the IRX-$\beta$
  relationship on the basis of our results for sources with stellar
  masses of $>10^{9.75}$ $M_{\odot}$ and $<10^{9.75}$ $M_{\odot}$,
  respectively.  If the dust temperature increases towards high
  redshift as indicated by Bethermin et al.\ (2015), the upper bound
  on these regions would be given by the dotted red and green regions.
  Our results are consistent with the IR emission from high-mass
  ($>10^{9.75}$ $M_{\odot}$) $z\sim2$-3 galaxies exhibiting an SMC
  IRX-$\beta$ relation.  However, for lower-mass ($<10^{9.75}$
  $M_{\odot}$) galaxies, our results suggest lower infrared excesses,
  less even than expected for an SMC dust law.\label{fig:irxbeta23}}
\end{figure}

\begin{figure*}
\epsscale{1.1}
\plotone{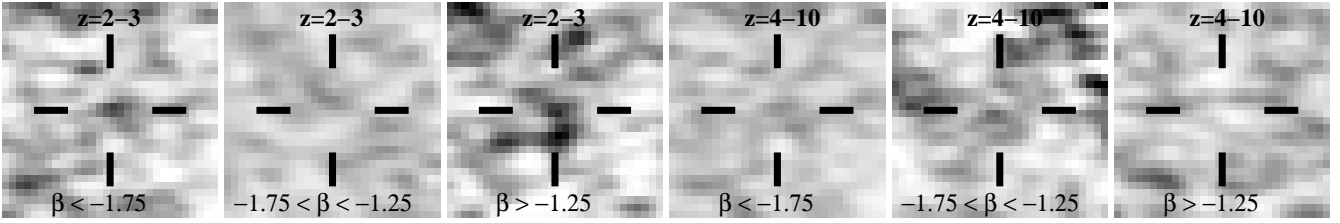}
\caption{Stacked 1.2$\,$mm-continuum images (9''$\times$9'') for
  $z=2$-3 and $z=4$-10 galaxies falling in different bins of
  $UV$-continuum slope $\beta$.  All sources that are individually
  detected at $\geq4\sigma$ are not included in the presented stack
  results.  Only the most massive ($>10^{9.75}$ $M_{\odot}$) sources
  are included in our $z=2$-3 stacks, while our $z=4$-10 stacks
  include sources over the full mass range (due to the small number of
  sources with $>10^{9.75}$ $M_{\odot}$).  In the stacks, sources are
  weighted according to the expected 1.2$\,$mm-continuum flux
  (assuming $L_{IR} \propto L_{UV}$) and according to the inverse
  square of the noise.  The 3 individually-detected sources (at
  $>$4$\sigma$) are not included in the presented stack
  results.\label{fig:betastack}}
\end{figure*}

\begin{figure}
\epsscale{1.15}
\plotone{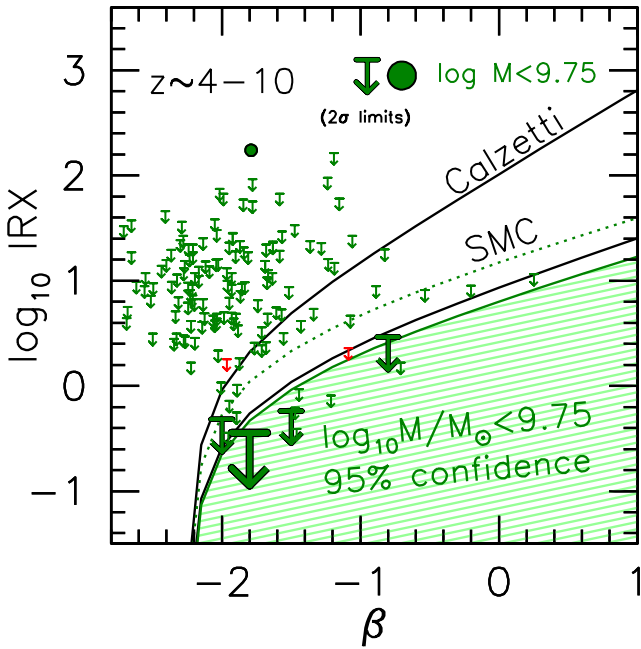}
\caption{Stacked constrants on the infrared excess in $z=4$-10
  galaxies versus $\beta$.  Similar to Figure~\ref{fig:irxbeta23} but
  for galaxies in the redshift range $z=4$-10.  We only present
  results for the lower-mass subsample, as we find only 2 $>10^{9.75}$
  $M_{\odot}$ galaxies over the 1 arcmin$^2$ ASPECS region and those 2
  sources are not detected.  Our stack results (indicated by the
  larger downward-pointing arrows which express the $2\sigma$ upper
  limits) strongly suggest that the infrared excess for the typical
  lower-mass $<10^{9.75}$ $M_{\odot}$ galaxy is low, even below that
  expected for an SMC dust law.  The very large downward pointing
  arrow is as in Figure~\ref{fig:irxbeta23}, but for $z=4$-10
  galaxies.  The light-green-shaded region gives our derived
  constraints (95\% confidence intervals) on the IRX-$\beta$
  relationship for $z=4$-10 galaxies with all but the highest stellar
  masses ($<10^{9.75}$ $M_{\odot}$).  The dotted green line indicates
  the upper bound on this region, if the dust temperature is much
  higher at $z=4$-10 than at $z\sim1.5$ (i.e., 44-50 K as suggested by
  the results of Bethermin et al.\ 2015). \label{fig:irxbeta410}}
\end{figure}

\subsubsection{IRX versus Stellar Mass}

We begin by looking at the average inferred infrared excesses of
$z=2$-10 galaxies as a function of the stellar mass.  Segregating our
samples in terms of stellar mass certainly is a logical place to
start.  Not only is there strong support in the literature for such a
correlation at lower redshifts (e.g., Pannella et al.\ 2009, 2015;
Reddy et al.\ 2010), but there is evidence for this correlation being
present in our own limited samples (see \S3.2).

In Figure~\ref{fig:massstack}, we show the stacked 1.2$\,$mm-continuum
observations of $z=2$-3 and $z=4$-10 galaxies in four different bins
of stellar mass: $>$10$^{9.75}$ $M_{\odot}$, 10$^{9.25}$-10$^{9.75}$
$M_{\odot}$, 10$^{8.75}$-10$^{9.25}$ $M_{\odot}$, and $<$10$^{8.75}$
$M_{\odot}$.  For these stacks, we weight sources according to the
square of the expected signal in the 1.2$\,$mm-continuum observations
(assuming $L_{IR} \propto L_{UV}$) and the inverse square of the noise
[in $\mu$Jy], i.e., $(L_{UV}/\sigma(f_{1.2mm}))^2$.\footnote{This
  weighting factor is just equal to the inverse square of the expected
  noise in a measurement of the infrared excess (noting that
  $L_{IR}\propto f_{1.2mm}$ and that the fractional uncertainty in
  $L_{UV}$ is negligible relative to that in $L_{IR}$ for all sources
  considered in this study).  We remark that photometric redshift
  errors should also have an impact on the uncertainties in
  $L_{IR}/L_{UV}$ and hence impact the weighting, but such
  uncertainties are small, given our stacks do not generally yield
  detections.}

The implied constraints on IRX as a function of stellar mass are
presented in Figure~\ref{fig:irxsm}, Table~\ref{tab:irx}, and
Table~\ref{tab:irxsm} from Appendix D for both our $z=2$-3 and
$z=4$-10 samples.  Significantly enough, the only mass bin where we
find a detection is for $>$10$^{9.75}$ $M_{\odot}$ galaxies at
$z=2$-3.  This is not surprising since 6 of the 11 sources that
compose this mass bin show tentative individual detections
($\gtrsim$2$\sigma$) in our ALMA observations.  All the other masss
bins are consistent with the infrared excess showing an approximate
$2\sigma$ upper limit of $IRX$$\sim$0.4 for $<$10$^{9.75}$ $M_{\odot}$
galaxies.

Making use of the collective constraints across our $z\sim2$-10
sample, we find an approximate $2\sigma$ upper limit on the infrared
excess of 0.4 for lower-mass ($<10^{9.75}$ $M_{\odot}$) galaxies.
This suggests that dust emission from faint UV-selected sources is
typically small.

In our stacking experiments, we also compute a constraint on the flux
at 1.2$\,$mm relative to the flux in the $UV$-continuum.  For these
results, sources are weighted according to the square of their
$UV$-continuum fluxes and inversely according to the noise in the ALMA
1.2$\,$mm observations.  Making use of all sources in our $z\sim2$-3
and $z\sim4$-10, $<10^{9.75}$ $M_{\odot}$ samples, we find a $2\sigma$
upper limit of 20 and 44, respectively, on the ratio of fluxes at
1.2$\,$mm and in the $UV$-continuum.

The impact of this result is illustrated in Figure~\ref{fig:showsed},
by comparing current constraints against several possible SED
templates at $z\sim2$-3 and $z\sim4$-10.  The result provides
information on the overall shape of the spectral energy distribution
that is independent of the assumed SED template.

\subsubsection{IRX versus $\beta$}

Next we subdivide our $z=2$-10 samples in terms of their
$UV$-continuum slopes.  Given evidence that the infrared excess
depends significantly on $\beta$ at $z\sim0$ (M99) and also at
$z\sim2$ (Reddy et al.\ 2006, 2010; Daddi et al.\ 2007; Pannella et
al.\ 2009), we want to quantify this dependence in our own sample.  We
split our results by stellar mass (i.e., $<10^{9.75}$ $M_{\odot}$ and
$>10^{9.75}$ $M_{\odot}$) motivated by the results of the previous
section.

We examine the IRX-$\beta$ relation for $z=2$-3 sources with
$>10^{9.75}$ $M_{\odot}$ in Figure~\ref{fig:irxbeta23},
Table~\ref{tab:irx}, and Table~\ref{tab:irxbeta} from Appendix D using
three different bins in $\beta$.  The only source from the present
ASPECS sample that shows a prominent X-ray detection (XDFU-2397246112)
is excluded.  Stacks of the ALMA continuum images at the positions of
the sources are provided in Figure~\ref{fig:betastack}, after
excluding those sources detected at $>$4$\sigma$.

Given the small sample size, it is difficult to compute accurate
uncertainties on the IRX-$\beta$ relationship at $z\sim2$-3, but the
large red solid circles give our best estimates.  The present
constraints appear most consistent with an SMC IRX-$\beta$
relationship.

We formalize this analysis by calculating the region of the
IRX-$\beta$ plane preferred at 68\% confidence.  For this, we compare
the derived IRX-$\beta$ relationship with what would be predicted
based on dust laws with various slopes $dA_{UV}/d\beta$ (where
$IRX=10^{0.4(dA_{UV}/d\beta)(\beta+2.23)}-1$).  The result we obtain
for $dA_{UV}/d\beta$ is 1.26$_{-0.36}^{+0.27}$ ($1.26_{-0.91}^{+0.49}$
at 95\% confidence) and is presented in Figure~\ref{fig:irxbeta23} as
a light-red shaded region.  It is most consistent with an SMC dust law
(where $dA_{UV}/d\beta \sim 1.1$).

\begin{deluxetable}{ccc}
\tablecolumns{3}
\tabletypesize{\footnotesize}
\tablecaption{Present constraints on the IRX-$\beta$ relationship\label{tab:abeta}}
\tablehead{\colhead{Sample} & \colhead{Mass Range} & \colhead{$dA_{UV}/d\beta$}}
\startdata
\multicolumn{3}{c}{Current Determinations}\\
$z\sim2$-3 & $>10^{9.75}$ $M_{\odot}$ & 1.26$_{-0.36}^{+0.27}$\\
$z\sim2$-3 & $<10^{9.75}$ $M_{\odot}$ & $<$1.22\tablenotemark{a}\\
$z\sim4$-10 & $<10^{9.75}$ $M_{\odot}$ & $<$0.97\tablenotemark{a,b}\\\\
\multicolumn{3}{c}{Canonical IRX-$\beta$ Relations}\\
\multicolumn{2}{c}{Meurer / Calzetti} & 1.99 \\
\multicolumn{2}{c}{SMC} & $\sim$1.10 
\enddata
\tablenotetext{a}{Upper limits are 2$\sigma$.}
\tablenotetext{b}{Upper limit is corrected for the expected noise in the derived $\beta$ values.}
\end{deluxetable}

We also quantify the IRX-$\beta$ relationship for the lower-mass
sources at $z=2$-3 and present the stack results in
Figure~\ref{fig:irxbeta23} as $2\sigma$ upper limits (\textit{downward
  green arrows}) and also in Table~\ref{tab:irxbeta} from Appendix D.
The limits are much lower than the constraints we obtained for the
highest-mass galaxies considered here and indicate that the
IRX-$\beta$ relationship depends on the stellar mass of the sources.
As with our higher mass sample, we use the stacked constraints to
constrain the IRX-$\beta$ relation (shown with the light-green-shaded
region in Figure~\ref{fig:irxbeta23}), finding $dA_{UV}/d\beta$ to
$<$1.22 at 95\% confidence.

We also derive constraints on the IRX-$\beta$ relationship for our
$z=4$-10 sample.  This sample only contain 2 galaxies with stellar
masses in excess of $10^{9.75}$ $M_{\odot}$ -- neither of which are
detected in our ALMA observations -- so we do not consider a
higher-mass subsample of galaxies.  The image stamps showing the stack
results are presented in Figure~\ref{fig:betastack}, while the
infrared excess derived from these stack results are presented in
Table~\ref{tab:irxbeta} from Appendix D and in
Figure~\ref{fig:irxbeta410}.  As in Figure~\ref{fig:irxbeta23}, we
determine the implications of these constraints for the IRX-$\beta$
relationship and present the result in Figure~\ref{fig:irxbeta410}
using a green-shaded contour.  The $2\sigma$ upper limit we derive for
$dA_{UV}/d\beta$ is 0.87.  Again, our derived constraints on
$dA_{UV}/d\beta$ suggests that dust emission from lower-mass
$<10^{9.75}$ $M_{\odot}$ galaxies is generally quite small.

Uncertainties in the measured $UV$-continuum slopes $\beta$ also have
an impact on these results, as scatter towards redder colors could
cause us to include intrinsically blue sources in the reddest $\beta$
bins.  Since sources with the reddest colors are expected to have
these colors due to dust extinction, these bins have significantly
more leverage in determining the value for $dA_{UV}/d\beta$ we derive.
Noise has a particularly significant impact on the $\beta$'s derived
from the highest redshift sources in our samples, i.e., particularly
at $z=7$-8.

To determine the impact of noise on the values we derive for
$dA_{UV}/d\beta$, we perturbed the best-estimate $\beta$ values for
individual sources by our uncertainty estimates on each $\beta$
determination, rebinned the sources as for our fiducial results, and
then rederived the $dA_{UV}/d\beta$ for the perturbed data set.  We
repeated this process 10$\times$ and we found that the $2\sigma$ upper
limit on $dA_{UV}/d\beta$ decreased on average by 0.1 for our
low-mass, $z=4$-10 sample, but did not have a noticeable impact on the
$dA_{UV}/d\beta$ value we derived for our $z=2$-3 sample.

We therefore correct the $1\sigma$ upper limit we derive on
$dA_{UV}/d\beta$ by 0.0 and 0.1 for our low-mass $z=2$-3 and $z=4$-10
samples, respectively.  This translates into $2\sigma$ upper limits on
$dA_{UV}/d\beta$ of 1.22 and 0.97 for our low-mass samples at $z=2$-3
and $z=4$-10, respectively.  All of the present constraints on
$dA_{UV}/d\beta$ for sources at different redshifts and with different
stellar masses are summarized in Table~\ref{tab:abeta}.

The conclusions here can be impacted by our assumed dust temperatures.
If we assume monotonically higher dust temperatures at $z\geq3$ such
that $T_d \sim 44$-50 K at $z\sim4$-6, then the $2\sigma$ upper limit
on the IRX-$\beta$ relation increases by 0.37 dex (\textit{shown as a
  dotted line}), consistent with an SMC IRX-$\beta$ relationship.
Previously, Ouchi et al.\ (1999) had argued on the basis of SCUBA
850$\mu$m observations over the Hubble Deep Field North (Hughes et
al.\ 1998) that $z\gtrsim3$ star-forming galaxies could be consistent
with the $z\sim0$ M99 relation only if the dust temperature was
$\gtrsim$40 K.

The present results are not especially different from IRX vs. $\beta$
resulted by Capak et al.\ (2015) for a small sample of $z\sim5$-6
galaxies where most of the sources in their sample lie below the SMC
relation, but are clearly lower than the Coppin et al.\ (2015) results
where IRX was found to be $\sim$ 8 for $z\sim3$ sources with $\beta$'s
of $\sim-2$ to $-1.5$.  The Coppin et al.\ (2015) results were based
on a deep stack of SCUBA-2 (Holland et al.\ 2013) Cosmology Legacy
Survey data (Geach et al.\ 2013) over the UKIDSS-UDS field (Lawrence
et al.\ 2007).  The explanation for differences relative to the Coppin
et al.\ (2015) results is not entirely clear.\footnote{Given the broad
  SCUBA-2 beam, it is possible that the far-IR stacks Coppin et
  al.\ (2015) create of the bluer sources include some flux from their
  bright neighbors.  Also the Coppin et al.\ (2015) stack results
  surely include more massive $z\sim3$ galaxies than in the present
  narrow field probe, and since IRX is typically much larger for more
  massive galaxies at a given $\beta$ (e.g. Reddy et al.\ 2006), this
  could contribute to the observed differences.}

\subsubsection{IRX versus Apparent Magnitude in the Rest-Frame $UV$}

Lastly, we look at the infrared excess as a function of the apparent
magnitude of sources in the rest-frame $UV$.  Knowing the dependence
of the infrared excess on the apparent magnitude is valuable, given
the relevance of this variable to source selection and also its close
connection to the SFR (if dust obscuration is low).

Again we break up our samples into two different redshift bins $z=2$-3
and $z=4$-10.  We consider a bright sample $m_{UV,AB}<25$ ($z=2$-3)
and $m_{UV,AB}<26$ ($z=4$-10) and a faint sample $m_{UV,AB}>25$
($z=2$-3) and $m_{UV,AB}>26$ ($z=4$-10).  As with our other stack
results, we weight the signal from individual sources to maximize the
signal in our measurement of the infrared excess.  Our stack results
for the different bins in apparent magnitude are presented in
Table~\ref{tab:irxmuv} from Appendix D.

Only the brightest ($H_{160,AB}<25$) $z=2$-3 galaxies show a detection
in our stack results.  This is consistent with IRX being positively
correlated with the star formation rate observed in the rest-frame
$UV$ for galaxies.  There has been substantial discussion in previous
work (e.g., Reddy et al.\ 2006) regarding a general correlation of IRX
with the SFR, though this correlation appears to show strong evolution
as a function of redshift (Reddy et al.\ 2010; Dom{\'{\i}}nguez et
al.\ 2013) such that $z\sim2$-3 galaxies show much less extinction at
a given SFR than at $z\sim0$.

\subsection{Sensitivity of Results to the Dust Temperature}

Essentially nothing is known about the typical dust temperature for
sub-L$^*$ star-forming galaxies at $z\sim2$-10.  While many
star-forming galaxies have measured dust temperatures of $\sim$25-30 K
(Magnelli et al.\ 2014; Elbaz et al.\ 2011; Genzel et al.\ 2015),
there are many faint individually detected sources which have much
higher dust temperatures (Sklias et al.\ 2014), i.e., $\sim$40-50 K.
Moreover, the dust temperature is known to depend on its sSFR relative
to that median value on the main sequence, ranging from values of
$\sim$20 K, $\sim$30 K, and $\sim$40 K depending on whether a galaxy
is below, on, and above the main sequence, respectively (Elbaz et
al.\ 2011; Genzel et al.\ 2015).

Uncertainties in the dust temperature of lower-mass, $z\geq2$ galaxies
are important since the results we derive depend significantly on the
form of the far-IR SED we assume.  To illustrate, if we assume that
the dust temperatures are lower than 35 K, it would imply lower IR
luminosities (and stronger upper limits on the luminosities).  On the
other hand, if we assume that the dust temperature of sub-L$^*$
galaxies at $z\sim2$-10 is higher than 35 K, it would imply higher IR
luminosities (and weaker upper limits on the IR luminosity) for
$z\geq2$ sources probed by the ASPECS program.  The latter possibility
would appear to be a particularly relevant one to consider in light of
recent results from Magdis et al.\ (2012) and Bethermin et al.\ (2015)
which have suggested that the mean intensity in the radiation field
(and hence the typical dust temperature) of galaxies with moderate to
high masses increases substantially towards higher redshifts, i.e., as
$(1+z)^{0.32}$ (\S3.1.3) and therefore in the range 44-50 K at
$z=4$-10.

For convenience, we provide a table in Appendix C, indicating how the
derived luminosities would change depending on the SED template or
dust temperature assumed.  Included in this table are dust
temperatures ranging from 25 K to 45 K and also empirical SED
templates for M51, M82, Arp 220, and NGC6946 from Silva et
al.\ (1998).  The typical amplitude of these dependencies is a factor
of 3 at $z\sim2$-3, a factor of 2 at $z\sim6$, and a factor of $<$1.5
at $z\sim8$-10.

While we would expect some uncertainties in the infrared excesses or
IR luminosities we derive from the ALMA data, we have verified that
the derived values are nonetheless plausible in the redshift range
$z\sim2$-3 by comparing with independent estimates made from the MIPS
24$\mu$m observations and using a prescription from Reddy et
al.\ (2010) to convert these 24$\mu$m fluxes to IR luminosities
(Appendix B).  The IR luminosities we derive for the few detected
sources agree to within 0.3 dex of the ALMA-estimated luminosities
(Table~\ref{tab:mipsalma} from Appendix B) if we adopt a fiducial dust
temperature of 35 K.  Even better agreement is obtained if we adopt
higher values for the dust temperature.

\begin{figure*}
\epsscale{1.1}
\plotone{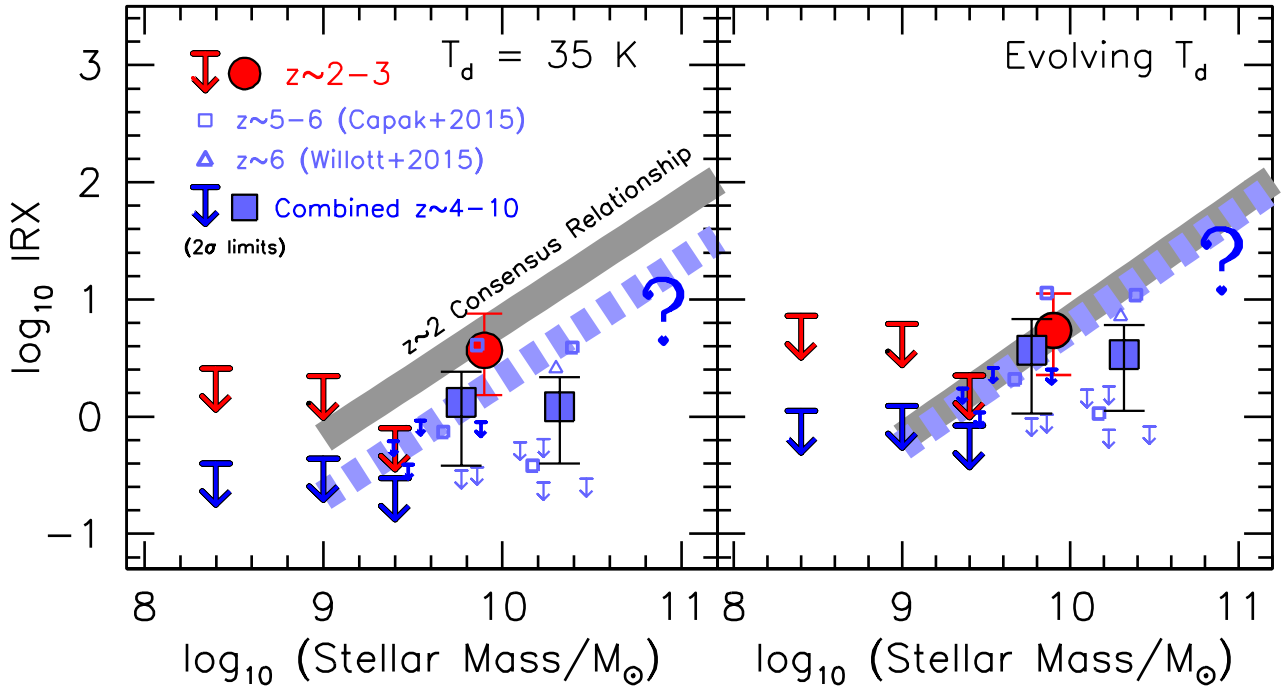}
\caption{Constraints on the infrared excess as a function of stellar
  mass including select results from the literature (\textit{large
    solid blue squares}) assuming a fixed dust temperature of 35 K
  (\textit{left panel}) and a dust temperature that monotically
  increases towards higher redshift as found by Bethermin et
  al.\ (2015: \textit{right panel}).  Large downward-pointing arrows
  indicate $2\sigma$ upper limits on the infrared excess for the
  average sources, after combining results from the literature with
  the current constraints.  The specific results for the infrared
  excess from individual sources in the Capak et al.\ (2015:
  \textit{open blue squares}) and Willott et al.\ (2015: \textit{open
    blue triangles}) studies are explicitly shown.  Blue downward
  arrows indicate $2\sigma$ upper limits on the inferred infrared
  excesses seen in five sources from the Capak et al.\ (2015) study.
  The thick light blue dotted line shows one potential IRX-stellar
  mass relationship at $z\sim4$-10 that roughly fits the available
  constraints.  For fixed dust temperature (\textit{left panel}), this
  is possible by shifting the consensus IRX-stellar mass relationship
  at $z\sim2$-3 (\textit{thick gray line}) to lower values of the dust
  extinction (and the observational constraints may support an even
  larger shift than the 0.5 dex shift presented) while for a
  monotonically increasing dust temperature (\textit{right panel}) no
  evolution in the consensus IRX-stellar mass relationship is required
  to match current high-redshift results.\label{fig:irxsmlit}}
\end{figure*}

\subsection{Synthesis of the Present Results with Earlier $z=5$-6 Results from ALMA}

Finally, we combine the current constraints on the infrared excess
with those available in the literature to construct a more complete
picture of the impact of dust obscuration on the overall energy output
from star-forming galaxies at $z=4$-10.

We focus in particular on constraints available from ALMA on luminous
$z=4$-10 galaxies due to the limited number of sources within the 1
arcmin$^2$ footprint of ASPECS.  We focus on 12 fairly luminous
$z=5$-7 galaxies originally identified as part of wide-area surveys in
the rest-frame $UV$ (e.g. Willott et al.\ 2013) and recently examined
with ALMA by Capak et al.\ (2015) and Willott et al.\ (2015).  Results
from those studies are particularly useful, since the
1.2$\,$mm-continuum fluxes, $UV$ luminosities, and estimates of the
stellar mass for individual sources are available.  6 of 12 galaxies
from those two studies show tentative detections in the available ALMA
data.

Combining our own results with those from Capak et al.\ (2015) and
Willott et al.\ (2015) -- self-consistently converting the observed
ALMA fluxes to IR luminosities -- we present our IRX versus stellar
mass constraints in Figure~\ref{fig:irxsmlit} with the solid blue
squares and large blue upper limits.  For context, we compare these
results with the consensus IRX-stellar mass relationship we derive
from various results on IRX-mass found at $z\sim2$-3 (Reddy et
al.\ 2010; Whitaker et al.\ 2014, {\'A}lvarez-M{\'a}rquez et
al.\ 2016: see Appendix A).

For the case of a fixed dust temperature of 35 K, we find we can
approximately match the current constraints at $z=4$-10 (light thick
dotted blue line in Figure~\ref{fig:irxsmlit}) using the consensus
IRX-stellar mass relationship at $z\sim2$-3, if sources of a given
stellar mass exhibit IR luminosities at least $\sim$0.5 dex lower than
at $z\sim2$-3.  This would suggest lower dust extinctions at
high-redshift at a fixed stellar mass.  

However, we should also look at how evolution in the dust temperature
could impact the results.  If the dust temperature exhibited a
monotonic increase towards higher redshift, e.g., as found by Magdis
et al.\ (2012) and Bethermin et al.\ (2015), we would infer much
higher (by $\sim$0.4-0.5 dex) IR luminosities for the detected sources
from the three samples considered here.  This would translate into
similarly higher infrared excesses at $z=4$-10 and which would be
plausibly consistent with IRX-stellar mass relationship at $z=0$-3
(\textit{right} panel of Figure~\ref{fig:irxsmlit}), suggesting no
significant evolution in this relationship from $z\sim6$ to $z\sim0$.

Pannella et al.\ (2015) found no strong evidence for evolution in the
IRX-stellar mass relation to $z\sim3.5$.  Recent results on the
average infrared excess for bright $z\sim3$-5 galaxies by Coppin et
al.\ (2015), where IRX $\sim$ 5-6 (drawing values from their Table 2)
for sources with $UV$ SFRs of $\sim$18-33 (corresponding to a stellar
mass of $\log_{10} (M/M_{\odot}) \sim$9.8: Duncan et al.\ 2014), are
also consistent with no evolution in the IRX-stellar mass relation to
$z\sim5$ (assuming minimal biases from stacking).  While one might
expect some evolution in this relationship due to the observed
evolution in the mass-metallicity relation (e.g., Erb et al.\ 2006a),
it is possible that higher amounts of gas and ISM mass in $z\gtrsim2$
galaxies could compensate for the lower metal content to produce a
relatively unevolving IRX-stellar mass relation (Tan et al.\ 2014).

\section{Discussion}

\subsection{Is Dust Emission from Lower-Mass Galaxies Really Negligible 
at $z\gtrsim3$?}

The results we obtained in the previous section imply that dust
emission from lower-mass galaxies is not large, particularly relative
to the emission from galaxies in the rest-frame $UV$ and as apparent
at 1.2$\,$mm.  The relative energy output in the IR from the
``average'' $<10^{9.25}$ $M_{\odot}$ and $<10^{9.75}$ $M_{\odot}$
galaxy in our HUDF selection is estimated to be less than 42\% and
32\% (95\% confidence), respectively, of what galaxies emit at
rest-frame $UV$ wavelengths (Table~\ref{tab:irxsm} from Appendix D).

\subsubsection{Comparison with $z\sim2$ Spectroscopic Results}

One can obtain a quick check on these results by inspecting the
results from spectroscopy, particularly measurements of the Balmer
decrement in $z\sim2$ galaxies.  Encouragingly, the decrements seen in
results from the MOSDEF program (Kriek et al.\ 2015) seem consistent
with lower-mass, lower-SFR systems showing low dust extinction.  This
is perhaps most clearly seen inspecting Figure 20 of Reddy et
al.\ (2015) where galaxies with SFRs $\lesssim$10 $M_{\odot}$/yr show
an H$\alpha$/H$\beta$ flux ratio of approximately $\sim$3, very close
to the intrinsic ratio.  This points to little dust extinction in
galaxies with lower SFRs.

\subsubsection{Do Examples of Low-Mass, but IR Luminous $z\gtrsim2$ Galaxies Exist?}

The present results prompt us to consider whether well-known examples
of $z\gtrsim2$ galaxies in the literature stand in significant
violation of these findings, i.e., whether there are sources which are
far-IR luminous, despite having lower stellar masses.

\begin{figure}
\epsscale{1.1}
\plotone{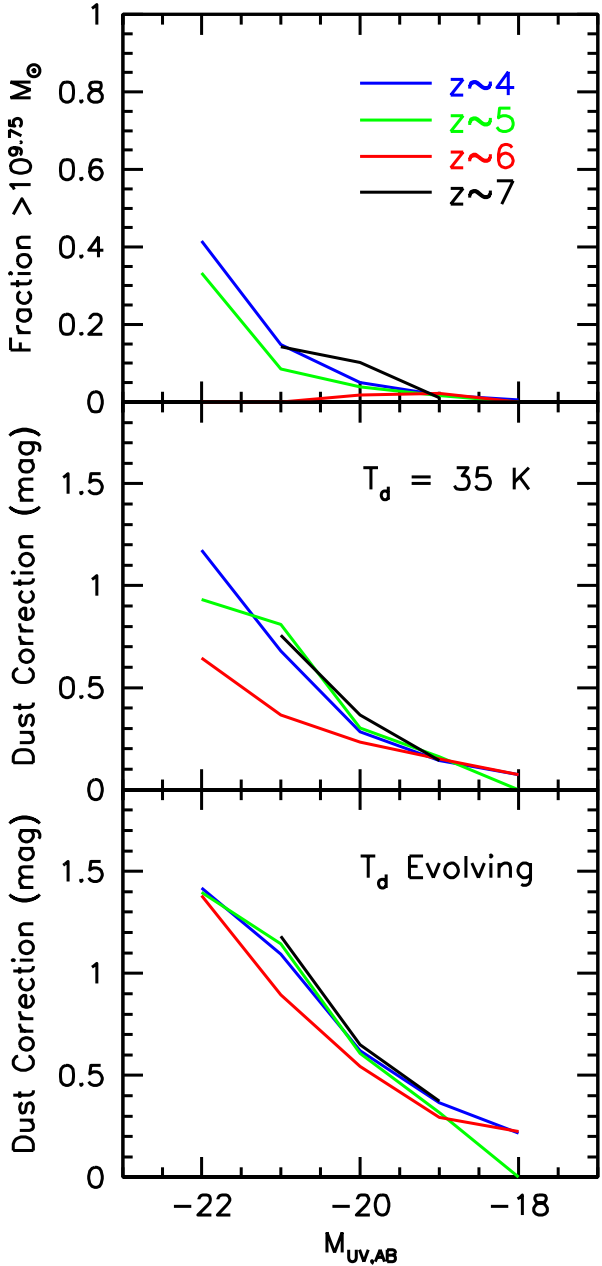}
\caption{(\textit{upper panel}) Fraction of $z\sim4$ (\textit{blue
    line}), $z\sim5$ (\textit{green line}), $z\sim6$ (\textit{red
    line}), and $z\sim7$ (\textit{black line}) galaxies (\textit{red
    line}) in a given 0.5-mag bin of $UV$ luminosity estimated to have
  a stellar mass in excess of $10^{9.75}$ $M_{\odot}$ where the dust
  extinction appears to be especially significant.  (\textit{middle
    panel}) Estimated correction for dust extinction versus $UV$
  luminosity for $z\sim4$ (\textit{blue}), $z\sim5$ (\textit{green
    line}), $z\sim6$ (\textit{red line}), and $z\sim7$ (\textit{black
    line}).  For sources with stellar masses in excess of $10^{9.75}$
  $M_{\odot}$, dust extinction is computed using the M99 IRX-$\beta$
  relation, while for those with stellar masses lower than $10^{9.75}$
  $M_{\odot}$, dust extinction is computed based on the consensus
  $z\sim2$-3 IRX-stellar mass relation (but shifted lower by 0.5 dex).
  (\textit{lower panel}) Similar to middle panel, but computing dust
  extinction for the lowest-mass galaxies assuming no evolution in the
  IRX-stellar mass relation from $z\sim0$-3 (appropriate if the dust
  temperature monotonically increases to high redshifts).  See
  Figure~\ref{fig:irxsmlit}.\label{fig:dustc}}
\end{figure}

\begin{deluxetable}{ccc}
\tabletypesize{\footnotesize}
\tablecaption{Estimated dust corrections to apply the $UV$ luminosity density results to various faint-end limits\label{tab:dustcorr}}
\tablehead{
\colhead{} & \multicolumn{2}{c}{$\textrm{log}_{10}$ Dust Correction} \\
\colhead{Sample} & \colhead{($>$0.05 $L_{z=3}^{*}$)\tablenotemark{a}} & \colhead{($>$0.03 $L_{z=3}^{*}$)\tablenotemark{a}}}\\
\startdata
\multicolumn{3}{c}{Assuming $T_d$ = 35 K (fixed)}\\
$z\sim3$ & 0.37\tablenotemark{*} & 0.34\tablenotemark{*}\\
$z\sim4$ & 0.15 & 0.14\\
$z\sim5$ & 0.16 & 0.14\\
$z\sim6$ & 0.09 & 0.07\\
$z\sim7$ & 0.04 & 0.03\\
$z\sim8$ & 0.04 & 0.03\\
\\
\multicolumn{3}{c}{Assuming Evolving $T_d$\tablenotemark{b}}\\
$z\sim3$ & 0.37\tablenotemark{*} & 0.34\tablenotemark{*}\\
$z\sim4$ & 0.27 & 0.25\\
$z\sim5$ & 0.27 & 0.24\\
$z\sim6$ & 0.21 & 0.18\\
$z\sim7$ & 0.09 & 0.07\\
$z\sim8$ & 0.08 & 0.06
\enddata
\tablenotetext{*}{For uniquely the $z\sim3$ sample, we make use of the
  finding by e.g. Reddy \& Steidel (2004) and Reddy et
  al.\ (2010) that the average infrared excess for galaxies brighter
  than 25.5 mag at $z\sim3$ is a factor of $\sim$5.}
\tablenotetext{a}{The specified limits 0.05 $L_{z=3}^{*}$ and 0.03 $L_{z=3}^{*}$ correspond to faint-end limits of $-17.7$ and $-17.0$, respectively, which is the limiting luminosity to which $z\sim7$ and $z\sim10$ galaxies can be found in current probes (Schenker et al.\ 2013; McLure et al.\ 2013; Ellis et al.\ 2013; Oesch et al.\ 2013; Bouwens et al.\ 2015).}
\tablenotetext{b}{We adopt $T_d = ((1+z)/2.5)^{0.32} (35 K)$ for the evolution following Bethermin et al.\ (2015).  See \S3.1.3.}
\end{deluxetable}

Perhaps the most prominent source which potentially stands in
violation of these general findings is the bright $z\sim7.5$ galaxy
A1689-zD1 initially identified behind Abell 1689 by Bradley et
al.\ (2008) with an estimated stellar mass of $\sim$1.6$\times10^{9}$
$M_{\odot}$ (Watson et al.\ 2015).  Despite its low mass, Knudsen et
al.\ (2016) report a 12$\sigma$ detection of the source in far-IR
continuum observations with ALMA, implying an infrared excess of
$\sim$3 for the source (Watson et al.\ 2015).  The reported IRX is
much higher than our $2\sigma$ upper limits we can set on the stack
results, suggesting that such a result (if true) is atypical for the
$\ll$10$^{9.75}$ $M_{\odot}$ population.

There are also 5 sources out of 122 (4\%) followed up by the ALESS
survey with ALMA (da Cunha et al.\ 2016) with estimated stellar masses
$<$$10^{9.75}$ $M_{\odot}$ which nevertheless show moderately high IR
luminosities $\sim$$10^{10.5}$ to $\sim$$10^{11.6}$ $L_{\odot}$.
Likewise, seven sources out of the 48 sources (15\%) from the
AzTEC/ASTE survey (Scott et al.\ 2010), also identified based on their
IR properties, have stellar masses below $10^{9.75}$ $M_{\odot}$.
Additionally, from the Capak et al.\ (2015) sample of $\sim$10 $z=5$-6
galaxies, there is one source (HZ4) below our $10^{9.75}$ $M_{\odot}$
threshold (stellar mass of $10^{9.67\pm0.21}$ $M_{\odot}$) which
nonetheless has a moderately high IR luminosity ($10^{11.13}$
$L_{\odot}$).

As more ALMA constraints become available for lower-mass $z\geq2$
galaxies, it will be important to see if any other sources are found
to be so bright in the far-IR.

\subsection{Prescription for the Average Infrared Excess in Star-forming 
Galaxies at $z\gtrsim3$:}

\begin{figure*}
\epsscale{1.1} \plotone{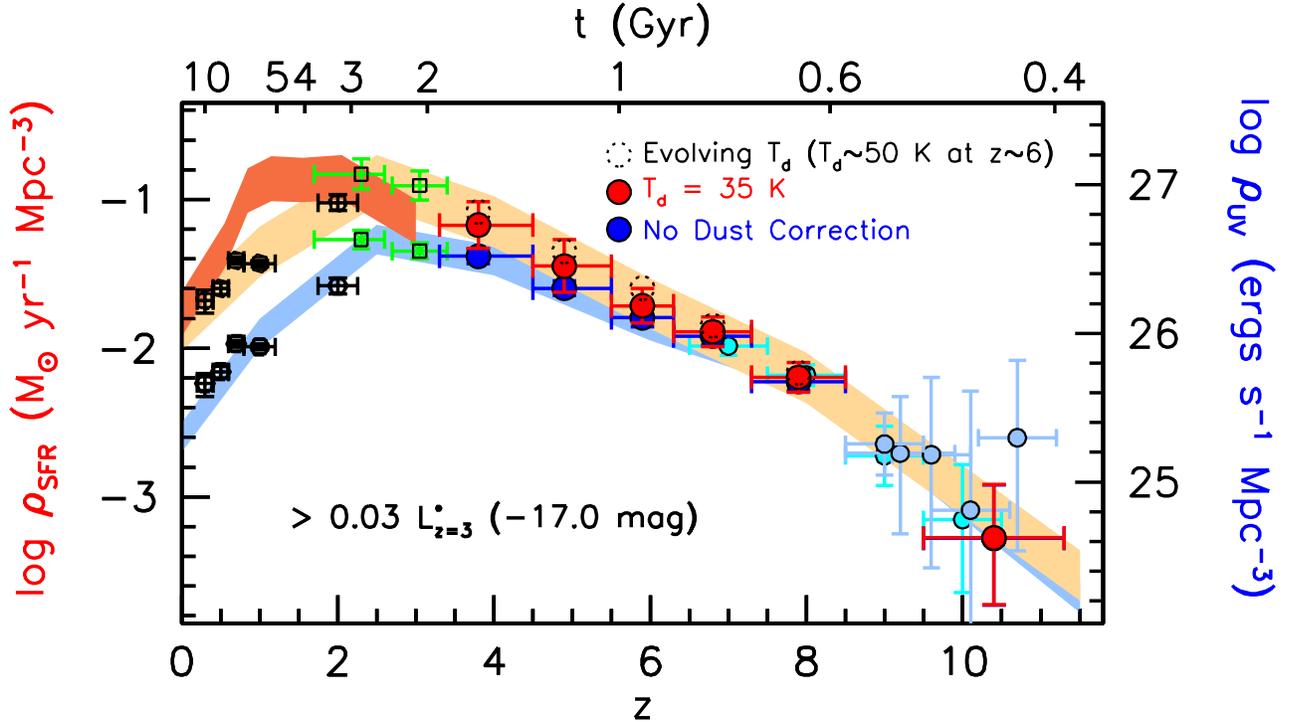}
\caption{Updated determinations of the derived SFR (\textit{left
    axis}) and $UV$ luminosity (\textit{right axis}) densities versus
  redshift (\S5.4).  The left axis gives the SFR densities we would
  infer from the measured luminosity densities, assuming the Madau et
  al.\ (1998) conversion factor relevant for star-forming galaxies
  with ages of $\gtrsim10^8$ yr (see also Kennicutt 1998).  The right
  axis gives the $UV$ luminosities we infer integrating the present
  and published LFs to a faint-end limit of $-17$ mag (0.03
  $L_{z=3}^{*}$) -- which is the approximate limit we can probe to
  $z\sim8$ in our deepest data set.  The upper and lower set of points
  (\textit{red and blue circles, respectively}) and shaded regions
  show the SFR and $UV$ luminosity densities corrected and uncorrected
  for the effects of dust extinction.  The dust correction we utilize
  relies on an IRX-$\beta$ relation intermediate between SMC and
  Calzetti for the highest mass galaxies in the present samples, i.e.,
  $>10^{9.75}$ $M_{\odot}$, but an IRX-stellar mass relation for
  lower-mass sources (\S4.2).  The dotted black open circles indicate
  the dust-corrected SFR densities, assuming an unevolving IRX-stellar
  mass relation (appropriate if the dust temperature increases
  monotonically towards high redshift: see Figure~\ref{fig:irxsmlit}).
  The dark red shaded region include the contribution from IR bright
  sources (Magnelli et al.\ 2009, 2011, 2013).  Also shown are the SFR
  densities at $z\sim2-3$ from Reddy et al.\ (2009: \textit{green
    crosses}), at $z\sim0$-2 from Schiminovich et al.\ (2005:
  \textit{black hexagons}), at $z\sim7$-9 from McLure et al.\ (2013)
  and Ellis et al.\ 2013: \textit{cyan solid circles}), and
  $z\sim9$-11 from CLASH (Bouwens et al.\ 2014b; Coe et al.\ 2013;
  Zheng et al.\ 2012: \textit{light blue circles}) and Oesch et
  al.\ (2013: \textit{light blue circles}).  The $z\sim9$-11
  constraints on the $UV$ luminosity density have been adjusted
  upwards to a limiting magnitude of $-17.0$ mag assuming a faint-end
  slope $\alpha$ of $-2.0$ (consistent with our constraints on
  $\alpha$ at both $z\sim7$ and at $z\sim8$).\label{fig:sfz}}
\end{figure*}

Synthesizing the results from our own program with those from other
programs, we can derive an approximate expression for the average
infrared excess in star-forming galaxies.  For sources with stellar
masses in excess of $10^{9.75}$ $M_{\odot}$, we find that the
IRX-$\beta$ relationship is most consistent with an SMC IRX-$\beta$
relation.  While many previous studies (e.g., Reddy et al.\ 2006,
2010; Daddi et al.\ 2007) found evidence that the highest-mass sources
followed a Calzetti et al.\ (2000) or M99 IRX-$\beta$ relation
implying more obscured star formation, here we are probing a smaller
volume, and the highest-mass sources from the present study may not be
especially dissimilar from the lowest-mass sources in many previous
studies.  For sources with stellar masses of $10^{10}$ $M_{\odot}$ and
less, many previous studies also found evidence for an SMC IRX-$\beta$
relation in $z\sim2$ galaxies, e.g., Baker et al.\ (2001), Reddy et
al.\ (2006: Figure 10), Reddy et al.\ (2010), and Siana et al.\ (2008,
2009).

Indeed, the present ALMA results are interesting in that they allow us
to extend these analyses into an even lower mass regime than was
generally studied before.  Despite some dependence on the assumed SED
template, our results suggest that dust emission from these lower-mass
sources is much less significant than for even our high-mass
subsample, i.e., with an IRX not larger than 0.40 ($2\sigma$) assuming
$T_d \sim 35$ K.  Even if we conservatively adopt a modified blackbody
SED with an evolving dust temperature similar to that found by
Bethermin et al. (2015), our results imply that the infrared excess
for lower-mass galaxies is not larger than 0.94 ($2\sigma$).

These results recommend to us a relatively simple recipe for the
infrared excess of star-forming galaxies at $z\gtrsim2$.  For galaxies
with stellar masses of $>10^{9.75}$ $M_{\odot}$, we make use of the an
IRX-$\beta$ relationship intermediate between the SMC and Calzetti et
al.\ (2000) dust laws:\footnote{The dust-free $UV$-continuum slope
  $\beta$ would plausibly be bluer at earlier times due to a younger
  average age of the stellar population (Wilkins et al.\ 2013;
  Castellano et al.\ 2014), potentially decreasing the $-$2.23
  intrinsic slope by $\sim$0.2 to $\sim$$-2.5$.}
\begin{equation}
A_{UV} = 1.5 (\beta + 2.23)
\label{eq:irxhighmass}
\end{equation}
For galaxies with stellar masses below $10^{9.75}$ $M_{\odot}$, the
dust extinction is much lower, as demonstrated e.g. by our own
results.  For such systems, we postulate that IRX can be derived more
reliably by using the correlation between IRX and stellar mass.
Utilizing the IRX-stellar mass relationship from the previous section
(keeping in mind the $\pm$0.2 dex uncertainties), we propose that
\begin{equation}
\log_{10} IRX = \log_{10} [M/M_{\odot}]-9.67
\label{eq:irxlowmass}
\end{equation}
where $M$ is the inferred stellar mass (assuming a fixed dust
temperature of 35 K).  If the dust temperature evolves with cosmic
time as found by Bethermin et al.\ (2015), i.e., as $(1+z)^{0.32}$
(\S3.1.3), the latter expression could plausibly be replaced by
$\log_{10} IRX = \log_{10} [M/M_{\odot}]-9.17$, i.e., approximately
the same relationship as at $z=0$-3 (see \S3.5).  Encouragingly
enough, the IRX-stellar mass prescription we apply in the low-mass
regime (Eq.~\ref{eq:irxlowmass}) gives very similar estimates for the
dust corrections in the high-mass regime ($>10^{9.75}$ $M_{\odot}$) as
we find using our primary prescription (which relies on an IRX-$\beta$
relationship).  As such, there is a basic consistency to the present
approach (despite some arbitrariness in how one parameterizes IRX in
terms of various physical variables, i.e., stellar mass, $\beta$, or
even the star formation rate itself).

With future data -- including both deeper and wider continuum mosaics
with ALMA -- it should be possible to improve on this prescription.
Particularly valuable will be observations at bluer wavelengths
(closer to the peak of the far-IR emission: see Figure 1) and
complementary information from other probes, i.e., stacks of the PACS
fluxes, X-ray, and near-IR spectra for even lower-mass sources.  In
addition, a measurement of the Balmer decrement out to $z\sim6$ should
soon be possible with JWST.

\begin{deluxetable*}{ccccccc}
\tablewidth{13cm}
\tablecolumns{7}
\tabletypesize{\footnotesize}
\tablecaption{$UV$ Luminosity Densities and Star Formation Rate Densities to $-17.0$ AB mag (0.03 $L_{z=3} ^{*}$)\label{tab:sfrdens}}
\tablehead{
\colhead{Lyman} & \colhead{} & \colhead{$\textrm{log}_{10} \mathcal{L}$} & \colhead{Dust} & \multicolumn{3}{c}{$\textrm{log}_{10}$ SFR density} \\
\colhead{Break} & \colhead{} & \colhead{(ergs s$^{-1}$} & \colhead{Correction} & \multicolumn{3}{c}{($M_{\odot}$ Mpc$^{-3}$ yr$^{-1}$)} \\
\colhead{Sample} & \colhead{$<z>$} & \colhead{Hz$^{-1}$ Mpc$^{-3}$)\tablenotemark{a}} & \colhead{(dex)\tablenotemark{b}} & \colhead{Uncorrected} & \colhead{Corrected} & \colhead{Incl. ULIRG\tablenotemark{b}}}
\startdata
\multicolumn{7}{c}{M99 IRX-$\beta$ (as in Bouwens et al.\ 2015)}\\
B & 3.8 & 26.52$\pm$0.06 & 0.42 & $-$1.38$\pm$0.06 & $-$1.00$\pm$0.13 & $-$0.96$\pm$0.13\\
V & 4.9 & 26.30$\pm$0.06 & 0.35 & $-$1.60$\pm$0.06 & $-$1.26$\pm$0.12 & $-$1.25$\pm$0.12\\
i & 5.9 & 26.10$\pm$0.06 & 0.25 & $-$1.80$\pm$0.06 & $-$1.55$\pm$0.13 & $-$1.55$\pm$0.13\\
z & 6.8 & 25.98$\pm$0.06 & 0.23 & $-$1.92$\pm$0.06 & $-$1.69$\pm$0.07 & $-$1.69$\pm$0.07\\
Y & 7.9 & 25.67$\pm$0.06 & 0.15 & $-$2.23$\pm$0.06 & $-$2.08$\pm$0.07 & $-$2.08$\pm$0.07\\
\\
\multicolumn{7}{c}{Fiducial Estimates: Assuming $T_d$ = 35 K (fixed)}\\
U & 3.0 & 26.55$\pm$0.06 & 0.44 & $-$1.35$\pm$0.03 & $-$1.01$\pm$0.09 & $-$0.91$\pm$0.09\\
B & 3.8 & 26.52$\pm$0.06 & 0.21 & $-$1.38$\pm$0.06 & $-$1.24$\pm$0.13 & $-$1.17$\pm$0.13\\
V & 4.9 & 26.30$\pm$0.06 & 0.15 & $-$1.60$\pm$0.06 & $-$1.46$\pm$0.12 & $-$1.45$\pm$0.12\\
i & 5.9 & 26.10$\pm$0.06 & 0.08 & $-$1.80$\pm$0.06 & $-$1.73$\pm$0.13 & $-$1.72$\pm$0.13\\
z & 6.8 & 25.98$\pm$0.06 & 0.03 & $-$1.92$\pm$0.06 & $-$1.89$\pm$0.07 & $-$1.89$\pm$0.07\\
Y & 7.9 & 25.67$\pm$0.06 & 0.03 & $-$2.23$\pm$0.06 & $-$2.20$\pm$0.07 & $-$2.20$\pm$0.07\\
J & 10.4 & 24.62$_{-0.45}^{+0.36}$ & 0.00 & $-3.28$$_{-0.45}^{+0.36}$ & $-3.28$$_{-0.45}^{+0.36}$ & $-3.28$$_{-0.45}^{+0.36}$\\
\\
\multicolumn{7}{c}{Assuming Evolving $T_d$\tablenotemark{c}}\\
B & 3.8 & 26.52$\pm$0.06 & 0.30 & $-$1.38$\pm$0.06 & $-$1.13$\pm$0.13 & $-$1.08$\pm$0.13\\
V & 4.9 & 26.30$\pm$0.06 & 0.25 & $-$1.60$\pm$0.06 & $-$1.36$\pm$0.12 & $-$1.35$\pm$0.12\\
i & 5.9 & 26.10$\pm$0.06 & 0.20 & $-$1.80$\pm$0.06 & $-$1.61$\pm$0.13 & $-$1.60$\pm$0.13\\
z & 6.8 & 25.98$\pm$0.06 & 0.07 & $-$1.92$\pm$0.06 & $-$1.85$\pm$0.07 & $-$1.85$\pm$0.07\\
Y & 7.9 & 25.67$\pm$0.06 & 0.06 & $-$2.23$\pm$0.06 & $-$2.17$\pm$0.07 & $-$2.17$\pm$0.07
\enddata
\tablenotetext{a}{Integrated down to 0.03 $L_{z=3}^{*}$.  Based upon
  LF parameters in Table 2 of Bouwens et al.\ (2015) (see \S6.1).  The
  SFR density estimates assume $\gtrsim100$ Myr constant SFR and a
  Salpeter IMF (e.g., Madau et al.\ 1998).  Conversion to a Chabrier
  (2003) IMF would result in a factor of $\sim$1.8 (0.25 dex) decrease
  in the SFR density estimates given here.}
\tablenotetext{b}{This factor includes both the impact of dust on the
  $z=3$-8 $UV$ luminosity densities and also the contribution of
  far-IR bright ($>10^{12}$ $L_{\odot}$) galaxies which might be
  missed in typical Lyman-break galaxy probes or which might have
  their IR luminosities underestimated (Reddy et al.\ 2006; Reddy \&
  Steidel 2009).}
\tablenotetext{c}{We adopt $T_d = ((1+z)/2.5)^{0.32} (35 K)$ for the
  evolution following Bethermin et al.\ (2015).  See \S3.1.3.}

\end{deluxetable*}

\section{Implications of these Results}

\subsection{Inferred Dust Corrections for $z\gtrsim3$ Samples}

The purpose of this section is to determine the approximate correction
we should apply to the observed $UV$ luminosity densities to correct
for dust extinction and therefore obtain the star-formation rate
density.

We base our dust extinction estimates on the large catalog of $z=4$-10
galaxies from Bouwens et al.\ (2015) in the CANDELS GOODS-North,
GOODS-South, and Early Release Science (Windhorst et al.\ 2011)
fields.  Critically, each of the $z\sim2$-10 galaxies in these
galaxies possess individually-estimated stellar masses and
$UV$-continuum slopes $\beta$, all derived on the basis of the deep
{\it HST} and {\it Spitzer}/IRAC photometry available over the
GOODS-North and South (Labb{\'e} et al.\ 2015).  Stellar mass, in
particular, is an important variable to establish given its utility
for predicting the IR luminosity and infrared excess for individual
galaxies.  In addition, as we saw in \S3.3.2, the estimated stellar
mass clearly impacts the dependence of IRX on $\beta$.

We estimate the dust correction in individual 0.5-mag bins of $UV$
luminosity.  For each bin, we first consider what fraction of galaxies
have stellar masses in excess of $10^{9.75}$ $M_{\odot}$.  For
galaxies in this mass range, we compute the estimated dust correction
based on IRX-$\beta$ relation intermediate between SMC and Calzetti et
al.\ (2000) and utilizing the $\beta$ distribution measured for such
high mass galaxies.  In making use of the observed $\beta$'s to derive
the correction for galaxies in a given luminosity bin, we either make
use of the full distribution of $\beta$'s derived for individual
sources (where the typical uncertainty in the $\beta$'s measured for
individual sources is $<$0.3) or make use of a model distribution
(where the typical uncertainty in $\beta$ is $>$0.3).  The
$UV$-continuum slopes $\beta$ we measure for individual sources are
estimated based on a power-law slope fit to its $UV$-continuum fluxes
(e.g., Castellano et al.\ 2012) avoiding those flux measurements which
could be impacted by IGM absorption or rest-frame optical
$\gtrsim$3500\AA$\,$light.\footnote{The inclusion of photometric
  constraints on the $UV$-continuum even to $\sim$3000$\AA$ is
  expected to have only a minor impact on the derived $\beta$
  ($\Delta\beta \lesssim 0.2$) given the general power-law-like shape
  of the $UV$ continuum (e.g., see Appendix A in Wilkins et
  al.\ 2016a).}

However, for galaxies with estimated stellar masses $<$$10^{9.75}$
$M_{\odot}$, we derive the dust correction from the prescription given
in Eq.~\ref{eq:irxlowmass}.

The top panel of Figure~\ref{fig:dustc} shows the fraction of galaxies
in our $z\sim4$, $z\sim5$, $z\sim6$, and $z\sim7$ samples whose
estimated stellar masses exceed $10^{9.75}$ $M_{\odot}$.  As expected,
the fraction of sources with $>$10$^{9.75}$ $M_{\odot}$ masses is
relatively high for the brightest sources in the rest-frame $UV$, but
decreases rapidly faintward of $-21$ mag and is $<$3\% at $-19$ mag.

In the middle panel of Figure~\ref{fig:dustc}, we present the dust
corrections we estimate for our samples based on our ALMA results and
using the prescription that we describe above.  The dust correction we
estimate is only particularly significant at $M_{UV,AB}<-21$ mag and becomes
negligible faintward of $-$20 mag.  In the bottom panel of
Figure~\ref{fig:dustc}, we present an alternate estimate of the dust
correction assuming that IRX-stellar mass relation does not evolve
from $z\sim0$-3 to $z\sim7$ (partially motivated by the findings in
the right panel of Figure~\ref{fig:irxsmlit}).

We can determine the approximate impact of dust corrections on the
inferred SFR densities at $z>3$ by multiplying the $UV$ LF by the
inferred dust corrections and integrating to specific faint-end
limits.  For convenience, these dust corrections are presented in
Table~\ref{tab:dustcorr}.  For the brightest ($<$25.5 mag) galaxies in
the rest-frame $UV$ at $z\sim3$, we assume (e.g. Reddy \& Steidel
2004) that the average dust correction for UV bright ($<$25.5 mag)
galaxies at $z\sim3$ is $\sim$5.

For the case of an evolving IRX-stellar mass relation, the corrections
are smaller (by $\sim$0.2 dex) than earlier correction factors (e.g.,
Madau \& Dickinson 2014; Bouwens et al.\ 2015).  This can be seen by
comparing the dust extinction estimates in the middle set of rows in
Table~\ref{tab:sfrdens} (to be presented in \S5.2) with the top set of
rows.

However, if the IRX-stellar mass relation does not evolve
significantly from $z\sim0$ to $z\sim6$ (i.e., as in the right panel
to Figure~\ref{fig:irxsmlit}), our estimated dust corrections are only
slightly smaller ($\sim$0.1 dex) than inferred in earlier work.
Previously, Capak et al.\ (2015) had also considered the impact of new
ALMA results for the implied star-formation rate densities at
$z\sim4$-7, finding suggestive evidence for smaller dust corrections
than had previously been utilized.

\subsection{Implied Star Formation Rate Densities at $z\geq3$}

We apply the dust corrections we derived in the previous sections to
the $UV$ luminosity densities integrating the $UV$ LF of Bouwens et
al.\ (2015) to 0.05 $L_{z=3}^{*}$ ($-$17.7 mag) and to 0.03
$L_{z=3}^{*}$ ($-$17.0 mag).  Both the dust corrections and $UV$ LFs
were derived over a similar range in $UV$ luminosity, so this process
is self consistent.

As in previous work, the $UV$ luminosity densities are converted into
SFR densities using canonical Madau et al.\ (1998) and Kennicutt
(1998) relationships:
\begin{equation}
L_{UV} = \left( \frac{\textrm{SFR}}{M_{\odot} \textrm{yr}^{-1}} \right) 8.0 \times 10^{27} \textrm{ergs}\, \textrm{s}^{-1}\, \textrm{Hz}^{-1}\label{eq:mad}
\end{equation}
This relationship assumes a Salpeter (1955) IMF and a constant SFR for
100 million years.  We also apply these dust corrections to the Reddy
\& Steidel (2009) and McLure et al.\ (2013) LF results.

Our quantitative results for the corrected and uncorrected SFR
densities at $z\sim3$-10 are presented in Table~\ref{tab:sfrdens}.  In
deriving the corrected SFR densities, we take the uncertainty in the
dust correction to be equal to the difference in the estimated dust
correction adopting a fixed dust temperature of 35 K and allowing for
an evolving dust temperature as implied by the Bethermin et
al.\ (2015) results.  We also present the derived SFR densities along
with many previous estimates (Schiminovich et al.\ 2006; Reddy \&
Steidel 2009; McLure et al.\ 2013) in Figure~\ref{fig:sfz}.

Of course, in computing the total SFR density from galaxies, we must
account not only for the contribution from $UV$-selected galaxies, but
also for the more massive, far-infrared bright sources where standard
dust corrections are not effective or which are sufficiently faint in
the $UV$ to be entirely missed in standard LBG searches (estimated to
occur when galaxies have IR luminosities $>10^{12}$ $L_{\odot}$ by
Reddy et al.\ 2008).  Such sources are known to contribute a
substantial fraction of the SFR density at $z\sim0$-2 (Karim et
al.\ 2011; Magnelli et al.\ 2013; Madau \& Dickinson 2014).  We
account for this contribution by using published IR LFs to integrate
up those galaxies with IR luminosities $>10^{12}$ $L_{\odot}$.  While
one might be concerned this would ``double count'' the SFR coming from
specific sources, it has been argued (e.g. Reddy et al.\ 2008) that
the full SFR from such sources would not be properly accounted for in
$UV$-selected samples (due to reasons specified at the beginning of
this paragraph).

\begin{figure}
\epsscale{1.13}
\plotone{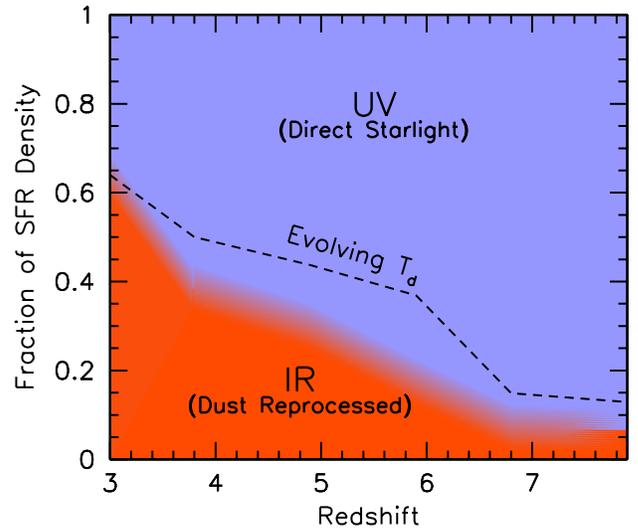}
\caption{Fraction of the SFR density that would be directly observed
  in the IR and in the rest-frame UV according to our fiducial SFR
  density estimates (Table~\ref{tab:sfrdens}).  The dashed line gives
  the dividing line for our fiducial evolving $T_d$ scenario (35 K
  $\times$ $((1+z)/2.5)^{0.32}$: see \S3.1.3).  The present figure
  represents an update to Figure 12 from Bouwens et al.\ (2009).
  Similar to the conclusions from Bouwens et al.\ (2009), we find that
  most of the SFR density at $z\gtrsim 3.5$ appears to be directly
  observable in the rest-frame $UV$ (see also Burgarella et al.\ 2013;
  Dunlop et al.\ 2016; Bourne et al.\ 2016).\label{fig:irfrac}}
\end{figure}

We utilize the results of Magnelli et al.\ (2013) at $z\sim0$-2 (which
build on the results of Caputi et al.\ 2007 and Magnelli et al.\ 2009,
2011), Reddy et al.\ (2008) at $z\sim3$, Daddi et al.\ (2009) and
Mancini et al.\ (2009) at $z\sim4$, Dowell et al.\ (2014) at $z\sim5$,
and Riechers et al.\ (2013) at $z\sim6$ (see also Wang et al.\ 2009,
Boone et al.\ 2013, and Asboth et al.\ 2016).  All together these
results suggest obscured SFR densities of 0.025, 0.01, 0.001, and
$<0.0003$ $M_{\odot} \textrm{yr}^{-1} \textrm{Mpc}^{-3}$ at $z\sim3$,
$z\sim4$, $z\sim5$, and $z\sim6$, respectively.

\begin{figure*}
\epsscale{0.9}
\plotone{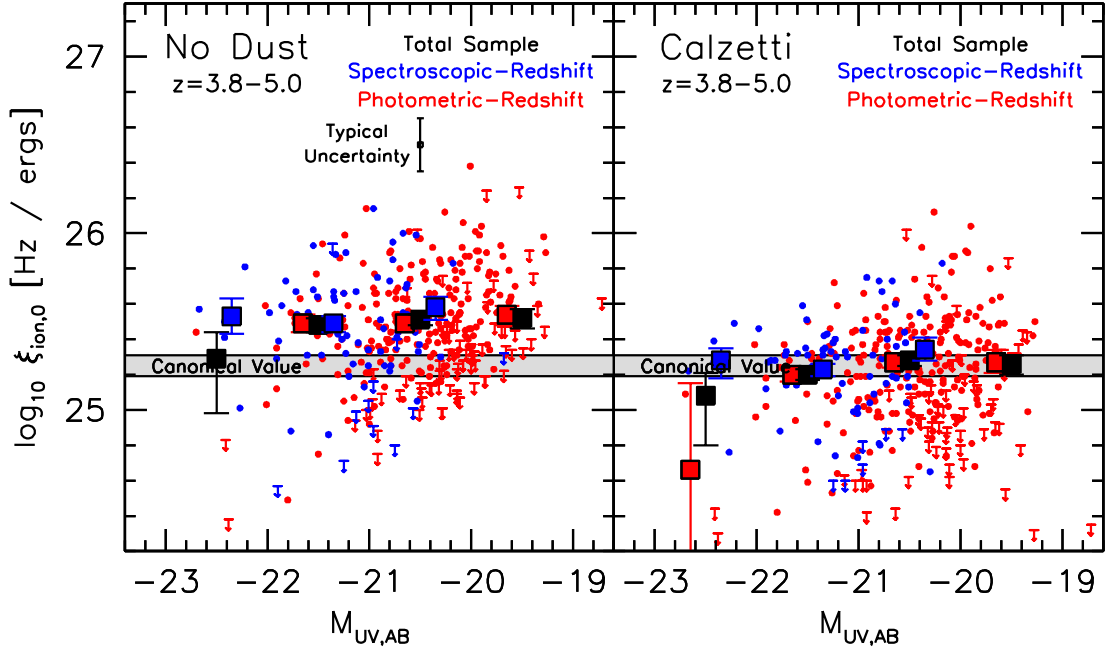}
\caption{(\textit{left}) New estimates of the Lyman-continuum photon
  production efficiencies $\xi_{ion}$ in $z=4$-5 galaxies (following
  Bouwens et al.\ 2015) using IRAC-based H$\alpha$ luminosities and
  HST-based $UV$ luminosities.  Sources where spectroscopic redshifts
  or well-determined photometric redshifts place the H$\alpha$ line in
  a specific IRAC band are indicated by the blue and red points,
  respectively.  $1\sigma$ upper limits are included on this diagram
  with downward arrows in cases where the H$\alpha$ emission line is
  not detected at $1\sigma$ in the photometry.  The solid red and blue
  squares indicate the mean value of $\xi_{ion}$ for red and blue
  colored points, while the solid black square indicate the mean
  values combining the spectroscopic and photometric-redshift selected
  samples.  The grey band indicates the Lyman-continuum photon
  production efficiencies $\xi_{ion}$ assumed in typical models (e.g.,
  Madau et al.\ 1999; Kuhlen \& Faucher-Gigu{\`e}re 2012; Robertson et
  al.\ 2013).  The black error bar near the top of the left panel
  indicate the typical uncertainties in the derived $\xi_{ion}$'s.
  (\textit{right}) Estimated Lyman-continuum photon production
  efficiencies $\xi_{ion}$ if a Calzetti extinction law is assumed.
  The derived values are $\sim$0.24 dex lower and were previously
  presented by Bouwens et al.\ (2015).  If dust extinction is as low
  in $z\sim4$-5 galaxies as suggested using our fiducial $T_d\sim35$ K
  results, it would imply that $z\sim4$-5 galaxies produce up to
  $\sim$1.8$\times$ as many ionizing photons per unit $UV$ luminosity
  as expected in conventional models.\label{fig:xi}}
\end{figure*}

We combine these SFR densities with those we derived by correcting the
$UV$ LFs at $z=3$-10 to present our best estimates for the SFR density
at $z=3$-10 in Table~\ref{tab:sfrdens} and Figure~\ref{fig:sfz}
alternatively assuming a fixed dust temperature $T_d\sim35$ K and
supposing that the dust temperature monotonically increases towards
high redshift as found by Bethermin et al.\ (2015).  For context, we
also present in Table~\ref{tab:sfrdens}, the SFR density Bouwens et
al.\ (2015) estimated based on the M99 IRX-$\beta$ relationship and
making use of the observed $\beta$ distribution at $z\sim4$-10.

Figure~\ref{fig:irfrac} shows the fraction of the SFR density that
would be directly observable in the rest-frame UV and also in the IR.
This figure is an update to Figure 12 from Bouwens et al.\ (2009).
Similar to the findings from Bouwens et al.\ (2009), we find that most
of the SFR density at $z>4.5$ would be observable in the rest-frame
UV.

\subsection{Implications for the Lyman Continuum Photon Production Efficiencies}

Another consequence of our new ALMA results is for the interpretation
of the prominent H$\alpha$ emission lines inferred in galaxies at
$z\sim4$-5 based on the observed {\it Spitzer}/IRAC $3.6\mu$m and
$4.5\mu$m fluxes (e.g., Schaerer \& de Barros 2009; Shim et al.\ 2011;
Stark et al.\ 2013).  

As first noted by Shim et al.\ (2011), the total H$\alpha$
luminosities of galaxies at $z\sim4$-5 are in excess of what one might
expect based on their luminosities in the rest-frame $UV$.  The excess
is conservatively as large as a factor of 2 using the general
$z\sim4$-5 LBG selections (Smit et al.\ 2016) but was earlier reported
to be a factor of 6 (Shim et al.\ 2011) using samples selected based
on their emission line properties.  One explanation for these high
luminosities would be if the selected galaxies predominantly had young
stellar ages (given that the Ly$\alpha$ emission was prominent in the
spectroscopic sample considered by Shim et al.\ 2011) or if dust
extinction preferentially had a larger impact on the observed
$UV$-continuum fluxes than it did the observed H$\alpha$ fluxes
(Marmol-Queralto et al.\ 2016; Smit et al.\ 2016).

Neither mechanism appears to provide a fully satisfactory explanation
for the high H$\alpha$ fluxes in $z\sim5$ galaxies.  Dust extinction
falls off rapidly towards lower masses, but the H$\alpha$ EWs in these
sources remain essentially unchanged (or possibly increase towards
lower mass: see Smit et al.\ 2016).  Similarly, young ages for the
$z\sim4$-5 galaxy population cannot provide an explanation, as a high
H$\alpha$ to $UV$-continuuum ratio is observed for typical
star-forming galaxies at $z\sim4$-5 (Marmol-Queralto et al.\ 2016;
Smit et al.\ 2016).\footnote{Shim et al.\ (2011) had speculated that
  young ages might explain the high H$\alpha$-to-UV-continuum ratios
  observed in $z\sim4$-5 galaxies they studied which showed strong
  Ly$\alpha$ emission.  However, it is now clear that such ratios
  apply both to sources which show Ly$\alpha$ in emission and those
  which do not.}

This essentially forces us to conclude that the observed
$H\alpha$-to-$UV$-continuum luminosity ratio must be intrinsic and
that star-forming galaxies indeed have very high H$\alpha$
luminosities relative to their luminosities in the $UV$ continuum,
particularly relative to conventional stellar population models (i.e.,
Bruzual \& Charlot 2003).  It has been speculated that such could be
achieved due to binarity or rapid rotation in massive stars, allowing
them to output large amounts of ionizing radiation tens of million of
years after an initial burst of star formation (e.g., Yoon et
al.\ 2006; Eldridge \& Stanway 2009, 2012; Levesque et al. 2012; de
Mink et al.\ 2013; Kewley et al. 2013; Leitherer et al. 2014;
Sz{\'e}csi et al.\ 2015; Gr{\"a}fener et al.\ 2015).  Changes to the
IMF might also be a possibility (e.g., if there are a larger number of
high-mass stars), but are disfavored given the general agreement
between the observed stellar mass density and the integrated SFR
densities (e.g. Stark et al.\ 2013).

These high $H\alpha$-to-$UV$-continuum ratios have implications for
the Lyman-continuum photon production efficiencies $\xi_{ion}$ used in
reionization modeling.  In Bouwens et al.\ (2015), we pioneered the
use of the relative luminosities in H$\alpha$ and the $UV$-continuum
to estimate these ratios.  Our results were consistent with the
high-end assumed values for this efficiency in the literature, but
were moderately sensitive to our assumptions about the dust
extinction.  Specifically, we derived an intrinsic value of
$\log_{10}$ $\xi_{ion}$/[Hz ergs$^{-1}$] of 25.27$_{-0.03}^{+0.03}$
and 25.33$_{-0.03}^{+0.02}$ assuming a Calzetti and SMC dust law
(where the systematic errors are likely 0.06 dex: see \S3.3 of Bouwens
et al.\ 2016).  These compare to various canonical values ranging from
25.20 to 25.30 in the literature (e.g., Madau et al.\ 1999; Kuhlen \&
Faucher-Gigu{\`e}re 2012; Robertson et al.\ 2013).

The present ALMA results suggest that the dust extinction in
$z\sim4$-5 galaxies is likely to be quite low.  It is therefore
interesting to estimate the mean Lyman-continuum photon production
efficiency assuming that dust extinction is zero, repeating the
estimates made in Bouwens et al.\ (2015).  This is almost certainly an
extreme case, as even lower mass galaxies likely show a small amount
of dust extinction.  Our new results for this efficiency factor
$\xi_{ion}$ are presented in Figure~\ref{fig:xi} as a function of $UV$
luminosity.  We compare the case of no dust extinction with the case
that all galaxies exhibit Calzetti et al.\ (2000) dust extinction and
using similar dust extinction for nebular lines and the $UV$
continuum.

The impact is fairly dramatic.  The Lyman-continuum photon production
efficiency $\log_{10} \xi_{ion}$ [Hz ergs$^{-1}$] we derive is
25.51$_{-0.03}^{+0.03}$, which is $\sim$1.6$\times$ higher than the
Bouwens et al.\ (2015) estimates assuming a Calzetti et al.\ (2000) or
SMC dust law.  This is very close to the efficiency
25.53$_{-0.06}^{+0.06}$ Bouwens et al.\ (2015) had previously derived
for $z\sim4$-5 galaxies with the bluest $UV$ continuum slopes $\beta$
($\beta<-2.3$) and very close to the values suggested by the stellar
population models including the impact of binary stars on the
evolution (e.g., Stanway et al.\ 2016; Wilkins et al.\ 2016b) or
suggested by models including stellar rotation (e.g., Topping \& Shull
2015).

Such a production efficiency is equivalent to $z\sim5$ star-forming
galaxies producing $\sim$1.8$\times$ as many ionizing photons per $UV$
continuum photon, as expected from standard stellar population models
(where we take to be $10^{25.2}$ to $10^{25.3}$ [Hz ergs$^{-1}$]
consistent with the use in the literature).  In calculating this
efficiency, we have assumed that the escape fraction is 0.  If we
assume that the escape fraction from galaxies is sufficient to
reproduce observed constraints on the ionizing emissivity of the
universe at $z\sim4.4$ (Becker \& Bolton 2013), this would translate
into a 0.02-dex higher mean $\xi_{ion}$.

To determine the impact of this higher efficiency factor on
reionization modeling, we take advantage of the modeling results from
Bouwens et al.\ (2015).  Bouwens et al.\ (2015) demonstrate that
ionizing emissivity derived from observed $z\geq6$ galaxies matches
that inferred from other observations (Planck, Ly$\alpha$ emission
fractions, etc.) if the following condition applies:
\begin{equation}
f_{esc}\xi_{\rm ion} f_{corr}(M_{lim}) (C/3)^{-0.3} = ~~~~~~~~~~~~~~~~~~~~
\label{eq:convf}
\end{equation}
\begin{displaymath}
~~~~~~~~~~~~~~~~~~10^{24.50\pm0.10} \textrm{s}^{-1}/(\textrm{ergs s}^{-1}\textrm{Hz}^{-1})
\end{displaymath}
where $f_{esc}$ is the escape fraction, $C=<\rho_H ^2>/<\rho_H>^2$ is
the clumping factor and where $f_{corr} (M_{lim}) =
10^{0.02+0.078(M_{lim}+13)-0.0088(M_{lim}+13)^2}$ corrects the $UV$
luminosity density $\rho_{UV}(z=8)$ derived to a faint-end limit of
$M_{lim}=-13$ mag to account for different faint-end cut-offs
$M_{lim}$'s.  The above constraint is essentially identical to what
Robertson et al.\ (2013) derive based on the available observations,
but the above expression also shows the approximate dependence on the
faint-end cut-off $M_{lim}$ to the LF and the clumping factor $C$.

\begin{deluxetable*}{ccccc}
\tablecolumns{5}
\tabletypesize{\footnotesize}
\tablecaption{Present Inferences for the IRX-$\beta$ and IRX-Stellar Mass Relations for $z\sim2$-10 Galaxies (if $T_d\sim35$ K)\label{tab:summary}}
\tablehead{
\colhead{Redshift Range} & \colhead{Stellar Mass Range} & \colhead{Sample} & \colhead{IRX-$\beta$ Relation} & \colhead{IRX-Stellar Mass Relation}}\\
\startdata
$z\sim2$-3 & $\log_{10} M/M_{\odot} > 9.75$ & ASPECS & Consistent with SMC  & Consistent with Consensus $z\sim2$-3 Relation \\
& & & \\
$z\sim2$-3 & $\log_{10} M/M_{\odot} < 9.75$ & ASPECS & Below SMC & Consistent with Consensus $z\sim2$-3 Relation \\
$z\sim4$-10 & $\log_{10} M/M_{\odot} > 9.75$ & ASPECS, Capak et & Consistent with SMC  & $\gtrsim$0.5-dex Below Consensus $z\sim2$-3 Relation\tablenotemark{b} \\
 & & al. (2015), Willott & or Below?\tablenotemark{a} \\
 & & et al.\ (2015) \\
$z\sim4$-10 & $\log_{10} M/M_{\odot} < 9.75$ & ASPECS & Below SMC & $\gtrsim$0.5-dex Below Consensus $z\sim2$-3 Relation\tablenotemark{b} 
\enddata
\tablenotetext{a}{Both the present results and those of Capak et al.\ (2015) are suggestive of the infrared excess matching the SMC IRX-$\beta$ relation or falling below it.}
\tablenotetext{b}{If the dust temperature monotonically increases towards higher redshift as found by Bethermin et al.\ (2015), the present results could be consistent with no evolution in the consensus IRX-stellar mass relation from $z\sim0$ to $z\sim6$.}
\end{deluxetable*}

If we apply such an efficiency to the observed $UV$ LFs and use fairly
standard assumptions (integrating the observed LFs down to $-13$ mag
and take the clumping factor $C$ equal to 3: e.g., Robertson et
al.\ 2013), these results would imply that galaxies can reionize the
universe if the escape fraction $f_{esc}$ is equal to $8\pm2$\%.  An
important corollary is that the escape fraction for $z\geq6$ galaxies
cannot be significantly higher than 8\%.  If it was higher, it would
imply a ionizing emissivity from galaxies that is higher than
observed.  Cosmic reionization would be complete at substantially
earlier times than $z\sim6$ (i.e., $z>6.5$).

\section{Summary} 

Here we make use of very sensitive (12.7$\mu$Jy/beam: $1\sigma$)
1.2$\,$mm observations to probe dust-enshrouded star formation from
330 robust $z=2$-10, $UV$-selected galaxies located within a 1
arcmin$^2$ field within the HUDF.  The present ALMA observations,
taken as part of the ASPECS program, represent some of the
deepest-ever continuum observations at 1.2$\,$mm (see papers I and II
in the ASPECS series).

Thirty-five $z=2$-10 galaxies were expected to be detected at
$>$2$\sigma$ extrapolating the M99 $z\sim0$ IRX-$\beta$ relation to
$z\gtrsim2$ and assuming a modified blackbody SED with dust
temperature 35 K.  Alternatively, using the approximate IRX-stellar
mass relation at $z\sim2$-3, the detection of 15 $z=2$-10 galaxies was
expected.

In significant contrast to these expectations, only 6 $z=2$-10
Lyman-break galaxies show convincing evidence for being detected in
ASPECS, after accounting for the likely spurious sources at
$>$2$\sigma$.  Only three of these $z\gtrsim2$ sources are detected at
substantially greater significance than $3\sigma$ (see paper II
[Aravena et al.\ 2016a] and paper IV [Decarli et al.\ 2016]).

The six detected $z=2$-10 galaxies are amongst the 13 sources from
ASPECS with inferred stellar masses $>10^{9.75}$ $M_{\odot}$.  No
other sources show a significant detection (after accounting for the
expected number of spurious $>$2$\sigma$ detections).

The fraction of high-mass $z=2$-10 galaxies detected (at $>$2$\sigma$)
is therefore 46\% (6 out of 13) for stellar mass estimates
$>10^{9.75}$ $M_{\odot}$.  If we exclude the five $>10^{9.75}$
$M_{\odot}$ sources with poor 1.2$\,$mm flux sensitivity (i.e.,
$>$17$\mu$Jy/beam rms), the detection fraction increases to
63$_{-17}^{+14}$\% in the $>10^{9.75}$ mass bin.  These results point
to stellar mass as being perhaps the best predictor of IR luminosity
in $z\gtrsim2$ galaxy samples.

Subdividing our samples of $z\gtrsim2$ galaxies into different bins of
$UV$ luminosity, stellar mass, and $UV$-continuum slope $\beta$ and
stacking the ALMA continuum observations, we only find a meaningful
detection in the stack results for $z=2$-3 Lyman-Break galaxies with
stellar masses $>10^{9.75}$ $M_{\odot}$.  Below $10^{9.75}$
$M_{\odot}$, we find a $2\sigma$ upper limit of 0.32 on the infrared
excess for such galaxies, i.e., $L_{IR}/L_{UV} < 0.32$ (see also
Papers II and V of this series: Aravena et al.\ 2016a, 2016b).

Combining the present results with previous ALMA results on
UV-selected samples, i.e., Capak et al.\ 2015, Willott et al.\ 2015,
we present the collective constraints on the observed infrared excess
versus stellar mass at $z=4$-6.  For $T_d = 35$ K, the results point
towards lower values of the infrared excess in $z>3$ galaxies (by
$\sim$0.5 dex) than in $z=2$-3 galaxies of comparable stellar mass.
However, if the dust temperature increases monotonically towards
higher redshift as found by Magdis et al.\ (2012) and Bethermin et
al.\ (2015), i.e., as $(1+z)^{0.32}$, the results are consistent with
an unevolving infrared excess versus stellar mass relation to
$z\sim6$.

We also examine the dependence of the infrared excess on $\beta$.  For
galaxies with stellar masses $>10^{9.75}$ $M_{\odot}$, the dependence
we find for the infrared excess on $UV$-continuum slope is most
consistent with an SMC IRX-$\beta$ relation.  However, for galaxies
with stellar masses $<$$10^{9.75}$ $M_{\odot}$, we derive $2\sigma$
upper limits on the infrared excess that lie below even the SMC
IRX-$\beta$ relationship.  These results suggest that dust emission
from lower-mass, $UV$-selected galaxies at $z>3$ is low ($<$40\%)
relative to emission in the rest-frame $UV$.

However, if the dust temperature is higher at $z\sim4$-10, i.e., $T_d
\sim 44$-50 K as suggested by the Magdis et al.\ (2012) and Bethermin
et al.\ (2015) results, the present limits would be $\sim$$3\times$
less stringent.

Based on our combined results with Capak et al.\ (2015) and Willott et
al.\ (2015), we present a crude prescription for the probable infrared
excesses in $z$$\geq$3 galaxies (\S4.2).  For galaxies with stellar
masses $>$$10^{9.75}$ $M_{\odot}$, we use an IRX-$\beta$ relationship
intermediate between M99 and SMC to estimate the probable infrared
excesses in galaxies.  However, for galaxies with stellar masses
$<$$10^{9.75}$ $M_{\odot}$, stellar mass appears to be the best guide.
Table~\ref{tab:summary} summarizes the implications of the ASPECS data
set for the IRX-$\beta$ and IRX-stellar mass relationship as a
function of redshift and stellar mass.

We apply this dust prescription to current state of the art catalogs
of $z=4$-8 galaxies to estimate the approximate dust extinction at a
given $UV$ luminosity.  From this, we derive the SFR density at
$z=3$-10 (Table~\ref{tab:sfrdens}).  For redshift-independent dust
temperatures $T_d \sim 35$ K, our results point towards lower SFR
density estimates at $z\gtrsim4$ than inferred in the past (by factors
of $\sim$2: see also Capak et al.\ 2015).  Nevertheless, if the dust
temperature is higher at $z\gtrsim4$ than $T_d\sim35$ K, the inferred
SFR density would still be lower (than previous estimates), but not by
as much.

The present results also help us to interpret the high
H$\alpha$-to-UV-continuum luminosity ratios in high-redshift galaxies
(e.g., Shim et al.\ 2011).  If dust does not majorly impact the
observed H$\alpha$ and $UV$-continuum fluxes from these galaxies, the
observed luminosity ratios are close to the intrinsic ones.
Star-forming galaxies at $z\sim5$ would then produce up to
$\sim$1.8$\times$ as many ionizing photons per $UV$ continuum photon,
as expected from standard stellar population models.

This would imply that star-forming galaxies can reionize the universe
even if the escape fraction is just 8$\pm$2\%.  In fact, if true, the
escape fraction cannot be higher than 8\% as it would imply a higher
ionizing emissivity from galaxies than is observed (e.g., Mitra et
al.\ 2013, 2015; Bouwens et al.\ 2015), requiring that the universe
finish reionization earlier than $z\sim6$ (i.e., $z>6.5$).

Given the small number of detected sources in our deep-continuum probe
and the large uncertainties on the IR luminosities of especially faint
sources, the ASPECS program represents but a first step.  With future
observations, we can probe dust emission from galaxies at even lower
masses in the high-redshift universe, while also probing much larger
areas to improve the overall statistics.

\acknowledgements

We thank an especially knowledgeable referee for their feedback, which
greatly improved our paper.  RJB acknowledges enlightening discussions
with Corentin Schreiber while writing this paper.  Seiji Fujimoto
provided us with some feedback on the public version of our paper
which greatly improved its overall clarity.  FW, IRS, and RJI
acknowledge support through ERC grants COSMIC–DAWN, DUSTYGAL, and
COSMICISM, respectively.  MA acknowledges partial support from
FONDECYT through grant 1140099.  IRS also acknowledges support from
STFC (ST/L00075X/1) and a Royal Society / Wolfson Merit award. Support
for RD and BM was provided by the DFG priority program 1573 ‘The
physics of the interstellar medium’. AK and FB acknowledge support by
the Collaborative Research Council 956, sub-project A1, funded by the
Deutsche Forschungsgemeinschaft (DFG).  FEB, LI, and JG-L acknowledge
support from CONICYT-Chile grants Basal-CATA PFB-06/2007.  FEB and
JG-L acknowledges support from FONDECYT Regular 1141218 (FEB,JG-L).
FEB also acknowledges support from ``EMBIGGEN'' Anillo ACT1101 (FEB)
and the Ministry of Economy, Development, and Tourism's Millennium
Science Initiative through grant IC120009, awarded to The Millennium
Institute of Astrophysics, MAS.  LI acknowledges Conicyt grants Anilo
ACT1417.  DR acknowledges support from the National Science Foundation
under grant number AST-⁠1614213 to Cornell University.  This paper
makes use of the ALMA data from the program 2013.1.00718.S. ALMA is a
partnership of ESO (representing its member states), NSF (USA) and
NINS (Japan), together with NRC (Canada), NSC and ASIAA (Taiwan), and
KASI (Republic of Korea), in cooperation with the Republic of
Chile. The Joint ALMA Observatory is operated by ESO, AUI/NRAO and
NAOJ.

\appendix

\begin{figure}
\epsscale{0.5} \plotone{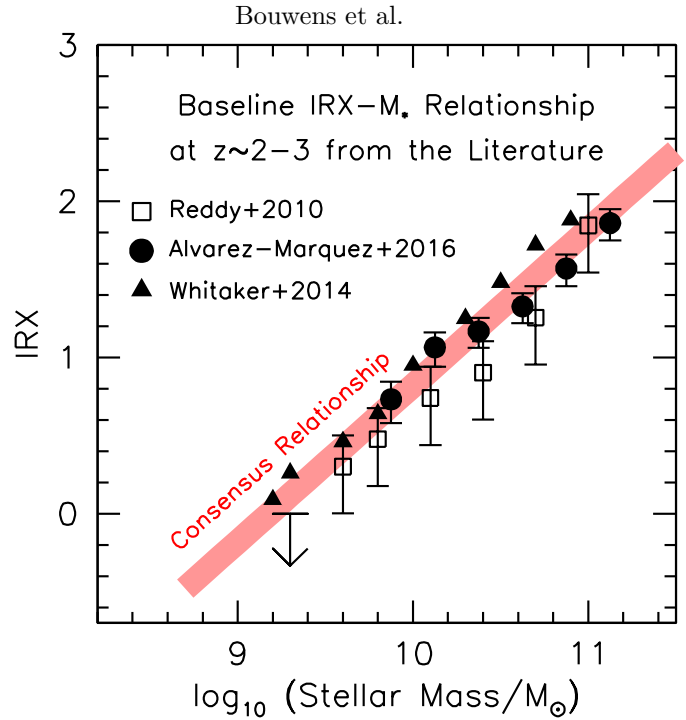}
\caption{Consensus relationship between the infrared excess of
  galaxies and their stellar mass at $z\sim2$-3.  The solid circles,
  open squares, and solid triangles show the results from
  {\'A}lvarez-M{\'a}rquez et al.\ (2016), Reddy et al.\ (2010), and
  Whitaker et al.\ (2014).  The thick light red line show the
  consensus relationship (Appendix A).\label{fig:lit}}
\end{figure}

\section{A.  Consensus Relationship between the Infrared Excess of
Galaxies at $z\sim2$-3 and their Inferred Stellar Masses}

There have been a large number of different measurements of the
infrared excess in galaxies as a function of their stellar mass from
$z\sim0$ to $z\sim3$ (e.g., Panella et al.\ 2009; Reddy et al.\ 2010;
Whitaker et al.\ 2014; {\'A}lvarez-M{\'a}rquez et al.\ 2016).  To
simplify the comparisons we make against these expectations, we aim to
derive a consensus determination of the infrared excess versus
apparent stellar mass that provide a reasonable representation of each
determination.  Towards this end, in Figure~\ref{fig:lit}, we present
three significant recent determinations of the infrared excess of
$z\sim2$-3 galaxies versus the stellar mass.  The results are
presented with the thick light red line and have the form:
\begin{equation}
\log_{10} IRX = \log_{10} [M/M_{\odot}]-9.17
\end{equation}
It is clear from this relationship provides an approximate match to
the results from each of the three studies (if some allowance is made
for modest systematics from one study to another).

\section{B.  Comparison Between MIPS-Inferred and ALMA-Inferred Infrared Luminosities}

Our ALMA continuum observations allow us to set important constraints
on dust emission from $z\gtrsim3$ galaxies.  However, these
constraints depend significantly on the assumptions we make regarding
the form of the far-IR SEDs.

As a sanity check on the present result, we derive independent
estimates of IR luminosities for $z=1.5$-3.0 galaxies in our selection
using deep 24$\mu$m MIPS observations over GOODS South.  The apparent
luminosity of $z\sim2$ galaxies in the 24$\mu$m band, i.e., rest-frame
8$\mu$m, is known to be well correlated with the IR luminosity of
distant galaxies.

To convert the measured 8$\mu$m luminosities of sources $(L_8)$ to the
equivalent luminosity in the IR, i.e., $L_{IR}$, we use the
prescription of Reddy et al.\ (2010).  For the highest luminosity
sources, the galaxy is assumed to be optically thick suggesting the
following formula:
\begin{equation}
\log[L_{IR}/L_{\odot}] = (0.95\pm0.10)\log_{10}[L_8/L_{\odot}]+(1.49\pm0.87)
\end{equation}
where we have modified the prescription from Reddy et al.\ (2010) to
be in terms of the IR luminosity instead of the SFR.  For lower
luminosity sources, it is more appropriate to assume that galaxies are
less optically thick.  Reddy et al.\ (2010) suggest the following
prescription in this case:
\begin{equation}
\log[L_{IR}/L_{\odot}] = (1.37\pm0.16)\log_{10}[L_8/L_{\odot}]-(3.01\pm1.34)
\end{equation}
The 24$\mu$m flux measurements we use for sources are from Whitaker et
al.\ (2014) and rely on the original MIPS data from the GOODS program.
They were derived using the same Mophongo software package (e.g., as
used in Labb{\'e} et al.\ 2006, 2010; Skelton et al.\ 2014) as we use
to perform IRAC photometry for samples in this paper.

The results are presented in Table~\ref{tab:mipsalma}, and it is clear
that there is a good correlation between the MIPS-derived IR
luminosities and the ALMA-derived luminosities assuming different dust
temperatures 35 K, 50 K, and also allowing for a monotonically
increasing dust temperature towards high redshift following the
results of Bethermin et al.\ (2015).

\begin{figure}
\epsscale{1.05}
\plotone{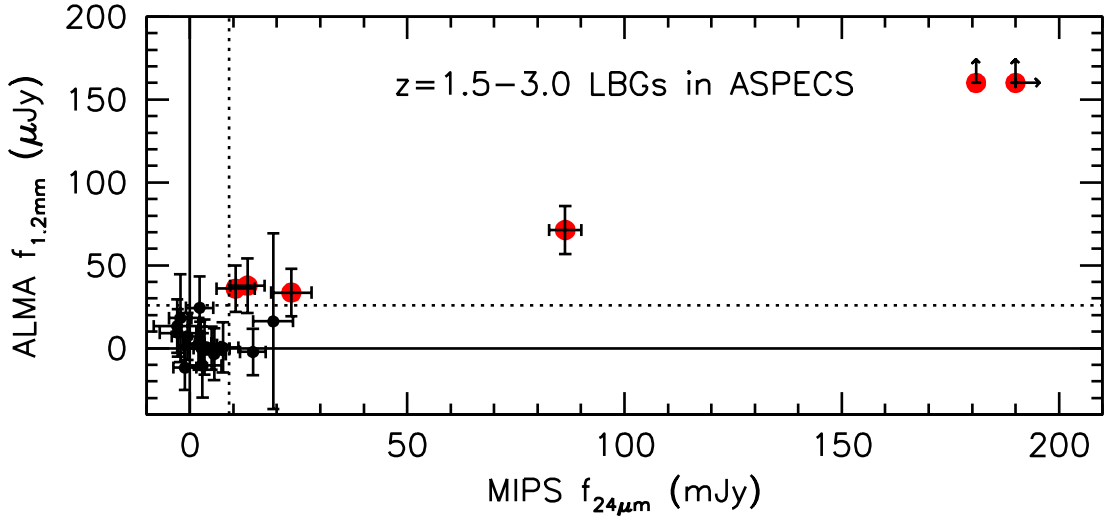}
\caption{Comparison of the observed MIPS 24$\mu$m fluxes (Whitaker et
  al.\ 2014) for sources in our $z\sim1.5$-3.0 LBG sample with the
  1.2$\,$mm continuum fluxes measured from ASPECS (solid black circles
  with $1\sigma$ error bars shown).  The larger solid red circles
  correspond to the sources we claim detections or tentative ALMA
  1.2$\,$mm detections.  The arrows indicate that the measured fluxes
  lie outside the bounds of the figure.  The horizontal and vertical
  dotted lines show the approximate flux levels where $2\sigma$
  detections are expected.  The MIPS 24$\mu$m fluxes exhibit a broad
  correlation with the ALMA 1.2$\,$mm fluxes for $z\sim1.5$-3.0
  galaxies, and those sources which are detected at $>2\sigma$ in the
  ALMA observations also tend to show significant $>$2$\sigma$
  detections in the MIPS observations (see
  Table~\ref{tab:mipsalma}).\label{fig:mips}}
\end{figure}

In general, the MIPS-derived IR luminosities we compute are $\sim$0.3
dex higher in general than our ALMA-derived IR luminosities, if we
adopt a fiducial dust temperature of 35 K, but agree better with these
luminosities if we allow the dust temperature to increase somewhat
towards higher redshift or adopt a dust temperature of 50 K.
Figure~\ref{fig:mips} illustrates the generally good correlation
between the observed 24$\mu$m fluxes and the ALMA 1.2$\,$mm fluxes.

Unfortunately, all aspects of this comparison are uncertain, from
source-to-source variations in $T_d$ (Elbaz et al.\ 2011; Genzel et
al.\ 2015; da Cunha et al.\ 2015), to the conversion from MIPS
24$\mu$m luminosities to IR luminosities, to the extension of these
results to fainter, lower-mass $z=2$-10 galaxies.

Because of these uncertainties, we persist with the 35 K dust
temperature preferred as a compromise between different studies and
compute IR luminosities using this assumption.  In making this
assumption, we should however realize that our results change by
$\sim$0.3-0.5 dex if the dust temperatures we have assumed are too low
or too high.

\begin{deluxetable}{ccccccc}
\tablecaption{Comparison between MIPS-Inferred and ALMA-Inferred IR Luminosities for Bright $z\sim1.5-3.0$ Galaxies Within ASPECS\label{tab:mipsalma}}
\tablehead{\colhead{} & \colhead{} & \colhead{} & \multicolumn{4}{c}{IR Luminosity ($10^{10}$ $L_{\odot}$)}\\
colhead{} & \colhead{} & \colhead{} & \colhead{} & \multicolumn{3}{c}{ALMA 1.2mm}\\
\colhead{ID} & \colhead{R.A.} & \colhead{DEC} & \colhead{MIPS $24\mu$m} & \colhead{$T_d=35$} & \colhead{Evolving $T_d$\tablenotemark{a}} & \colhead{$T_d=50$}}
\startdata
\multicolumn{7}{c}{Best Detected Sources in Our ALMA Observations}\\
XDFU-2370746171 & 03:32:37.07 & $-$27:46:17.1 & 20$\pm$5 & 6$\pm$2 & 9$\pm$3 & 25$\pm$8\\
XDFU-2397246112 & 03:32:39.72 & $-$27:46:11.2 & 82$\pm$1 & 50$\pm$5 & 54$\pm$5 & 222$\pm$22\\
XDFU-2373546453 & 03:32:37.35 & $-$27:46:45.3 & 57$\pm$2 & 14$\pm$3 & 17$\pm$4 & 61$\pm$13\\
XDFU-2385446340 & 03:32:38.54 & $-$27:46:34.0 & 242$\pm$4 & 97$\pm$2 & 152$\pm$3 & 401$\pm$8\\
XDFU-2365446123 & 03:32:36.54 & $-$27:46:12.3 & 5$\pm$2 & 7$\pm$3 & 9$\pm$4 & 30$\pm$13\\
XDFU-2384246348 & 03:32:38.42 & $-$27:46:34.8 & 12$\pm$7 & 6$\pm$2 & 10$\pm$3 & 25$\pm$8\\\\
\multicolumn{7}{c}{Other Sources over the ASPECS field}\\
XDFU-2378846451 & 03:32:37.88 & $-$27:46:45.1 & 0$\pm$1 & 0$\pm$3 & 0$\pm$4 & 0$\pm$13\\
XDFU-2379146261 & 03:32:37.91 & $-$27:46:26.1 & 1$\pm$3 & 0$\pm$2 & 0$\pm$3 & 0$\pm$8\\
XDFU-2379146261 & 03:32:37.91 & $-$27:46:26.1 & 3$\pm$4 & 0$\pm$2 & 0$\pm$3 & 0$\pm$8\\
XDFU-2393346236 & 03:32:39.33 & $-$27:46:23.6 & $-$1$\pm$2 & $-$2$\pm$2 & $-$3$\pm$3 & $-$8$\pm$8\\
XDFU-2387446541 & 03:32:38.74 & $-$27:46:54.1 & $-$2$\pm$2 & 3$\pm$4 & 5$\pm$7 & 12$\pm$16\\
XDFU-2366846484 & 03:32:36.68 & $-$27:46:48.4 & 1$\pm$1 & $-$2$\pm$4 & $-$2$\pm$5 & $-$9$\pm$17\\
XDFU-2370846470 & 03:32:37.08 & $-$27:46:47.0 & 2$\pm$1 & 0$\pm$3 & 0$\pm$4 & 0$\pm$13\\
XDFU-2358146436 & 03:32:35.81 & $-$27:46:43.6 & 9$\pm$3 & 3$\pm$10 & 4$\pm$12 & 13$\pm$43\\
XDFU-2369146348 & 03:32:36.91 & $-$27:46:34.8 & $-$0$\pm$1 & 2$\pm$3 & 2$\pm$4 & 9$\pm$13\\
XDFU-2374446154 & 03:32:37.44 & $-$27:46:15.4 & $-$0$\pm$1 & 1$\pm$3 & 1$\pm$4 & 4$\pm$13\\
XDFU-2363346155 & 03:32:36.33 & $-$27:46:15.5 & 2$\pm$2 & 0$\pm$3 & 0$\pm$4 & 0$\pm$13\\
XDFU-2382946284 & 03:32:38.29 & $-$27:46:28.4 & 4$\pm$1 & $-$0$\pm$3 & $-$0$\pm$4 & $-$0$\pm$13\\
XDFU-2366946210 & 03:32:36.69 & $-$27:46:21.0 & 1$\pm$1 & 0$\pm$3 & 0$\pm$4 & 0$\pm$13\\
XDFU-2369146023 & 03:32:36.91 & $-$27:46:02.3 & $-$1$\pm$3 & 3$\pm$3 & 4$\pm$4 & 13$\pm$13
\enddata
\tablenotetext{a}{Assuming dust temperature $T_d$ evolves as
  (35 K)$((1+z)/2.5)^{0.32}$ (Bethermin et al.\ 2015) such that $T_d
  \sim 44$-50 K at $z\sim4$-6.  See \S3.1.3.}
\end{deluxetable}

\section{C.  Converting the Present Results to that Appropriate for Other Assumed SEDs}

The IR luminosities and obscured SFRs we quote depend on the form
we assume for the far-IR SED.  To help our audience convert the
present results to the equivalent results for other assumed far-IR
SEDs, we have calculated multiplicative factors to allow for such
conversions.

We provide these conversion factors in Table~\ref{tab:convfact}.
Inspecting the results in this table, we can see that our $z\sim7$-10
results are the least sensitive to the assumed SED shape while our
$z\sim2$-3 results show the greatest dependence.  In general, the form
of the SED introduces a systematic uncertainty of $\sim$0.2 dex in the
overall results at $z\sim4$-10.

\section{D.  Comprehensive Presentation of Stack Results}

The purpose of this appendix is to provide a much more comprehensive
presentation of the stack results from ASPECS than is convenient for
the main text.  Tables~\ref{tab:irxsm}-\ref{tab:irxmuv} show our
results for $z\sim2$-10 samples split by stellar mass, $UV$-continuum
slope $\beta$, and apparent magnitude in the $UV$.  The stack results
are alternatively presented including or excluding individually
detected at $>$4$\sigma$ or which show evidence for an AGN
(XDFU-2397246112).

\begin{deluxetable*}{cccccccccc}
\tablewidth{0cm}
\tablecolumns{10}
\tabletypesize{\footnotesize}
\tablecaption{Multiplicative Factors to Convert Fiducial Results here for other model SEDs\label{tab:convfact}}
\tablehead{\colhead{} & \multicolumn{9}{c}{Conversion Factor to Apply to the Fiducial IR}\\
\colhead{} & \multicolumn{9}{c}{Luminosities and Limits ($L_{\odot}$) Derived Here\tablenotemark{a}}\\
\colhead{Far-IR SED Model} & \colhead{$z$$\sim$2} & \colhead{$z$$\sim$3} & \colhead{$z$$\sim$4} & \colhead{$z$$\sim$5} & \colhead{$z$$\sim$6} & \colhead{$z$$\sim$7} & \colhead{$z$$\sim$8} & \colhead{$z$$\sim$9} & \colhead{$z$$\sim$10}}
\startdata
35K modified blackbody\tablenotemark{b} (\textit{fiducial}) & 1.0 & 1.0 & 1.0 & 1.0 & 1.0 & 1.0 & 1.0 & 1.0 & 1.0 \\
Modified blackbody with evolving $T_d$\tablenotemark{b,c} & 1.3 & 1.8 & 2.2 & 2.6 & 2.8 & 2.7 & 2.3 & 1.9 & 1.5 \\
25K modified blackbody\tablenotemark{b} & 0.3 & 0.3 & 0.4 & 0.4 & 0.5 & 0.7 & 1.0 & 1.4 & 2.1 \\
30K modified blackbody\tablenotemark{b} & 0.5 & 0.6 & 0.6 & 0.6 & 0.7 & 0.8 & 0.9 & 1.0 & 1.2 \\
40K modified blackbody\tablenotemark{b} & 1.7 & 1.7 & 1.6 & 1.5 & 1.5 & 1.4 & 1.3 & 1.2 & 1.1 \\
45K modified blackbody\tablenotemark{b} & 2.8 & 2.6 & 2.5 & 2.3 & 2.1 & 1.9 & 1.7 & 1.5 & 1.3 \\
50K modified blackbody\tablenotemark{b} & 4.3 & 4.0 & 3.7 & 3.4 & 3.0 & 2.7 & 2.3 & 1.9 & 1.6 \\
NGC6946\tablenotemark{d} & 0.3 & 0.3 & 0.4 & 0.4 & 0.5 & 0.7 & 0.8 & 1.1 & 1.5 \\
M51\tablenotemark{d} & 0.3 & 0.3 & 0.4 & 0.4 & 0.5 & 0.7 & 0.8 & 1.1 & 1.4 \\
Arp220\tablenotemark{d} & 1.5 & 1.5 & 1.5 & 1.5 & 1.5 & 1.5 & 1.4 & 1.2 & 1.1 \\
M82\tablenotemark{d} & 2.4 & 2.4 & 2.4 & 2.3 & 2.3 & 2.2 & 2.1 & 1.9 & 1.7 
\enddata
\tablenotetext{a}{If the typical SED for star-forming galaxies at
  $z\gtrsim3$ are any of the following, the following multiplicative
  conversion factors should applied to fiducial IR luminosities
  presented throughout this paper.}
\tablenotetext{b}{Standard modified blackbody form (e.g., Casey 2012).}
\tablenotetext{c}{Assuming dust temperature $T_d$ evolves as (35 K)$((1+z)/2.5)^{0.32}$ (Bethermin et al.\ 2015).  See \S3.1.3.}
\tablenotetext{d}{Empirical SED template fits to these galaxies (Silva et al.\ 1998).}
\end{deluxetable*}

\begin{deluxetable*}{cccccccc}
\tablewidth{0cm}
\tablecolumns{8}
\tabletypesize{\footnotesize}
\tablecaption{Stacked Results: IRX versus Stellar Mass\label{tab:irxsm}}
\tablehead{
\colhead{} & \colhead{} & \colhead{$\log_{10}$} & \colhead{} & \colhead{Measured} & \colhead{Predicted $f_{1.2mm}$} & \colhead{} & \colhead{Measured}\\
\colhead{} & \colhead{\# of} & \colhead{M$_{wht}/$} & \colhead{} & \colhead{$f_{1.2mm}$ } & \colhead{flux [$\mu$Jy]} & \colhead{Measured} & \colhead{$f_{1.2mm}$/}\\
\colhead{Mass (M)} & \colhead{sources} & \colhead{M$_{\odot}$} & \colhead{$\beta_{wht}$} & \colhead{flux [$\mu$Jy]\tablenotemark{a,b}} & \colhead{Mass\tablenotemark{c,d}} & \colhead{IRX\tablenotemark{a,b,d}} & \colhead{$f_{UV}$\tablenotemark{a,b,e}}}
\startdata
\multicolumn{8}{c}{$z=2$-3} \\
\multicolumn{6}{c}{} & (if $T_d$ higher at $z>2$,\tablenotemark{f} \\
\multicolumn{6}{c}{} & multiply by $\sim$1.5-2.0$\times$)\\
$>10^{9.75} M_{\odot}$ & 11 & 10.0 & $-$1.4 & 104$_{-65}^{+91}$$\pm$7 & 261 & 3.80$_{-2.40}^{+3.61}$$\pm$0.19 & 120$_{-62}^{+92}$$\pm$6\\
$>10^{9.75} M_{\odot}$ & 10 & 9.9 & $-$1.5 & 103$_{-68}^{+98}$$\pm$7 & 254 & 3.68$_{-2.55}^{+3.86}$$\pm$0.19 & 112$_{-67}^{+94}$$\pm$6\\
(excluding AGN) \\
$>10^{9.75} M_{\odot}$ & 8 & 9.9 & $-$1.6 & 28$_{-4}^{+4}$$\pm$9 & 224 & 0.74$_{-0.15}^{+0.25}$$\pm$0.21 & 29$_{-3}^{+4}$$\pm$8\\
(excluding $\geq4\sigma$ \\
detected sources) \\
$10^{9.25} M_{\odot}$ - $10^{9.75} M_{\odot}$ & 11 & 9.4 & $-$1.6 & 8$_{-10}^{+5}$$\pm$9 & 68 & 0.12$_{-0.52}^{+0.29}$$\pm$0.40 & 4$_{-10}^{+8}$$\pm$11\\
$10^{8.75} M_{\odot}$ - $10^{9.25} M_{\odot}$ & 24 & 9.0 & $-$1.7 & 9$_{-4}^{+6}$$\pm$5 & 6 & 1.77$_{-1.02}^{+1.15}$$\pm$0.94 & 70$_{-35}^{+46}$$\pm$34\\
$< 10^{8.75} M_{\odot}$ & 116 & 8.4 & $-$1.9 & $-$6$_{-6}^{+6}$$\pm$4 & 1 & $-$1.41$_{-1.19}^{+1.35}$$\pm$0.90 & $-$57$_{-56}^{+55}$$\pm$38\\
$< 10^{9.75} M_{\odot}$ & 151 & 9.2 & $-$1.7 & 6$_{-6}^{+4}$$\pm$7 & 50 & 0.11$_{-0.42}^{+0.32}$$\pm$0.34 & 6$_{-10}^{+9}$$\pm$10\\\\
\multicolumn{8}{c}{$z=4$-10} \\
\multicolumn{6}{c}{} & (if $T_d$ higher at $z>2$,\tablenotemark{f} \\
\multicolumn{6}{c}{} & multiply by $\sim$2.5-3.0$\times$)\\
$M>10^{9.75} M_{\odot}$ & 2 & 9.8 & $-$1.7 & $-$4$_{-14}^{+9}$$\pm$13 & 118 & $-$0.49$_{-1.13}^{+0.69}$$\pm$0.71 & $-$41$_{-44}^{+56}$$\pm$45\\
$10^{9.25} M_{\odot}$ - $10^{9.75} M_{\odot}$ & 5 & 9.4 & $-$1.5 & 15$_{-8}^{+3}$$\pm$13 & 117 & 0.33$_{-0.17}^{+0.10}$$\pm$0.30 & 27$_{-26}^{+13}$$\pm$31\\
$10^{8.75} M_{\odot}$ - $10^{9.25} M_{\odot}$ & 19 & 9.0 & $-$1.9 & $-$1$_{-4}^{+10}$$\pm$11 & 29 & 0.13$_{-0.17}^{+0.44}$$\pm$0.27 & 23$_{-21}^{+18}$$\pm$34\\
$< 10^{8.75} M_{\odot}$ & 142 & 8.4 & $-$1.9 & $-$2$_{-1}^{+3}$$\pm$7 & 5 & $-$0.17$_{-0.32}^{+0.31}$$\pm$0.40 & $-$15$_{-57}^{+57}$$\pm$65\\
$< 10^{9.75} M_{\odot}$ & 166 & 9.0 & $-$1.8 & 4$_{-5}^{+5}$$\pm$7 & 56 & 0.14$_{-0.14}^{+0.15}$$\pm$0.18 & 21$_{-16}^{+11}$$\pm$22\\\\
\multicolumn{8}{c}{$z=2$-10} \\
\multicolumn{6}{c}{} & (if $T_d$ higher at $z>2$,\tablenotemark{f} \\
\multicolumn{6}{c}{} & multiply by $\sim$2.5$\times$)\\
$< 10^{9.75} M_{\odot}$ & 317 & 9.1 & $-$1.8 & 5$_{-4}^{+4}$$\pm$6 & 55 & 0.14$_{-0.13}^{+0.14}$$\pm$0.16 & 8$_{-7}^{+7}$$\pm$9\\
$< 10^{9.25} M_{\odot}$ & 301 & 8.8 & $-$1.9 & $-$1$_{-3}^{+4}$$\pm$7 & 19 & 0.04$_{-0.16}^{+0.18}$$\pm$0.21 & 15$_{-21}^{+21}$$\pm$19
\enddata
\tablenotetext{a}{This column presents stack results.  Each source is
  weighted according to the square of its expected 1.2$\,$mm signal in
  our continuum observations (assuming $L_{IR} \propto L_{UV}$) and the
  inverse square of the noise.  The weightings are therefore independent
  of stellar mass and $UV$-continuum slope $\beta$.}
\tablenotetext{b}{Both the bootstrap and formal uncertainties are quoted on the result (presented first and second, respectively).}
\tablenotetext{c}{The 1.2$\,$mm continuum flux predicted from the consensus $z\sim2$-3 IRX-stellar mass relationship weighting individual
sources in exactly the same way as for the measured 1.2$\,$mm continuum flux.  This column should therefore be directly comparable with
the column directly to the left, i.e., giving the measured flux.}
\tablenotetext{d}{Assuming a standard modified blackbody SED with dust
  temperature of 35 K and accounting for the impact of the CMB on the
  measured flux (da Cunha et al.\ 2013).}
\tablenotetext{e}{Results do not dependent on the assumed far-IR SED template.}
\tablenotetext{f}{The suggested multiplicative factors are for the
  scenario that the dust temperature $T_d$ evolves as (35
  K)$((1+z)/2.5)^{0.32}$ (Bethermin et al.\ 2015) such that
  $T_d\sim44$-50 K at $z\sim4$-6.  See \S3.1.3.}
\end{deluxetable*}

\begin{deluxetable*}{cccccccccc}
\tablewidth{0cm}
\tablecolumns{10}
\tabletypesize{\footnotesize}
\tablecaption{IRX versus $\beta$\label{tab:irxbeta}}
\tablehead{\colhead{} & \colhead{} & \colhead{$\log_{10}$} & \colhead{} & \colhead{Measured} & \multicolumn{2}{c}{Predicted} & \colhead{} & \colhead{} & \colhead{Measured}\\
\colhead{} & \colhead{\# of} & \colhead{M$_{wht}/$} & \colhead{} & \colhead{$f_{1.2mm}$ } & \multicolumn{2}{c}{$f_{1.2mm}$ [$\mu$Jy]} & \colhead{Measured} &  \colhead{Predicted} & \colhead{$f_{1.2mm}$/}\\
\colhead{$\beta$} & \colhead{sources} & \colhead{M$_{\odot}$} & \colhead{$\beta_{med}$} & \colhead{[$\mu$Jy]\tablenotemark{a,b}} & \colhead{Calz\tablenotemark{c,d}} & \colhead{SMC\tablenotemark{c,d}} & \colhead{IRX\tablenotemark{a,b,d}} & \colhead{IRX$_{SMC}$\tablenotemark{c}} & \colhead{$f_{UV}$\tablenotemark{a,b,e}}}
\startdata
\multicolumn{10}{c}{$z=2$-3 (All Masses)} \\
\multicolumn{7}{c}{} & (if $T_d$ higher at $z>2$,\tablenotemark{f} \\
\multicolumn{7}{c}{} & multiply by $\sim$1.5-2.0$\times$)\\
$-4.0 < \beta < -1.75$ & 90 & 9.6 & $-$1.9 & 20$_{-13}^{+3}$$\pm$10 & 49 & 12 & 0.43$_{-0.32}^{+0.12}$$\pm$0.24 & 0.34 & 20$_{-8}^{+5}$$\pm$10\\
$-1.75 < \beta < -1.25$ & 51 & 9.5 & $-$1.4 & 10$_{-7}^{+7}$$\pm$8 & 183 & 19 & 0.51$_{-0.47}^{+0.47}$$\pm$0.45 & 1.27 & 12$_{-15}^{+13}$$\pm$12\\
$-1.25 < \beta$ & 21 & 10.1 & $-$1.0 & 176$_{-122}^{+156}$$\pm$8 & 512 & 71 & 6.69$_{-4.55}^{+6.14}$$\pm$0.26 & 2.70 & 161$_{-103}^{+158}$$\pm$7\\
$-1.25 < \beta$  & 20 & 10.1 & $-$1.0 & 175$_{-129}^{+150}$$\pm$8 & 494 & 70 & 6.47$_{-4.90}^{+5.88}$$\pm$0.26 & 2.62 & 151$_{-102}^{+163}$$\pm$7\\
(excluding AGN) \\
$-1.25 < \beta$  & 18 & 10.1 & $-$1.1 & 32$_{-32}^{+4}$$\pm$12 & 368 & 62 & 0.87$_{-0.94}^{+0.17}$$\pm$0.35 & 2.14 & 29$_{-17}^{+4}$$\pm$11\\
(excluding $\geq4\sigma$ \\
individual detections)\\
All & 162 & 9.8 & $-$1.5 & 82$_{-49}^{+63}$$\pm$6 & 145 & 37 & 2.96$_{-1.80}^{+2.53}$$\pm$0.16 & 1.41 & 92$_{-49}^{+63}$$\pm$5\\\\

\multicolumn{9}{c}{$z=2$-3 ($>10^{9.75} M_{\odot}$)} \\
$-4.0 < \beta < -1.75$ & 1 & 9.8 & $-$2.0 & 26$_{-0}^{+0}$$\pm$14 & 48 & 13 & 0.54$_{-0.00}^{+0.00}$$\pm$0.29 & 0.49 & 23$_{-0}^{+0}$$\pm$13\\
$-1.75 < \beta < -1.25$ & 2 & 9.8 & $-$1.4 & 22$_{-16}^{+11}$$\pm$13 & 163 & 24 & 1.31$_{-0.94}^{+0.67}$$\pm$0.72 & 2.32 & 40$_{-17}^{+1}$$\pm$19\\
$-1.25 < \beta$ & 8 & 10.1 & $-$1.0 & 179$_{-121}^{+135}$$\pm$8 & 519 & 72 & 6.79$_{-4.51}^{+5.38}$$\pm$0.26 & 4.72 & 164$_{-91}^{+147}$$\pm$7\\
(excluding AGN) \\
$-1.25 < \beta$ & 5 & 10.1 & $-$1.1 & 33$_{-33}^{+3}$$\pm$12 & 376 & 63 & 0.90$_{-0.82}^{+0.12}$$\pm$0.36 & 2.13 & 29$_{-23}^{+2}$$\pm$11\\
(excluding $\geq4\sigma$ \\
individual detections)\\\\
\multicolumn{10}{c}{$z=2$-3 ($<10^{9.75} M_{\odot}$)} \\
$-4.0 < \beta < -1.75$ & 89 & 9.2 & $-$1.9 & 8$_{-13}^{+5}$$\pm$10 & 51 & 11 & 0.19$_{-0.75}^{+0.40}$$\pm$0.44 & 0.48 & 15$_{-21}^{+12}$$\pm$16\\
$-1.75 < \beta < -1.25$ & 49 & 9.3 & $-$1.4 & 2$_{-3}^{+4}$$\pm$10 & 196 & 16 & $-$0.01$_{-0.36}^{+0.36}$$\pm$0.58 & 1.24 & $-$4$_{-8}^{+8}$$\pm$15\\
$-1.25 < \beta$ & 13 & 9.0 & $-$1.0 & $-$6$_{-9}^{+12}$$\pm$9 & 78 & 10 & $-$0.14$_{-3.45}^{+5.04}$$\pm$2.11 & 2.58 & 22$_{-47}^{+92}$$\pm$51\\
\multicolumn{10}{c}{$z=4$-10 (All Masses)} \\
\multicolumn{7}{c}{} & (if $T_d$ higher at $z>2$,\tablenotemark{f} \\
\multicolumn{7}{c}{} & multiply by $\sim$2.5-3.0$\times$)\\
$-4.0 < \beta < -1.75$ & 123 & 8.9 & $-$2.0 & $-$2$_{-3}^{+5}$$\pm$8 & 40 & 9 & 0.05$_{-0.13}^{+0.22}$$\pm$0.23 & 0.26 & 9$_{-19}^{+18}$$\pm$28\\
$-1.75 < \beta < -1.25$ & 29 & 9.4 & $-$1.5 & 14$_{-8}^{+4}$$\pm$12 & 310 & 44 & 0.33$_{-0.14}^{+0.11}$$\pm$0.29 & 1.19 & 37$_{-10}^{+16}$$\pm$31\\
$-1.25 < \beta$ & 12 & 9.4 & $-$1.0 & $-$14$_{-4}^{+8}$$\pm$9 & 141 & 23 & $-$1.39$_{-0.32}^{+0.41}$$\pm$0.90 & 2.53 & $-$89$_{-75}^{+3}$$\pm$58\\
All & 168 & 9.1 & $-$1.8 & 4$_{-4}^{+5}$$\pm$6 & 145 & 23 & 0.10$_{-0.13}^{+0.13}$$\pm$0.18 & 0.70 & 9$_{-16}^{+15}$$\pm$19\\\\
\multicolumn{10}{c}{$z=2$-10 ($<10^{9.75} M_{\odot}$)} \\
\multicolumn{7}{c}{} & (if $T_d$ higher at $z>2$,\tablenotemark{f} \\
\multicolumn{7}{c}{} & multiply by $\sim$2.5$\times$)\\
$-4.0 < \beta < -1.75$ & 211 & 8.9 & $-$2.0 & 0$_{-4}^{+5}$$\pm$7 & 43 & 9 & 0.07$_{-0.18}^{+0.20}$$\pm$0.21 & 0.31 & 13$_{-16}^{+10}$$\pm$14\\
$-1.75 < \beta < -1.25$ & 78 & 9.3 & $-$1.5 & 12$_{-7}^{+4}$$\pm$10 & 287 & 39 & 0.27$_{-0.17}^{+0.11}$$\pm$0.26 & 1.20 & 3$_{-9}^{+8}$$\pm$13\\
$-1.25 < \beta$ & 24 & 8.8 & $-$0.8 & $-$5$_{-4}^{+3}$$\pm$8 & 141 & 19 & $-$0.74$_{-1.49}^{+1.32}$$\pm$1.20 & 3.24 & 13$_{-48}^{+75}$$\pm$49
\enddata
\tablenotetext{a}{This column presents stack results.  Each source is
  weighted according to the square of its expected 1.2-$mm$ signal in
  our continuum observations (assuming $L_{IR} \propto L_{UV}$) and the
  inverse square of the noise.  The weightings are therefore independent
  of stellar mass and $UV$-continuum slope $\beta$.}
\tablenotetext{b}{Both the bootstrap and formal uncertainties are quoted on the result (presented first and second, respectively).}
\tablenotetext{c}{The 1.2$\,$mm continuum flux predicted using the M99 or SMC IRX-$\beta$ relationship weighting individual
sources in exactly the same way as for the measured 1.2$\,$mm continuum flux, so these two quantities should be directly comparable.}
\tablenotetext{d}{Assuming a standard modified blackbody SED with dust
  temperature of 35 K and accounting for the impact of the CMB on the
  measured flux (da Cunha et al.\ 2013).}
\tablenotetext{e}{Results do not dependent on the assumed far-IR SED template.}
\tablenotetext{f}{The suggested multiplicative factors are for the scenario that the dust temperature $T_d$ evolves as (35 K)$((1+z)/2.5)^{0.32}$ (Bethermin et al.\ 2015).  See \S3.1.3.}
\end{deluxetable*}

\begin{deluxetable*}{cccccccccc}
\tablewidth{0cm}
\tablecolumns{10}
\tabletypesize{\footnotesize}
\tablecaption{IRX versus Apparent Magnitude in the Rest-frame $UV$ ($m_{UV,AB}$)\label{tab:irxmuv}}
\tablehead{
\colhead{} & \colhead{} & \colhead{$\log_{10}$} & \colhead{} & \colhead{Measured } & \multicolumn{3}{c}{Predicted} & \colhead{} & \colhead{}\\
\colhead{} & \colhead{\# of} & \colhead{$M_{med}$/} & \colhead{} & \colhead{$f_{1.2mm}$ } & \multicolumn{3}{c}{$f_{1.2mm}$ [$\mu$Jy]} & \colhead{} & \colhead{$f_{1.2mm}$/}\\
\colhead{$m_{UV}$} & \colhead{sources} & \colhead{$M_{\odot}$} & \colhead{$\beta_{med}$} & \colhead{[$\mu$Jy]\tablenotemark{a}} & \colhead{Calz\tablenotemark{a}} & \colhead{SMC\tablenotemark{a}} & \colhead{Mass\tablenotemark{a}} & \colhead{IRX\tablenotemark{a}} & \colhead{$f_{UV}$\tablenotemark{a}}}
\startdata
\multicolumn{10}{c}{$z=2$-3} \\
\multicolumn{8}{c}{} & (if $T_d$ higher at $z>2$,\tablenotemark{b} \\
\multicolumn{8}{c}{} & multiply by $\sim$1.5-2.0$\times$)\\
$<25$ & 12 & 9.9 & $-$1.5 & 95$_{-56}^{+79}$$\pm$7 & 275 & 40 & 239 & 3.45$_{-2.09}^{+3.23}$$\pm$0.18 & 100$_{-52}^{+69}$$\pm$5\\
$<25$ (excluding AGN) & 11 & 9.9 & $-$1.5 & 94$_{-56}^{+83}$$\pm$7 & 266 & 40 & 233 & 3.34$_{-2.13}^{+3.27}$$\pm$0.18 & 93$_{-55}^{+75}$$\pm$5\\
$<25$ (excluding $\geq4\sigma$ & 9 & 9.8 & $-$1.7 & 26$_{-4}^{+4}$$\pm$8 & 163 & 29 & 201 & 0.71$_{-0.13}^{+0.19}$$\pm$0.20 & 24$_{-5}^{+4}$$\pm$6\\
detected sources)\\
25-31 & 150 & 9.3 & $-$1.5 & $-$1$_{-3}^{+3}$$\pm$4 & 115 & 14 & 46 & $-$0.20$_{-0.49}^{+0.44}$$\pm$0.45 & $-$4$_{-20}^{+19}$$\pm$18\\
All & 162 & 9.8 & $-$1.5 & 82$_{-49}^{+64}$$\pm$6 & 253 & 37 & 213 & 2.96$_{-1.83}^{+2.56}$$\pm$0.16 & 92$_{-48}^{+62}$$\pm$5\\\\
\multicolumn{10}{c}{$z=4$-10} \\
\multicolumn{8}{c}{} & (if $T_d$ higher at $z>2$,\tablenotemark{b} \\
\multicolumn{8}{c}{} & multiply by $\sim$2.5-3.0$\times$)\\
$<26$ & 7 & 9.1 & $-$1.9 & $-$3$_{-3}^{+6}$$\pm$10 & 64 & 13 & 43 & $-$0.09$_{-0.13}^{+0.17}$$\pm$0.26 & $-$11$_{-23}^{+19}$$\pm$26\\
26-31 & 160 & 8.8 & $-$1.8 & 4$_{-4}^{+4}$$\pm$5 & 56 & 11 & 16 & 0.20$_{-0.24}^{+0.26}$$\pm$0.30 & 30$_{-41}^{+43}$$\pm$48\\
All & 168 & 9.1 & $-$1.8 & 4$_{-5}^{+5}$$\pm$6 & 145 & 23 & 60 & 0.10$_{-0.14}^{+0.14}$$\pm$0.18 & 9$_{-19}^{+14}$$\pm$19
\enddata
\tablenotetext{a}{Calculated identically to the columns in Table~\ref{tab:irxsm}, but using the subdivisions of sources indicated in the rows of this table.}
\tablenotetext{b}{The suggested multiplicative factors are for the
  scenario that the dust temperature $T_d$ evolves as (35
  K)$((1+z)/2.5)^{0.32}$ (Bethermin et al.\ 2015) such that
  $T_d\sim44$-50 K at $z\sim4$-6.  See \S3.1.3.}
\end{deluxetable*}
\end{document}